\documentclass[10pt,letterpaper,twoside,fleqn]{article}
\usepackage{amsmath}
\usepackage{amssymb}
\usepackage{amstext}
\usepackage{theorem}
\usepackage{mathrsfs}
\usepackage{graphics}
\usepackage{ulem}
\usepackage{moredefs}
\usepackage[mla]{lips}
\usepackage{csquotes}
\usepackage{epstopdf}
\usepackage[pdftex]{graphicx}
\usepackage{epsfig}
\usepackage{caption}
\usepackage{subcaption}

\numberwithin{equation}{section} \numberwithin{figure}{section}

\theorembodyfont{\rmfamily}
\newtheorem{theorem}{Theorem}[section]

\theorembodyfont{\rmfamily}
\newtheorem{lemma}[theorem]{Lemma}

\theorembodyfont{\rmfamily}
\newtheorem{corollary}[theorem]{Corollary}

\theorembodyfont{\rmfamily}
\newtheorem{proposition}[theorem]{Proposition}

\theorembodyfont{\rmfamily}
\newtheorem{defi}[theorem]{Definition}

\theorembodyfont{\rmfamily}
\newtheorem{remark}[theorem]{Remark}

\theorembodyfont{\rmfamily}

\theorembodyfont{\rmfamily}

\newenvironment{lquote}{\begin{list}{}{}\item[]}{\end{list}}

\newcommand{\be}{\begin{eqnarray}}
\newcommand{\ee}{\end{eqnarray}}
\newcommand{\nn}{\nonumber}

\newcommand{\unity}{{\setlength{\unitlength}{1em}
\begin{picture}(0.75,0.75)
\put(0,0){$1$} \put(.38,0){\line(0,1){0.65}}
\end{picture}}}

\newcommand{\Done}{${}$\rightline{$\blacksquare$}\\ }
\newcommand{\done}{${}$\rightline{$\square$}\\ }
\newcommand{\bit}{\begin{it}}
\newcommand{\eit}{\end{it}}

\newcommand{\ea}{\mbox{$e^{-ia_f}$}}
\newcommand{\eb}{\mbox{$e^{b_f}$}}

\newcommand{\unif}{uniformly for $z\in M$}

\newcommand{\lima}{\mbox{$\lim\limits_{\alpha\downarrow0}$}}
\newcommand{\psia}{\mbox{$\widetilde{\psi}_{\alpha,f}$}}

\title{Instability of pre-existing resonances under a small constant electric field\footnotemark[3] \footnotetext[3]{This is a prior version (preprint version of March 29, 2014) of the article ``Instability of Pre-Existing Resonances Under a Small Constant Electric Field'', Annales Henri Poincar\a'e 16, 2783 - 2835 (2015), \linebreak DOI 10.1007/s00023-014-0389-2. The final publication is available at http://link.springer.com}}
\author{I. Herbst\footnote{University of Virginia, Department of Mathematics} \and J. Rama{\footnotemark[1]\ \footnote{Supported by the Deutsche Forschungsgemeinschaft (DFG), research grants RA 2020/1-1 and RA 2020/1-2.}}}
\date{March 29, 2014}

\begin{document}
\maketitle

\begin{abstract}
\noindent Two simple model operators are considered which have pre-existing resonances. A potential corresponding to a small electric field, $f$, is then introduced and the resonances of the resulting operator are considered as $f \rightarrow 0$. It is shown that these resonances are not continuous in this limit.  It is conjectured that a similar behavior will appear in more complicated models of atoms and molecules. Numerical results are presented.
\end{abstract}

\section{Introduction and Results}

It has long been known that atomic bound states below the continuum turn into resonances when subjected to a weak constant electric field. The resonances move continuously as a function of the field strength, $f$, for small $|f|$ and converge to the original bound state as $f\to 0$; see for example \cite{CRA, GG1,GG2, HaSi, He, HeSi1, HeSi2,  R, YTS} and references given there. A simple, natural question arises: Is the same true for pre-existing atomic resonances? More precisely, suppose an atom has a resonance in the lower half complex plane, just below and near the continuous spectrum. The resonance might be for example a shape resonance or due to a broken symmetry.  Suppose the resonance is defined, using for example the dilation analytic framework, as a pole in matrix elements of the resolvent between dilation analytic vectors (in the sense of Definition \ref{defi:resonance} below). Suppose then a term is added to the Hamiltonian representing an electric field of strength $f>0$. One can then ask whether there exist new resonance poles near the original atomic resonance $r_0$ and whether the new poles converge to $r_0$ as $f\downarrow0$.
At first one might expect that methods entirely similar to those used for the problem of bound states turning into resonances should be sufficient to treat this problem. But a hint as to the difficulty which is encountered is the following: Let
\be H_0(f,\theta):=e^{-2i\theta}p^2+fxe^{i\theta}\hspace{0.5cm}(\theta,x\in\mathbb{R},\ f>0)\nn\ee
be the complex dilated Hamiltonian in $L^2(\mathbb{R})$ without potential, which (by \cite[Theorem II.1]{He}) has empty spectrum if $|\theta|\in(0,\frac{\pi}{3})$. Then if $f>0$ and $z$ is in the numerical range $\{z\in\mathbb{C}\,|\,\textrm{Im\,}(e^{-i\theta}z)<0\}$ of $H_0(f,\theta)$ one has \be \|(H_0(f,\theta)-z)^{-1}\|\geq c_1\,e^{c_2/f}\nn\ee for some strictly positive constants $c_1$ and $c_2$; see \cite[Proposition II.6]{He}.  (This lower bound is true in any dimension if $x$ is replaced with a component of $x$.) Thus from a technical standpoint there is a loss of control of relevant resolvents for the spectral parameter near the original resonance.

In this paper we consider two simple models with pre-existing resonances. For these models we prove (see Theorem \ref{theorem:resonances_instable} and Theorem \ref{theorem:resonance_instable_Mod2}) that pre-existing resonances, when perturbed by a constant electric field of strength $f>0$ (DC Stark effect), are unstable in the weak field limit $f\downarrow0$. Roughly speaking, this is the fact that the analytically continued resolvent does not converge to the resolvent obtained by first taking $f$ to 0 and then analytically continuing. In other words, taking the limit $f\downarrow0$ does not commute with analytic continuation to the second Riemann sheet. In Section \ref{section:numerical_analysis} we present some numerical results illustrating this instability result.  Furthermore, for one of our models (Friedrichs model) we show that if the direct current field is replaced by an alternating current field (AC Stark effect) pre-existing resonances are stable in the weak field limit (see Theorem \ref{theorem:AC}). We conjecture similar instability results in the DC case for more complicated models of atoms and molecules.\bigskip\\
\indent We introduce some notation: Let
\be S_\gamma:=\{z\in\mathbb{C}\,|\,|\textrm{Im\,}z|<\gamma\}\quad(\gamma>0),
\hspace{0.5cm}\mathbb{C}_\pm:=\{z\in\mathbb{C}\,|\,\textrm{Im\,}z\gtrless0\}\label{strip}\ee and $\overline{\mathbb{C}_\pm}$ be the closure of $\mathbb{C}_\pm$.  By $(\,\cdot\,,\,\cdot\,)_X$ we denote an inner product in a linear space $X$. Let
$\stackrel{\wedge}{g}$ denote the Fourier transform of a function $g$, and $\stackrel{\vee}{g}$ its inverse, provided they exist. The symbol $\sqrt{\cdot}$ denotes the principal branch of the square root with branch cut $(-\infty,0)$. In particular, $\sqrt{z}:=\big|\,|z|^{1/2}\big|\;e^{\pm i\frac{\phi}{2}}$ $(z\in\mathbb{C}_\pm)$ for some $\phi\in(0,\pi)$.\smallskip\\

Our models use analyticity in the following sense:
\begin{defi}\label{defi:dilation_analytic_vector}${}$
\begin{enumerate}\item
The unitary group of dilations on $L^2(\mathbb{R})$ is defined by
\be (U(\theta)\psi)(x):= e^{\theta/2}\psi(e^{\theta}x)\hspace{0.5cm}(\theta\in\mathbb{R},\ \psi\in L^2(\mathbb{R})).\label{unitary_group}\ee
If $U(\theta)\psi$ has an $L^2$-valued analytic extension (in the variable $\theta$) to a strip $S_{\theta_0}$ for some $\theta_0>0$, the function $\psi$ is called  dilation analytic (in angle $\theta_0$). \\
We denote the dense subspace of $L^2$-functions which are dilation analytic in angle $\theta_0$ by $\mathcal{D}_{\theta_0}$.
\item
Let $k_0>0$,
\begin{align}
\mathcal{T}_{k_0}:=\big\{\psi\in L^2(\mathbb{R})\,\big|&\,\stackrel{\wedge}{\psi} \textrm{ has an analytic extension to the strip }S_{k_0}\nn\\ &
\textrm{ and }\sup\limits_{k\in S_{k_0}}|\stackrel{\wedge}{\psi}(k)|<\infty\big\}.\label{defiTk_0}
\end{align}
\end{enumerate}
\end{defi}
We remark that for any $k_0>0$, $\mathcal{T}_{k_0}$ is a dense subspace of $L^2(\mathbb{R})$. Note that \eqref{defiTk_0} is just a pointwise analyticity condition on $\stackrel{\wedge}{\psi}$ and not an $L^2$-condition (as needed in the classical Paley-Wiener Theorem for $L^2$-functions; see, e.g., \cite[Theorem IX.13]{RS2}). In particular, \eqref{defiTk_0} does not imply exponential decay of $\psi$. \\
Appendix \ref{dilation analytic vectors} contains some results on boundedness of dilation analytic vectors in a sector.

\subsection{Model I: Friedrichs model}\label{section:first_model}
\subsubsection{Friedrichs model with DC Stark effect}
In the first model the Hilbert space is
\be
\mathscr{H}:=L^2(\mathbb{R})\oplus\mathbb{C}\label{Hilbert_space}\ee with inner product
\be (u,u)_{\mathscr{H}}:=(u_1,u_1)_{L^2}+(u_2,u_2)_{\mathbb{C}}\hspace{0.5cm}(u:=\begin{pmatrix}u_1\\ u_2\end{pmatrix}\in L^2(\mathbb{R})\oplus\mathbb{C}).\nn\ee
We start out with a Hamiltonian with a simple eigenvalue, 1, embedded in its continuous spectrum:
\be H_0:=\begin{pmatrix}p^2 & 0 \\[1.5ex] 0 & 1 \end{pmatrix}\,,
\label{H_0}\ee
where $p$ denotes the self-adjoint realization of $-i d/dx$ in $L^2(\mathbb{R})$.
We then add a small rank-2 perturbation which removes the eigenvalue and turns it into a nearby resonance (in the sense of Definition \ref{defi:resonance} below):
\be H_\varphi:=\begin{pmatrix}p^2 & M_\varphi \\[1.5ex] (\varphi\,,\,\cdot\,)_{L^2} & 1 \end{pmatrix}\hspace{0.5cm}(\varphi\in\mathcal{D}_{\theta_0}\cup\mathcal{T}_{k_0}\textrm{ for some $\theta_0,k_0>0$})\,,\label{H}\ee
where $\theta_0$ and $k_0$ depend (in a certain sense specified below) on the size of $\varphi$. $M_\varphi:\mathbb{C}\to L^2(\mathbb{R})$  denotes the multiplication operator generated by the function $\varphi$.

We now add the external electric field to the problem to obtain the Hamiltonian \be H_\varphi(f):=\begin{pmatrix} p^2+f x & M_\varphi \\[1.5ex] (\varphi\,,\,\cdot\,)_{L^2} & 1  \end{pmatrix}\hspace{0.5cm}(x\in\mathbb{R},\ f\geq0,\ \varphi\in\mathcal{D}_{\theta_0}\cup\mathcal{T}_{k_0})\,.\label{H(f)}\ee We are interested in $f>0$ small.\\

\noindent The operators $H_0$ and $H_\varphi(f)$ $(f\geq0)$, $H_\varphi(0)=H_\varphi$ are essentially self adjoint on $C_0^{\infty}(\mathbb{R})\oplus\mathbb{C}$.

With the abbreviations \be R_f(z):=(p^2+fx-z)^{-1}\,,\quad F_{f,\varphi}(z):=1-z-(\varphi\,,\,R_f(z)\varphi)_{L^2}\label{F_f}\ee
we compute for later use
\be \hspace{-1cm}(H_\varphi(f)-z)^{-1}=\begin{pmatrix}R_f(z)\big(1 +F_{f,\varphi}(z)^{-1}(\varphi\,,\,R_f(z)\,\cdot\,)_{L^2}\varphi\big) & -F_{f,\varphi}(z)^{-1}M_{R_f(z)\varphi}\\[1.5ex] -F_{f,\varphi}(z)^{-1}(\varphi\,,\,R_f(z)\,\cdot\,)_{L^2} & F_{f,\varphi}(z)^{-1}  \end{pmatrix},\label{resolvent}
\ee where $\textrm{Im\,}z\neq0$, $\varphi\in\mathcal{D}_{\theta_0}\cup\mathcal{T}_{k_0}$ and $f\geq0$.

There are many ways to see resonances (see, e.g., \cite{S4}). In this model we use the following definition:

\begin{defi}\label{defi:resonance}
\begin{lquote}
Let $\kappa_0>0$ and $0<\vartheta_0<\pi/3$.  Assume \eqref{Hilbert_space} through  \eqref{H(f)}.
Fix $\varphi\in\mathcal{D}_{\theta_0}\cup\mathcal{T}_{k_0}$, where $k_0\geq\kappa_0$ and $\theta_0\geq\vartheta_0$. Let \begin{align}&\hspace{-1cm}\Omega_f:=\left\{\begin{array}{l@{\,,\quad }l}\{z\in\mathbb{C}\,|\,z=|z|e^{i\theta},\ \theta\in\mathbb{R},\ 0\leq|\theta|<2\vartheta_0\}\cap\mathbb{C}_-&\varphi\in\mathcal{D}_{\theta_0}\textrm{ and }f=0\\ S_{\kappa_0}\cap\mathbb{C}_-&\varphi\in\mathcal{T}_{k_0}\textrm{ and }f=0\\
\mathbb{C}_- & \varphi\in\mathcal{D}_{\theta_0}\cup\mathcal{T}_{k_0}\textrm{ and }f>0\end{array}\right..\nn\\
&\label{Omega}\end{align}
\end{lquote}
\begin{enumerate}
\item Let $f\geq0$. A number $\widetilde{z}$ in $\mathbb{C}_-$ is defined to be a resonance of $H_\varphi(f)$, if for some $\Psi:=\begin{pmatrix}\psi \\ u \end{pmatrix}\in L^2(\mathbb{R})\oplus\mathbb{C}$, with $\psi\in\mathcal{D}_{\vartheta_0}$ if $\varphi\in\mathcal{D}_{\theta_0}$ and $\psi\in\mathcal{T}_{\kappa_0}$ if $\varphi\in\mathcal{T}_{k_0}$, the meromorphic extension of the resolvent matrix element $(\Psi\,,\,(H_\varphi(f)-z)^{-1}\Psi)_{\mathscr{H}}$ $(z\in\mathbb{C}_+)$ to the region $\Omega_f\cup\overline{\mathbb{C}_+}\backslash(-\infty,0]$ has a pole at $\widetilde{z}$.
\item For any function $g_{f,\varphi}(\cdot)$ (depending on given $\varphi$ and $f$) which is analytic in $\mathbb{C}_+$ the symbol $g_{f,\varphi}^{\mathbf{c},\Omega_f}(\cdot)$ denotes the analytically (or meromorphically) continued function to the region $\Omega_f\cup\overline{\mathbb{C}_+}\backslash(-\infty,0]$.
\end{enumerate}
\end{defi}

\begin{remark}\label{remark:1a}
\begin{lquote} The region $\Omega_f$ $(f\geq0)$, defined in \eqref{Omega}, is a region of analyticity for the continued resolvent matrix element of the 1-dimensional Stark operator, i.e., for  $(u\,,\,(p^2+fx-\,\cdot\,)^{-1}v)_{L^2}^{\mathbf{c},\Omega_f}$ for any $u,v$ in $\mathcal{T}_{k_0}$ or in $\mathcal{D}_{\theta_0}$. If $f>0$, then the function $(u\,,\,(p^2+fx-\,\cdot\,)^{-1}v)_{L^2}^{\mathbf{c},\Omega_f}$ is entire (see \cite[Theorem III.4]{He} for the dilation analytic case $\mathcal{D}_{\theta_0}$ and Section \ref{subsection:infinitely_many} for both cases, $\mathcal{D}_{\theta_0}$ and $\mathcal{T}_{k_0}$). The bound $0<\vartheta_0<\pi/3$ in Definition \ref{defi:resonance} for the dilation analytic case comes from the fact that the dilated operator $U(\theta)(p^2+fx)U(\theta)^{-1}\stackrel{\eqref{unitary_group}}{=}e^{-2\theta}(p^2+fe^{3\theta}x)$ has a compact resolvent for $\textrm{Im\,}(fe^{3\theta})\neq0$, thus in particular for $0<|\textrm{Im\,}\theta|=:\vartheta_0<\pi/3$ but not for the endpoints of these intervals, i.e., not for $\theta=0$, $\theta=\pm i\pi/3$; see \cite[Theorem II.3 b)]{He}.
\end{lquote}
\end{remark}

\noindent Using \eqref{resolvent}, for any $\Psi$ as in Definition \ref{defi:resonance} and $z\in\mathbb{C}_+$, one finds
 \begin{align}(\Psi\,,&\,(H_\varphi(f)-z)^{-1}\Psi)_\mathscr{H}\nn\\
=&\ (\psi,R_f(z)\psi)_{L^2}+F_{f,\varphi}(z)^{-1}(\varphi,R_f(z)\psi)_{L^2}(\psi,\varphi)_{L^2}
-u F_{f,\varphi}(z)^{-1}(\psi,R_f(z)\varphi)_{L^2}\nn\\ &-\overline{u}F_{f,\varphi}(z)^{-1}(\varphi,R_f(z)\psi)_{L^2}+|u|^2 F_{f,\varphi}(z)^{-1}\,,\nn\end{align}
where $R_f$ and $F_{f,\varphi}$ are defined in \eqref{F_f}.
Now, since (by Remark \ref{remark:1a}) $\Omega_f$ is a region of analyticity for the resolvent matrix elements $(\psi,R_f(\cdot)\psi)_{L^2}$, $(\varphi,R_f(\cdot)\varphi)_{L^2}$  $(\varphi,R_f(\cdot)\psi)_{L^2}$ and $(\psi,R_f(\cdot)\varphi)_{L^2}$ one sees that the only way poles in $\Omega_f$ of the continued matrix element $(\Psi\,,\,(H_\varphi(f)-\,\cdot\,)^{-1}\Psi)_\mathscr{H}$ can come into play is as zeros in $\Omega_f$ of the continued function $F_{f,\varphi}$. Since this is true for all such $\Psi$, the Definition \ref{defi:resonance} of resonances is actually independent of the choice of $\Psi$. That said, we get the following proposition:

\begin{proposition}\label{proposition:resonance=solution}
\begin{lquote}Let $\theta_0$ and $k_0$ be as in Definition \ref{defi:resonance}. Let $\varphi\in\mathcal{D}_{\theta_0}\cup\mathcal{T}_{k_0}$, $f\geq0$. Let $H_\varphi(f)$ and $F_{f,\varphi}$ be given by \eqref{H(f)} and \eqref{F_f}. The resonances of $H_\varphi(f)$ are precisely the solutions of
\be F_{f,\varphi}^{\mathbf{c},\Omega_f}(z)=0\label{zeros=resonances}\ee in $\Omega_f$. ($F_{f,\varphi}^{\mathbf{c},\Omega_f}$ denotes the analytic extension of $F_{f,\varphi}$ in the sense of Definition \ref{defi:resonance}.)\\
The real zeros of $F_{f,\varphi}^{\mathbf{c},\Omega_f}$ are eigenvalues of the self-adjoint operator $H_\varphi(f)$.
\end{lquote}
\end{proposition}

\noindent We need the following important property of $F_{f,\varphi}^{\mathbf{c},\Omega_f}$, $f>0$:
\begin{proposition}\label{proposition:infinitely_many_zeros}
\begin{lquote} Let $\theta_0$ and $k_0$ be as in Definition \ref{defi:resonance}. Let $\varphi\in\mathcal{T}_{k_0}\cup\mathcal{D}_{\theta_0}$.
Let $H_\varphi(f)$ and $F_{f,\varphi}$ be given by \eqref{H(f)} and \eqref{F_f}. Then for all $f>0$ the function $F_{f,\varphi}$ has an extension to an entire function of finite order (cf. Definition \ref{defi:finite_order}). This entire extension $F_{f,\varphi}^{\mathbf{c},\Omega_f}$ has infinitely many zeros.
\end{lquote}
\end{proposition}

\noindent\textbf{Proof:}\quad See Section \ref{subsection:infinitely_many}.\medskip\\

A consequence of Proposition \ref{proposition:resonance=solution} and  Proposition \ref{proposition:infinitely_many_zeros} is the following corollary:

\begin{corollary}\label{corollary:infinitely_many_resonances}
\begin{lquote}Suppose the same hypotheses as in Proposition \ref{proposition:infinitely_many_zeros}. Suppose in addition that $\varphi$ is in the domain of $\langle p\rangle^{(1+\varepsilon)/2}$ for some $\varepsilon>0$ (as an operator in $L^2(\mathbb{R})$), where $\langle p\rangle:=(1+|p|^2)^{1/2}$.
Then for all $f>0$ the operator $H_\varphi(f)$ has infinitely many resonances.
\end{lquote}
\end{corollary}

\noindent\textbf{Proof:}\quad See Appendix \ref{appendix:infinitely_many_resonances}.\medskip\\
In this model the existence of pre-existing resonances (i.e., resonances of $H_\varphi(0)$) is shown by the following proposition. We further collect some additional useful information about the spectrum of $H_\varphi(0)$:

\begin{proposition}\label{proposition:existence_resonance}
\begin{lquote} Let $\theta_0$ and $k_0$ be as in Definition \ref{defi:resonance}. Let $\varphi\in\mathcal{D}_{\theta_0}\cup\mathcal{T}_{k_0}$ and $H_\varphi(0)$ be given by \eqref{H}. Then:
\begin{enumerate}
\item $H_\varphi(0)$ has an eigenvalue $\lambda<0$ if and only if
$$\int\limits_{\mathbb{R}}\frac{|\stackrel{\wedge}{\varphi}(k)|^2}{(k^2+|\lambda|)(1+|\lambda|)}\;dk=1\,.$$
\item $H_\varphi(0)$ has an eigenvalue $\lambda\geq0$ if and only if $\varphi$ is in the domain of $(p^2-\lambda)^{-1}$ and $(\varphi\,,\,(p^2-\lambda)^{-1}\varphi)_{L^2}+\lambda-1=0$.
\item If $\stackrel{\wedge}{\varphi}(\pm\sqrt{\lambda})\neq0$ for $\lambda\in[1-\varepsilon\,,\,1+\varepsilon]$ with some $\varepsilon\in(0,1)$, then for sufficiently small $\varphi$ the operator $H_\varphi(0)$ has exactly one resonance close to 1. The smallness of $\varphi$ is measured by the size of $\varrho^{-1}|(\varphi\,,\,(p^2-z)^{-1}\varphi)^{\mathbf{c},\Omega_0}_{L^2}|$  for $|z-1|=\varrho$, where $\varrho$ is some number in $(0,\varepsilon)$.
\end{enumerate}
\end{lquote}
\end{proposition}
\enlargethispage{1cm}
\noindent\textbf{Sketch of proof:}\quad 1. and 2. follow from the eigenvalue equation for $H_\varphi(0)$ and the fact that $p^2$ does not have any eigenvalues.\\
3. can be seen as follows: By Proposition \ref{proposition:resonance=solution}, the zeros in $\mathbb{C}_-$ of
\be F_{0,\varphi}^{\mathbf{c},\Omega_0}(z)\stackrel{\eqref{F_f}}{=}1-z-(\varphi\,,\,(p^2-z)^{-1}\varphi)_{L^2}^{\mathbf{c},\Omega_0}\label{F00}\ee are precisely the resonances of $H_\varphi(0)$. Let $\partial B_\varrho(1)$ denote the boundary of $B_\varrho(1):=\{z\in\mathbb{C}\,|\,|1-z|\leq \varrho\}$, $\varrho>0$.
Let, for some $0<\varrho<\varepsilon$,
\be \frac{|(\varphi\,,\,(p^2-z)^{-1}\varphi)^{\mathbf{c},\Omega_0}_{L^2}|}{\varrho}<1\hspace{0.5cm}(z\in\partial B_\varrho(1))\,.\nn\ee  Then
\be |(\varphi\,,\,(p^2-z)^{-1}\varphi)^{\mathbf{c},\Omega_0}_{L^2}|<\varrho=|1-z|\hspace{0.5cm}(z\in\partial B_\varrho(1))\,.\nn\ee Thus by Rouch\a'e's Theorem (see, e.g., \cite{BN}) the functions $1-z-(\varphi\,,\,(p^2-z)^{-1}\varphi)^{\mathbf{c},\Omega_0}_{L^2}=F_{0,\varphi}^{\mathbf{c},\Omega_0}(z)$ and $1-z$ have the same number of zeros (namely exactly one) in $B_\varrho(1)$. This zero of $F_{0,\varphi}^{\mathbf{c},\Omega_0}$ is in $\mathbb{C}_-$: Note that $F_{0,\varphi}^{\mathbf{c},\Omega_0}$ has no zeros in $\mathbb{C}_+$, since $(H_\varphi(0)-\,\cdot\,)^{-1}$ is analytic in $\mathbb{C}_+$. Thus it suffices to show that there are no zeros of $F_{0,\varphi}^{\mathbf{c},\Omega_0}$ on the real axis in some sufficiently small neighborhood of 1. A brief calculation shows
\begin{align}(\varphi\,,\,(p^2-z)^{-1}\varphi)_{L^2}&=(\stackrel{\wedge}{\varphi}\,,\,(k^2-z)^{-1}\stackrel{\wedge}{\varphi})_{L^2}\nn\\
&=\frac{1}{2\sqrt{z}}\,\int\limits_{\mathbb{R}} \overline{\stackrel{\wedge}{\varphi}}(\overline{k})\stackrel{\wedge}{\varphi}(k)\Big(\frac{1}{k-\sqrt{z}}-\frac{1}{k+\sqrt{z}}\Big)\;dk\hspace{0.5cm}
(z\in\mathbb{C}_+)\,,
\label{reso00}
\end{align} where the integrand has singularities at $\pm\sqrt{z}$. By contour deformation (as shown in Figure \ref{fig:deform_gamma}) one gets the meromorphic extension of the resolvent matrix element \eqref{reso00}, which on the real line is given by
\be (\varphi\,,\,(p^2-z)^{-1}\varphi)^{\mathbf{c},\Omega_0}_{L^2}=\frac{i\pi}{2\sqrt{z}}\;\big(|\stackrel{\wedge}{\varphi}(+\sqrt{z})|^2+
|\stackrel{\wedge}{\varphi}(-\sqrt{z})|^2\big)+\mathcal{P}(z)\hspace{0.5cm}(z\in\mathbb{R})\,,\label{pv}\ee where
\be \mathcal{P}(z):=\lim\limits_{\varepsilon\downarrow 0}\frac{1}{2\sqrt{z}}\Big(\int\limits_{|k-\sqrt{z}|\geq\varepsilon} \frac{|\stackrel{\wedge}{\varphi}(k)|^2}{k-\sqrt{z}}\;dk -
\int\limits_{|k+\sqrt{z}|\geq\varepsilon} \frac{|\stackrel{\wedge}{\varphi}(k)|^2}{k+\sqrt{z}}\;dk \Big)\hspace{0.5cm}(z\in\mathbb{R})\,,\nn\ee
and $\mathcal{P}(z)\in\mathbb{R}$ for $z\in\mathbb{R}\backslash\{0\}$. Now assume that $F_{0,\varphi}^{\mathbf{c},\Omega_0}$ has a zero $\lambda$ in $[1-\varepsilon,1+\varepsilon]$. Then, by use of \eqref{pv} in \eqref{F00} with $z=\lambda$,
\be 1-\lambda-\mathcal{P}(\lambda)=\frac{i\pi}{2\sqrt{\lambda}}\;\big(|\stackrel{\wedge}{\varphi}(+\sqrt{\lambda})|^2+
|\stackrel{\wedge}{\varphi}(-\sqrt{\lambda})|^2\big)\,,\label{**}\ee where the l.h.s. is an element of $\mathbb{R}$ and the r.h.s. a priori is in $i\mathbb{R}$. Thus the r.h.s. of \eqref{**} is in $\mathbb{R}\cap i\mathbb{R}=\{0\}$, which implies $\stackrel{\wedge}{\varphi}(\pm\sqrt{\lambda})=0$. Contradiction.
\\\Done

\noindent The following theorem on the location of resonances of $H_\varphi(f)$ for small $f>0$ and their instability in the weak field limit is our main result in Section \ref{section:first_model}:

\begin{theorem}\label{theorem:resonances_instable}
\begin{lquote}  Let $k_0$ and $\theta_0$ be as in Definition \ref{defi:resonance}.
Let $\varphi\in\mathcal{D}_{\theta_0}\cup\mathcal{T}_{k_0}$. If $\varphi$ is in $\mathcal{D}_{\theta_0}\backslash\mathcal{T}_{k_0}$, suppose in addition that
\be \sup\big\{|\frac{d^n}{dk^n}\,\stackrel{\wedge}{\varphi}(e^{i\theta}k)|\,\big|\,k\in\mathbb{R}\backslash\{0\},\ \theta\in\mathbb{R},\ |\theta|<\min\{\theta_0,\frac{\pi}{3}\},\ n\leq2\big\}<\infty\,.\label{sup}\ee Let $H_\varphi(f)$, $f>0$, be given by \eqref{H(f)}.
If $\varphi\in\mathcal{D}_{\theta_0}\backslash\mathcal{T}_{k_0}$ let
\be M:=\{z\in\mathbb{C}_-\,|\, 0>\textrm{arg\,}\sqrt{z}\geq -\min\{\frac{\pi}{3},\theta_0\}+\varepsilon_1\,,\ \textrm{Re\,}\sqrt{z}\in[\varepsilon_2,a]\}\nn\ee and if $\varphi\in\mathcal{T}_{k_0}$ let
\begin{align} & M:=\{z\in\mathbb{C}_-\,|\, 0>\textrm{arg\,}\sqrt{z}\geq-\frac{\pi}{3}+\varepsilon_1,\ \textrm{Re\,}\sqrt{z}\in[\varepsilon_2,a],\ 0>\textrm{Im\,}\sqrt{z}\geq -k_0+\varepsilon_3\}\nn
\end{align} for any $\varepsilon_1,\varepsilon_2,\varepsilon_3>0$ sufficiently small and any $a>\varepsilon_2$.
Suppose there exists $\delta>0$ such that for all $z\in M$ \be |\stackrel{\wedge}{\varphi}(\sqrt{z})\overline{\stackrel{\wedge}{\varphi}}(-\overline{\sqrt{z}})|\geq\delta\,.\label{geq_delta}\ee Then:
\begin{align} \exists\, c_0>0\ \exists\, f_0>0\textrm{ sufficiently}&\ \textrm{small}\ \forall\, f\in(0,f_0]:\nn\\
&r \textrm{ is a resonance of }H_\varphi(f)\textrm{ in }M\Rightarrow |\textrm{Im\,}r|\leq c_0 f\,.\nn\end{align}
In particular, \be\textrm{Im\,}r\to0\hspace{0.5cm}(f\downarrow0)\,.\nn\ee Thus $r$ does not converge to any resonance of $H_\varphi(0)$ (given by \eqref{H}) as $f\downarrow 0$.
\end{lquote}
\end{theorem}

\noindent\textbf{Proof:}\quad See Section \ref{subsection:instability}.

\begin{remark}\label{remark:1}${}$
\begin{enumerate}
\item It follows from dilation analyticity that $\stackrel{\wedge}{\varphi}$ is analytic in $\{z\in\mathbb{C}\,|\, z=\pm|z|e^{i\theta},\ \theta\in\mathbb{R}$\linebreak $\textrm{with }|\theta|<\theta_0,\ z\neq0\}$. But \eqref{sup} is an additional assumption near $k=0$.
\item If $z\in M$, $\pm\sqrt{z}$ is in the region of analyticity for $\stackrel{\wedge}{\varphi}$.
\item Our proof of Theorem \ref{theorem:resonances_instable} uses power series expansions (in $\zeta$) of $\stackrel{\wedge}{\varphi}(\sqrt{z}+\zeta)$ and $\overline{\stackrel{\wedge}{\varphi}}(-\overline{\sqrt{z}}-\overline{\zeta})$ up to first order (e.g., in \eqref{use_of_power_exp1} and \eqref{use_of_power_exp2}). Presumably the results stated in Theorem \ref{theorem:resonances_instable} are true without the assumption that $\stackrel{\wedge}{\varphi}(\sqrt{z})\overline{\stackrel{\wedge}{\varphi}}(-\overline{\sqrt{z}})\neq0$. The proof would involve going to higher order in the power series for $\stackrel{\wedge}{\varphi}(\sqrt{z}+\zeta)$ and $\overline{\stackrel{\wedge}{\varphi}}(-\overline{\sqrt{z}}-\overline{\zeta})$.
\end{enumerate}
\end{remark}

\subsubsection{Friedrichs model with AC Stark effect}\label{subsubsection:AC_Stark}

If in $H_\varphi(f)$, given by \eqref{H(f)}, the direct current field $fx$ is replaced by an alternating electric field, then the behavior of pre-existing resonances of $H_\varphi(0)$ changes: In contrast to pre-existing resonances in the DC Stark effect (which are unstable by Theorem \ref{theorem:resonances_instable}), pre-existing resonances in the AC Stark effect are stable (in the sense of Theorem \ref{theorem:AC} below). This is reminiscent of the situation in classical mechanics that introducing a periodic change in parameters may cause an unstable equilibrium to become stable. An example is the unstable equilibrium for a pendulum (see, e.g., \cite{Arnold}).

The Hamiltonian in the AC field regime is given by
\be \hspace{-1cm}H_\varphi(t,f):=\begin{pmatrix} h(t,f)& M_\varphi \\[1.5ex] (\varphi\,,\,\cdot\,)_{L^2} & 1 \end{pmatrix}\,,\hspace{0.3cm}h(t,f):=p^2+fx\sin(\omega t)\hspace{0.3cm}(f\geq0,\ x,t\in\mathbb{R})\,,\label{H(f)_AC}\ee where $\omega>0$ denotes the frequency of an alternating electric field and $\varphi$ is in $\mathcal{D}_{\theta_0}$ for some $\theta_0\in(0,\frac{\pi}{2})$.

In order to define resonances in this time dependent setting, we adapt the method of \cite{Y1} and \cite{Y} (which uses Floquet theory) to the Friedrichs model \eqref{H(f)_AC}. (In \cite{Y} Yajima proved that eigenvalues of one-body Hamiltonians with a certain class of analytic potentials turn into resonances under the influence of an alternating electric field and that these resonances converge, in the sense of \cite[Theorem 3.2]{Y}, to the original eigenvalue as the field strength goes to zero.)

We introduce the unitary transformation on $\mathscr{H}\stackrel{\eqref{Hilbert_space}}{=}L^2(\mathbb{R})\oplus\mathbb{C}$
\be \widetilde{T}(t,f):\ \mathscr{H}\to \mathscr{H}\,,\hspace{0.5cm}
\widetilde{T}(t,f):=\begin{pmatrix}T(t,f) & 0 \\[1.5ex] 0 & 1 \end{pmatrix}\label{Ttilde}\ee for all $f\geq0$ and $t\in\mathbb{R}$, where
\be T(t,f):= e^{i2f\omega^{-2}\sin(\omega t)p}\,e^{-if\omega^{-1}\cos(\omega t)x}\label{T}\ee and $p:=-id/dx$ as operators in $L^2(\mathbb{R})$. Then, if
$i \partial_t\Psi = H_\varphi(t,f)\Psi$, one has
\be i\partial_t(\widetilde{T}(t,f)\Psi)=\widetilde{H}_\varphi(t,f)\widetilde{T}(t,f)\Psi\,,\label{1.23}\ee where
\be \widetilde{H}_\varphi(t,f):= (i\partial_t\,\widetilde{T}(t,f))\widetilde{T}(t,f)^{-1}+\widetilde{T}(t,f)H_\varphi(t,f)\widetilde{T}(t,f)^{-1}
\label{Htilde}\ee
and
\be \widetilde{T}(t,f)H_\varphi(t,f)\widetilde{T}(t,f)^{-1}=
\begin{pmatrix}T(t,f)h(t,f)T(t,f)^{-1} & T(t,f)\varphi \\[1.5ex] (\varphi\,,\,T(t,f)^{-1}\,\cdot\,)_{L^2} & 1 \end{pmatrix}\label{THT^-1}\ee
for all $f\geq0$, $t\in\mathbb{R}$. By \eqref{1.23}, for fixed $t$, the operators $H_\varphi(t,f)$ and $\widetilde{H}_\varphi(t,f)$ are unitarily equivalent. The purpose of $\widetilde{T}(t,f)$ is to transform the unbounded electric potential to zero, modulo a constant (w.r.t. $x$) shift. This can be seen in \eqref{htilde} below. (By a remark of Hunziker \cite[Chapter 7.3, Remark 3]{CyFKS}, one can interpret $\widetilde{T}(t,f)$ as the implementation of a gauge transformation via a unitary transformation on $\mathscr{H}$. The motivation of defining $\widetilde{T}(t,f)$ by \eqref{Ttilde} is essentially the same as for the operator $\widetilde{T}(t)$ given in \cite[Chapter 7.3]{CyFKS}.) A calculation shows
\begin{align} &T(t,f)h(t,f)T(t,f)^{-1}=\Big(\frac{f}{\omega}\;\cos(\omega t)+p\Big)^2+\Big(x+\frac{2f}{\omega^2}\;\sin(\omega t)\Big)f\sin(\omega t)\,,\label{1.21}\\
&i(\partial_t T(t,f))T(t,f)^{-1}=-\frac{2f}{\omega}\;\cos(\omega t)p-\Big(x+\frac{2f}{\omega^2}\;\sin(\omega t)\Big)f\sin(\omega t)\,.\label{1.21a}\end{align}

\noindent Combining \eqref{Htilde} and \eqref{THT^-1} gives
\be \widetilde{H}_\varphi(t,f)=\begin{pmatrix}\widetilde{h}(t,f) & T(t,f)\varphi \\[1.5ex] (T(t,f)\varphi\,,\,\cdot\,)_{L^2} & 1 \end{pmatrix}\,,\label{HtildeAC}\ee where
\be\widetilde{h}(t,f)&:=& (i\partial_t T(t,f))T(t,f)^{-1}+T(t,f)h(t,f)T(t,f)^{-1}\nn\\
&\stackrel{{\eqref{1.21}\atop \eqref{1.21a}}}{=}& p^2+\frac{f^2}{2\omega^2}\cos(2\omega t)+\frac{f^2}{2\omega^2}\label{htilde}
\ee
as an operator in $L^2(\mathbb{R})$.

 Using Howland's idea\footnote{
Roughly speaking, in \cite{How} Howland proves (among other things) that for time-dependent Hamiltonians $H(t)$ in a separable Hilbert space $\mathscr{H}$ there is a correspondence between the unitary propagator (i.e., the solution) of $i\partial_tu=H(t)u$ in $\mathscr{H}$ and the spectrum of the Floquet Hamiltonian $\widetilde{H}:=-i\partial_t+H(t)$ in the Hilbert space $L^2(\mathbb{R},\mathscr{H}):=\{u\,|\,u\textrm{ strongly measurable $\mathscr{H}$-valued, }\int_\mathbb{R}\|u(s)\|^2_{\mathscr{H}}\;ds<\infty\}$. As described in \cite{How}, interpreting $-i\partial_t$ as the quantum mechanical momentum, conjugate to the coordinate $t$, this correspondence  has a well known analog in classical mechanics: A classical (non energy-preserving) system with a time-dependent Hamilton function $\mathrm{H}(q_1,\ldots,q_n,p_1,\ldots,p_n,t)=:\mathrm{H}(t)$ can be transformed into a formally autonomous (energy-preserving) system by introducing the time $t$ as a new coordinate and the external energy $E$ as its conjugate momentum. This leads to the new Hamilton function $\widetilde{\mathrm{H}}(q_1,\ldots,q_n,t,p_1,\ldots,p_n,E)= E+\mathrm{H}(t)$. For more details we refer to, e.g., \cite{How}, \cite[Chapter 7.4]{CyFKS} or \cite[Chapter X.12]{RS2}.

For the time-periodic Stark Hamiltonian considered in \cite{Y} the Hilbert space $\mathscr{H}$ is $L^2(\mathbb{R}^3)$ and $L^2(\mathbb{R},\mathscr{H})$ can be replaced by $L^2(\mathbb{T}_T)\otimes L^2(\mathbb{R}^3)$, where $\mathbb{T}_T=\mathbb{R}\backslash T\mathbb{Z}$ and $T$ denotes the period.} from \cite{How} and adapting the outline in \cite[Introduction]{Y} to our matrix model \eqref{H(f)_AC}, we consider
\be K(f):=-i\partial_t + \widetilde{H}_\varphi(t,f)\hspace{0.5cm}(f\geq0\,,\ t\in\mathbb{R})\nn\ee as an operator in
\be \widetilde{\mathscr{H}}:=(L^2(\mathbb{R})\oplus\mathbb{C})\otimes L^2(\mathbb{T}_\omega)\,,\quad\textrm{where}\quad \mathbb{T}_\omega:=\mathbb{R}/\tau\mathbb{Z}\,,\quad \tau:=\frac{2\pi}{\omega}\,.\nn\ee Then
\begin{align} K(f)&=\unity_{L^2(\mathbb{R})\oplus\mathbb{C}}\otimes(-i\partial_t)+\widetilde{H}_\varphi(t,f)\otimes\unity_{L^2(\mathbb{T}_\omega)}
\hspace{0.5cm}(f\geq0\,,\ t\in\mathbb{R})\,,\nn\end{align} and $K(f)$ has a self-adjoint realization in $\widetilde{\mathscr{H}}$, which we also denote by $K(f)$. ($K(f)$ is called the Floquet Hamiltonian.) The self-adjoint operator $K(f)$ generates a unitary group $\{e^{-isK(f)}\}_{s\in\mathbb{R}}$. By Floquet theory,  $e^{-i\tau K(f)}$ is unitarily equivalent to the propagator\linebreak
$\widetilde{U}(t+\tau,t;f)\otimes\unity_{L^2(\mathbb{T}_\omega)}$ $(t\in\mathbb{R})$ over the period $\tau$, where
\be i\partial_t\widetilde{U}(t,s;f)=\widetilde{H}_\varphi(t,f)\widetilde{U}(t,s;f),\quad \widetilde{U}(t,t;f)=\unity\nn\ee for $t,s\in\mathbb{R}$.

We shall now define resonances by use of the dilation analytic machinery:  Define
\be V(\theta)^{\pm1}:=\begin{pmatrix}U(\theta)^{\pm1} & 0 \\[1.5ex] 0 & 1 \end{pmatrix}\hspace{0.5cm}(\theta\in S_{\theta_0}) \label{V}\ee as an operator in $\mathscr{H}\stackrel{\eqref{Hilbert_space}}{=}L^2(\mathbb{R})\oplus\mathbb{C}$, where $U(\cdot)$ denotes the group of dilations in $L^2(\mathbb{R})$ (cf. \eqref{unitary_group}). In particular,  $U(\theta)x U(\theta)^{-1}=e^\theta x$ $(x\in\mathbb{R},\ \theta\in S_{\theta_0})$. One gets
\begin{align} &V(\theta)\widetilde{H}_\varphi(t,f)V(\theta)^{-1}\stackrel{{\eqref{V}\atop \eqref{HtildeAC}}}{=}\begin{pmatrix}U(\theta)\widetilde{h}(t,f)U(\theta)^{-1}
 & U(\theta)T(t,f)\varphi \\[1.5ex] (U(\overline{\theta})T(t,f)\varphi\,,\,\cdot\,)_{L^2} & 1 \end{pmatrix}\,,\label{VHV}\\
&U(\theta)\widetilde{h}(t,f)U(\theta)^{-1}\stackrel{\eqref{htilde}}{=}e^{-2\theta}p^2+\frac{f^2}{2\omega^2}\,\cos(2\omega t)+\frac{f^2}{2\omega^2}
\label{UhU}\end{align} for all $\theta\in S_{\theta_0}$, $f\geq0$ and $t\in\mathbb{R}$. Similar to \cite[(1.9)]{Y}, we define
\begin{align} K(f,\theta):=-i\partial_t+V(\theta)\widetilde{H}_\varphi(t,f)V(\theta)^{-1}\hspace{0.5cm}(\theta\in S_{\theta_0},\ f\geq0,\ t\in\mathbb{R})\label{K}\end{align}
as an operator in $\widetilde{\mathscr{H}}$. In particular, \be K(0,0)= -i\partial_t+\begin{pmatrix}p^2 & \varphi \\[1.5ex] (\varphi\,,\,\cdot\,)_{L^2} & 1 \end{pmatrix}\stackrel{\eqref{H(f)_AC}}{=}-i\partial_t+H_\varphi(\,\cdot\,,f=0)\,.\nn\ee

\begin{defi}\label{defi:resonanceAC}
\begin{lquote}
Let $f\geq0$, $\theta\in S_{\theta_0}$ with $\textrm{Im\,}\theta>0$. The nonreal eigenvalues (which, by construction, for $\textrm{Im\,}\theta>0$ lie in $\mathbb{C}_-$) of $K(f,\theta)$ are defined to be the resonances of $H_\varphi(\,\cdot\,,f)$.
\end{lquote}
\end{defi}

\begin{theorem}\label{theorem:AC}
\begin{lquote} Let $\omega>0$ denote the frequency of an alternating electric field. Let $\theta_0\in(0,\frac{\pi}{2})$ and  $\varphi\in\mathcal{D}_{\theta_0}$. Assume in addition that there exists an $f_0>0$ sufficiently small such that for all $\beta\in[-f_0,f_0]$ and all $\theta\in S_{\theta_0}$ the function $e^{f_0|x|\omega^{-1}\sin\theta_0}\varphi(e^{\theta}x+2\beta\omega^{-2})$ $(x\in\mathbb{R})$ is analytic in the parameter $\theta$ and \be \|e^{f_0|\,\cdot\,|\omega^{-1}\sin\theta_0}\varphi(e^{\theta}\cdot+2\beta\omega^{-2})\|_2\leq C\label{phi_uniformly}\ee for some constant $C<\infty$, uniformly for $(\theta,\beta)\in S_{\theta_0}\times[-f_0,f_0]$. Let $H_\varphi(t,f)$ $(f\geq0,\ t\in\mathbb{R})$ be given by \eqref{H(f)_AC} and $K(f,\theta)$ $(f\geq0,\ \theta\in S_{\theta_0})$ by \eqref{K}.\\
Fix $\theta\in S_{\theta_0}$ with $\textrm{Im\,}\theta>0$. Suppose there exists an eigenvalue $r_0$ (in $\mathbb{C}_-$, not necessarily close to 1) of $K(0,\theta)$ of multiplicity $m$. Then for all $f>0$ sufficiently small there are exactly $m$ eigenvalues (counting multiplicities) of $K(f,\theta)$ close to $r_0$ and they all converge to $r_0$ as $f\downarrow0$.
(By Definition \ref{defi:resonanceAC}, $r_0$ is a resonance of $H_\varphi(\,\cdot\,,0)$ and the eigenvalues of $K(f,\theta)$ are resonances of $H_\varphi(\,\cdot\,,f)$.)
\end{lquote}
\end{theorem}

\noindent\textbf{Proof:}\quad See Section \ref{subsection:AC_proof}.

\subsection{Model II}\label{subsection:Model3}
In this model we consider the operator \be H_\varepsilon(f):=p^2+fx+V_\varepsilon\hspace{0.5cm}(\varepsilon\geq0,\ f\geq0,\ x\in\mathbb{R})\,,\label{H(f)_Mod2}\ee in $L^2(\mathbb{R})$, where $p$ denotes the self-adjoint realization of $-id/dx$ in $L^2(\mathbb{R})$ and \be V_\varepsilon:=(\varphi_\varepsilon\,,\cdot\,)\varphi_\varepsilon\label{V_epsilon_Mod2}\ee is a rank one self-adjoint operator in $L^2(\mathbb{R})$. (In this Section \ref{subsection:Model3} the inner product in $L^2(\mathbb{R})$ is denoted by $(\,\cdot\,,\,\cdot\,)$.) We shall construct $\varphi_\varepsilon$ in such a way that, for
$\varepsilon=0$, the operator $p^2+V_0$ will have eigenvalue 1 embedded in the continuous spectrum $[0,\infty)$ of $p^2$. For $\varepsilon>0$ this eigenvalue will turn into a resonance of $p^2+V_\varepsilon$ which we will then perturb by taking $f>0$ in \eqref{H(f)_Mod2}.\\ Our second model needs the following observation (similar to Proposition \ref{proposition:existence_resonance}):

\begin{proposition}
\begin{lquote} Let $H_2(\mathbb{R})$ denote the Sobolev space of second order. Let $p=-id/dx$, $\varphi\in L^2(\mathbb{R})$. Clearly, the operators $p^2$ and $T:=p^2+(\varphi\,,\cdot\,)\varphi$ with domain $\mathcal{D}(T)=\mathcal{D}(p^2)=H_2(\mathbb{R})$ are self-adjoint in $L^2(\mathbb{R})$. $T$ has an eigenvalue $\lambda>0$ if and only if \be\varphi\in\textrm{Ran\,}(p^2-\lambda)\quad\textrm{and}\quad (\varphi\,,\,(p^2-\lambda)^{-1}\varphi)=-1\,.\label{3}\label{2} \ee
\end{lquote}
\end{proposition}

\noindent\textbf{Proof:}\quad Let $\lambda>0$. If $p^2\psi+(\varphi\,,\,\psi)\varphi=\lambda\psi$ for some non-zero $\psi\in H_2(\mathbb{R})$, then $(p^2-\lambda)\psi=-(\varphi\,,\,\psi)\varphi$. Since $p^2$ has no eigenvalues we cannot have $(\varphi\,,\,\psi)=0$, thus $\varphi\in\textrm{Ran\,}(p^2-\lambda)$ and in addition
\be\psi=-(\varphi\,,\,\psi)(p^2-\lambda)^{-1}\varphi.\nn\ee Thus
\be (\varphi\,,\,\psi)=-(\varphi\,,\,\psi)(\varphi\,,\,(p^2-\lambda)^{-1}\varphi)\,,\nn\ee so $(\varphi\,,\,(p^2-\lambda)^{-1}\varphi)=-1$. Conversely, assuming \eqref{2} and defining \be \psi:=(p^2-\lambda)^{-1}\varphi\,,\nn\ee
we have \be (p^2-\lambda)\psi+(\varphi\,,\,\psi)\varphi = \varphi + (\varphi\,,\,(p^2-\lambda)^{-1}\varphi)\varphi=0.\nn\ee\Done

We shall now construct $V_\varepsilon$ and calculate the resonance of $H_\varepsilon(0)$:\quad Let $\theta_0>0$, $k_0>0$. Let $\varphi_0\in\mathcal{D}_{\theta_0}\cup\mathcal{T}_{k_0}$ (where $\mathcal{D}_{\theta_0}$ and $\mathcal{T}_{k_0}$ are given in Definition \ref{defi:dilation_analytic_vector}) with
\be\varphi_0=\frac{(p^2-1)\psi_0}{\sqrt{(\psi_0\,,\,(1-p^2)\psi_0)}}\nn\ee for some $\psi_0\in H_2(\mathbb{R})$ with
$(\psi_0\,,\,(1-p^2)\psi_0)<0$. Choose $\psi_0$ so that $\stackrel{\wedge}{\psi}_0$ has no zeros on the real axis which implies 1 is the only eigenvalue of $p^2+(\varphi_0\,,\,\cdot\,)\varphi_0$. Define
\be\varphi_\varepsilon:=\frac{(p^2-1+i\varepsilon)\psi_0}{\sqrt{(\psi_0\,,\,(1-p^2)\psi_0)}}\hspace{0.5cm}
(\varepsilon\geq0)\label{phi_epsilon_Mod2}\ee such that for $\varepsilon>0$, the self-adjoint operator
\be H_\varepsilon(0)=p^2+V_\varepsilon\nn\ee has no eigenvalues. Let
\be R_f(z):=(p^2+fx-z)^{-1}\hspace{0.5cm}(f\geq0\,,\ \textrm{Im\,}z\neq0)\,.\nn\ee
A short calculation shows that, for $f,\varepsilon\geq0$ and $z\in\mathbb{C}_+$,
\be (H_\varepsilon(f)-z)^{-1}=R_f(z)-F_{\varepsilon,f}^{-1}(z)
(\varphi_\varepsilon\,,\,R_f(z)\,\cdot\,)R_f(z)\varphi_\varepsilon\,,\nn\ee
where
\be F_{\varepsilon,f}(z):=1+(\varphi_\varepsilon\,,\,R_f(z)\varphi_\varepsilon)\,.\label{F_Mod2}\ee
For $f=0$ and $\varepsilon\geq0$ one gets
\begin{align} F_{\varepsilon,0}(z)&=1+(\varphi_\varepsilon\,,\,R_0(z)\varphi_\varepsilon)\nn\\
&=\frac{(z-1)\|\psi_0\|^2+(\varepsilon^2+(z-1)^2)(\psi_0\,,\,(p^2-z)^{-1}\psi_0)}
{|(\psi_0\,,\,(1-p^2)\psi_0)|}\hspace{0.5cm}(z\in\mathbb{C}_+)\,.\nn
\end{align}
In this model the resonances of $H_{\varepsilon}(f)$ (for $\varepsilon,f\geq0$ sufficiently small) are defined to be the zeros in $\mathbb{C}_-$ of $F_{\varepsilon,f}(\cdot)^{\mathbf{c},\Omega_f}$ (i.e., the zeros in $\mathbb{C}_-$ of the analytically continued $F_{\varepsilon,f}(\cdot)$ from the upper half complex plane into $\Omega_f\cup\overline{\mathbb{C}_+}\backslash(-\infty,0]$, where $\Omega_f$ is given by \eqref{Omega} and $F_{\varepsilon,f}$ by \eqref{F_Mod2}). (This definition is an analog of Proposition \ref{proposition:resonance=solution} for Model I.)

For $\varepsilon>0$ sufficiently small $H_\varepsilon(0)$ has exactly one resonance, $r_0$,  near $z=1$. (This can be seen by the same arguments as used in the proof of Proposition \ref{proposition:existence_resonance}, 3.) This resonance $r_0$ is a solution (in $\mathbb{C}_-$, near 1) of $F_{\varepsilon,0}(z)^{\mathbf{c},\mathbb{C}_-}=0$. One finds
\begin{align}
r_0=1-\frac{\varepsilon^2}{\|\psi_0\|^2}\Big(P.V.(\psi_0\,,\,(p^2-1)^{-1}\psi_0)+\frac{i\pi}{2}(|\stackrel{\wedge}{\psi}_0(1)|^2+
|\stackrel{\wedge}{\psi}_0(-1)|^2)\Big)+O(\varepsilon^4)\label{5}
\end{align} as $\varepsilon\downarrow0$.

As with Model I (Friedrichs model; cf. Theorem \ref{theorem:resonances_instable}) there is in general no convergence of resonances of $H_\varepsilon(f)$ to $r_0$ (or to any other resonance of $H_\varepsilon(0)$) as $f\downarrow 0$:

\begin{theorem}\label{theorem:resonance_instable_Mod2}
\begin{lquote}Let $\varphi_\varepsilon$ $(\varepsilon\geq0)$ be given by \eqref{phi_epsilon_Mod2}, where
$\psi_0\in\mathcal{T}_{k_0}\cup\mathcal{D}_{\theta_0}$ with $k_0,\theta_0>0$ as in Definition \ref{defi:resonance}. Suppose \be\int\limits_{|k|>\Lambda}|\stackrel{\wedge}{\psi}_0(k+i\alpha)|^2k^4\;dk<\infty\nn\ee for large $\Lambda$ and $|\alpha|<k_0$. If $\psi_0\not\in\mathcal{T}_{k_0}$ assume in addition that $\psi_0$ satisfies \eqref{sup}.
Let $H_\varepsilon(f)$, $V_\varepsilon$ $(\varepsilon,f\geq0)$ be given by \eqref{H(f)_Mod2} and \eqref{V_epsilon_Mod2}. Let $M$ be as in Theorem \ref{theorem:resonances_instable}.\\
Fix $\varepsilon>0$ sufficiently small. Suppose that there exists $\delta>0$ such that for all $z\in M$ $|\stackrel{\wedge}{\varphi}_\varepsilon(\sqrt{z}) \overline{\stackrel{\wedge}{\varphi}}_\varepsilon(-\overline{\sqrt{z}})|\geq\delta$. Then:
\begin{align} \exists\, c_0>0\ \exists\, f_0>0\textrm{ sufficiently}&\ \textrm{small}\ \forall\, f\in(0,f_0]:\nn\\
&r \textrm{ is a resonance of }H_\varepsilon(f)\textrm{ in }M\Rightarrow |\textrm{Im\,}r|\leq c_0 f\,.\nn\end{align}
In particular, \be\textrm{Im\,}r\to0\hspace{0.5cm}(f\downarrow0)\,.\nn\ee Thus $r$ does not converge to any resonance of $H_\varepsilon(0)$ as $f\downarrow 0$.
\end{lquote}
\end{theorem}

\noindent\textbf{Proof:}\quad Theorem \ref{theorem:resonance_instable_Mod2} follows from the estimates in Section \ref{section:Proofs}, used for the proof of Theorem \ref{theorem:resonances_instable}.\\\Done

\begin{remark}
\begin{lquote}
For small $\varepsilon>0$ it is easy to see from \eqref{phi_epsilon_Mod2} that $|\stackrel{\wedge}{\varphi}_\varepsilon(\sqrt{r_0}) \overline{\stackrel{\wedge}{\varphi}}_\varepsilon(-\overline{\sqrt{r_0}})|>0$ if $|\stackrel{\wedge}{\psi}_0(\sqrt{r_0}) \overline{\stackrel{\wedge}{\psi}}_0(-\overline{\sqrt{r_0}})|>0$.
\end{lquote}
\end{remark}

\section{Proofs}\label{section:Proofs}

\subsection{Proof of Proposition \ref{proposition:infinitely_many_zeros}}\label{subsection:infinitely_many}
Our proof of Proposition \ref{proposition:infinitely_many_zeros} needs the following Theorem \ref{theorem:infinitely_many} and Lemmata \ref{lemma:unitary_equivalence} - \ref{lemma:phi} below.

\begin{defi}\label{defi:finite_order}
\begin{lquote}Let $g$ be entire. The function $g$ is defined to be of finite order if there exist an $n\in\mathbb{R}$ and $R>0$ such that for all $z\in\mathbb{C}$ with $|z|\geq R$ \be |g(z)|\leq e^{|z|^n}\,.\nn\ee
\end{lquote}
\end{defi}

\begin{theorem}\label{theorem:infinitely_many}\cite[Theorem 16.13]{BN}
\begin{lquote} Let $g$ be an entire function of finite order. Then either $g$ has infinitely many zeros or $g(z)=P(z) e^{Q(z)}$  $(z\in\mathbb{C})$ for some polynomials $P$ and $Q$.
\end{lquote}
\end{theorem}

\begin{lemma}\label{lemma:unitary_equivalence}
\begin{lquote}Let $p=-id/dx$. For all $x,f\in\mathbb{R}$, $\phi\in S(\mathbb{R})$ one has
\be e^{-i\frac{p^{3}}{3f}}(p^2 + fx)\phi = fx e^{-i\frac{p^{3}}{3f}}\phi
\label{unitary_equivalence_fx}\,.\ee In particular, the unique extension of $p^2 + fx$ to an operator in $L^2(\mathbb{R})$ is unitarily equivalent to the unique extension of $fx$.
\end{lquote}
\end{lemma}

\noindent The proof of Lemma \ref{lemma:unitary_equivalence} is by Fourier transform.

\begin{lemma}\label{lemma:psi_f_entire}
\begin{lquote} Let $\stackrel{\wedge}{\varphi}\in L^2(\mathbb{R})$ be analytic in an open neighborhood of
\be \mathcal{A}:=\{\zeta\in\mathbb{C}\,|\,|\textrm{Re\,}\zeta|\geq N\,,\ |\textrm{Im\,}\zeta|\leq\alpha_0\}\nn\ee
for some $\alpha_0>0$, $N>0$ and let $\stackrel{\wedge}{\varphi}$ be bounded in $\mathcal{A}$. Let \be
\psi_f:=(e^{-i\frac{(\cdot)^3}{3f}}\stackrel{\wedge}{\varphi})^{\vee}\hspace{0.5cm}(f>0)\,.\label{psi_f_consequ}
\ee Then $\psi_f$, which a priori is only in $L^2(\mathbb{R})$, actually extends to an entire function (which we also denote by $\psi_f$). (In particular, $\psi_f$ has a real analytic realization on $\mathbb{R}$.) The entire function $\psi_f$ is of finite order $\leq2$. More precisely,
\be |\psi_f(x)|\leq a\,e^{b|x|^2}\hspace{0.5cm}(x\in\mathbb{C})\label{psi_order2_est}\ee for some finite real constants $a$ and $b$, possibly depending on $f$, $\varphi$ and $N$.
\end{lquote}
\end{lemma}
\noindent \textbf{Proof of Lemma \ref{lemma:psi_f_entire}:}\quad Lemma \ref{lemma:psi_f_entire} follows from Lemma \ref{lemma:psi_f=psi_contor} below and the estimate \eqref{finalest2} given in the proof of Lemma \ref{lemma:psi_f=psi_contor}.\\\Done

\begin{remark}
\begin{lquote} In distributional sense
\be \psi_f(x)=f^{1/3}\int\mathrm{Ai}(f^{1/3}(y-x))\varphi(y)\;dy\,,\nn\ee where $\mathrm{Ai}$ denotes the Airy function. $\mathrm{Ai}$ is an entire function (of order 3/2); see, e.g., \cite[Chapter 7.6]{Hoer}. We conjecture that $\psi_f$ actually is of order $3/2$.
\end{lquote}
\end{remark}

\begin{lemma}\label{lemma:psi_f=psi_contor}
\begin{lquote} Assume the same conditions and notation as in Lemma \ref{lemma:psi_f_entire}. Let $0<\alpha\leq\alpha_0$ and $f>0$.
Let
\begin{align}&\widetilde{\gamma}_{-N}:\ [0,\alpha]\to\mathbb{C}\,,\quad t\mapsto -N-i(\alpha-t)\,,\hspace{0.5cm}\widetilde{\gamma}_{+N}:\ [0,\alpha]\to\mathbb{C}\,,\quad t\mapsto N-it\,,\nn\\
&\gamma_\alpha:=\{t-i\alpha\,|\,t\in(-\infty,-N)\}\cup\widetilde{\gamma}_{-N}\cup(-N,N)\cup\widetilde{\gamma}_{+N}\cup\{t-i\alpha\,|\,t\in(N,\infty)\}\,.\nn
\end{align}
Then the contour integral \begin{align}
\widetilde{\psi}_{\alpha,f}(x):=& \frac{1}{\sqrt{2\pi}}\int\limits_{\gamma_\alpha} e^{-i\left(\frac{k^3}{3f}-kx\right)}\stackrel{\wedge}{\varphi}(k)\;dk\label{intgamma}
\end{align} exists for all $x\in\mathbb{C}$, is independent of $\alpha\in(0,\alpha_0]$ and $\widetilde{\psi}_{\alpha,f}$ is an entire function of finite order $\leq2$.  Furthermore, as an element in $L^2(\mathbb{R})$, $\widetilde{\psi}_{\alpha,f}$ coincides with $\psi_f$ (defined in \eqref{psi_f_consequ}).
\end{lquote}
\end{lemma}
\noindent Note that $\gamma_\alpha\subset\overline{\mathbb{C}_-}$ lies in the region of analyticity for $\stackrel{\wedge}{\varphi}$. The contour $\gamma_\alpha$ is sketched in Figure \ref{figure:gamma}.

\begin{lemma}\label{lemma:phi}
\begin{lquote}
Let $\varphi$ be as in Proposition \ref{proposition:infinitely_many_zeros} or Theorem \ref{theorem:resonances_instable}. Then $\stackrel{\wedge}{\varphi}$ fulfills the hypotheses of Lemma \ref{lemma:psi_f_entire}.
\end{lquote}
\end{lemma}

\noindent\textbf{Proof of Lemma \ref{lemma:phi}:}\quad For $\varphi\in\mathcal{T}_{k_0}$, the proof is by the definition of $\mathcal{T}_{k_0}$ (see Definition \ref{defi:dilation_analytic_vector}). For $\varphi\in\mathcal{D}_{\theta_0}$, combining Remark \ref{remark:also_phi_hat}, Remark \ref{remark:B2} and Proposition \ref{proposition:B2} implies the analyticity of $\stackrel{\wedge}{\varphi}$ in the union of sectors given in Proposition \ref{proposition:B2} and the estimate \be \stackrel{\wedge}{\varphi}(e^{i\theta}k)=O(|k|^{-1/2})\hspace{0.5cm}(|k|\to\infty)\,,\nn\ee uniformly for $\theta$ in any compact subset of $\{\theta\in\mathbb{R}\,|\,|\theta|<\theta_0\}$.\\\Done

\noindent\textbf{Proof of Lemma \ref{lemma:psi_f=psi_contor}:}\quad We write $\zeta:=\kappa-i\tau$ for $\zeta\in\overline{\mathbb{C}_-}$ and calculate \be
\frac{\zeta^3}{3}=\frac{(\kappa-i\tau)^3}{3}=\frac{\kappa^3}{3}-i\tau\kappa^2-\tau^2\kappa+\frac{(-i\tau)^3}{3}\,,\label{phase}\ee which implies the estimate
\begin{align}
\Big| e^{-i\left(\frac{(\kappa-i\tau)^3}{3f}-(\kappa-i\tau)x\right)} \Big|&=|e^{-i\frac{(\kappa-i\tau)^3}{3f}}|\, |e^{i(\kappa-i\tau)x}|\nn\\
&\leq e^{\frac{\tau^3}{3f}}\, e^{-\frac{\tau\kappa^2}{f}}\, e^{|\kappa|\,|x|}\, e^{\tau |x|} \hspace{0.5cm}(\kappa\in\mathbb{R}\,,\ \tau\geq0\,\ x\in\mathbb{C})\,.\label{est}
\end{align}

\noindent \bit Proof of existence, entireness and finite order\eit:\smallskip\\
The following piecewise estimates prove the existence of the integral in \eqref{intgamma} for all $x\in\mathbb{C}$:

\begin{figure}
    \centering
    \includegraphics[draft=false, width=10cm, angle=0]{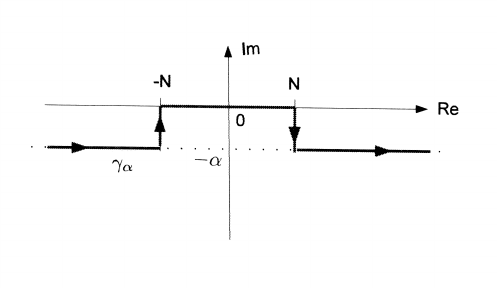}
    \caption{The contour $\gamma_\alpha$.}
    \label{figure:gamma}
\end{figure}

\noindent \bit Integral from $-N$ to $N$\eit:
\begin{align} \Phi(x)&:=\int\limits_{-N}^N e^{-i\left(\frac{k^3}{3f}-kx\right)}\stackrel{\wedge}{\varphi}(k)\;dk\ =\ \int\limits_{-N}^N e^{ikx} G(k) \;dk\hspace{0.5cm}(x\in\mathbb{R})\,,\label{Phi}\end{align} where \be G(k):=e^{-i\,\frac{k^3}{3f}}\stackrel{\wedge}{\varphi}(k)\hspace{0.5cm}(k\in\mathbb{R})\,.\nn\ee Since $\stackrel{\wedge}{\varphi}\in L^2(\mathbb{R})$, the function $G$ is in $L^2(\mathbb{R})$ and in particular in $L^2((-N,N))$. Thus (see, e.g., \cite[Chapter 19]{Rudin})
$\Phi(x)$ $(x\in\mathbb{R})$ has an entire extension and \be |\Phi(x)|\leq e^{N|x|}I_N\hspace{0.5cm}(x\in\mathbb{C})\,,\hspace{0.5cm}I_N:=\int\limits_{-N}^N|G(k)|\;dk <\infty\,.\label{intN}\ee

\noindent \bit Integrals along the paths $\widetilde{\gamma}_{\pm N}$\eit:\quad For all $x\in\mathbb{C}$ one gets
\be
\Big|\int\limits_{\widetilde{\gamma}_{+N}} e^{-i\left(\frac{k^3}{3f}-kx\right)}\stackrel{\wedge}{\varphi}(k)\;dk\Big|&=&\Big|-i\int\limits_0^\alpha e^{-i\left(\frac{(N-it)^3}{3f}-(N-it)x\right)}\stackrel{\wedge}{\varphi}(N-it)\;dt\Big|\nn\\
&\stackrel{\eqref{est}}{\leq}& \sup\limits_{\zeta\in\mathcal{A}}|\stackrel{\wedge}{\varphi}(\zeta)|\, e^{N|x|}\int\limits_0^\alpha e^{-\frac{tN^2}{f}} e^{\frac{t^3}{3f}} e^{t|x|} \;dt\nn\\
&\leq& \alpha \sup\limits_{\zeta\in\mathcal{A}}|\stackrel{\wedge}{\varphi}(\zeta)|\, e^{\frac{\alpha^3}{3f}} e^{\alpha |x|}\,e^{N|x|}\,.\label{intN+}
\ee
Similarly,
\begin{align}
\Big|\int\limits_{\widetilde{\gamma}_{-N}}& e^{-i\left(\frac{k^3}{3f}-kx\right)}\stackrel{\wedge}{\varphi}(k)\;dk\Big| = \Big|i\int\limits_0^\alpha e^{-i\left(\frac{(-N-i(\alpha-t))^3}{3f}-(-N-i(\alpha-t))x\right)}\stackrel{\wedge}{\varphi}(-N-i(\alpha-t))\;dt\Big|\nn\\
&\hspace{-0.2cm}\stackrel{\eqref{est}}{\leq} \sup\limits_{\zeta\in\mathcal{A}}|\stackrel{\wedge}{\varphi}(\zeta)|
e^{N|x|}
\int\limits_0^\alpha e^{\frac{(\alpha-t)^3}{3f}} e^{-\frac{(\alpha-t)N^2}{f}} e^{(\alpha-t)|x|}\;dt\nn\\
&= \sup\limits_{\zeta\in\mathcal{A}}|\stackrel{\wedge}{\varphi}(\zeta)|e^{N|x|}e^{\alpha|x|}e^{-\frac{\alpha N^2}{f}}e^{\frac{\alpha^3}{3f}}
\int\limits_0^\alpha e^{\frac{tN^2}{f}}e^{-t|x|}e^{-\frac{t\alpha^2}{f}}e^{\frac{t^2\alpha}{f}}e^{-\frac{t^3}{3f}} \;dt\nn\\
&\leq \alpha \sup\limits_{\zeta\in\mathcal{A}}|\stackrel{\wedge}{\varphi}(\zeta)|\, e^{\frac{\alpha^3}{3f}} e^{\frac{\alpha^3}{f}} e^{\alpha |x|}e^{N |x|}\hspace{0.5cm}(x\in\mathbb{C})\,.
\label{intN-}\end{align}

\noindent \bit Integrals from $-\infty-i\alpha$ to $-N-i\alpha$ and from $N-i\alpha$ to $\infty-i\alpha$\eit:
\be
\Big|\Big(\int\limits_{-\infty-i\alpha}^{-N-i\alpha}\hspace{-0.5cm}&+&\hspace{-0.5cm}\int\limits_{N-i\alpha}^{\infty-i\alpha}\Big) e^{-i\left(\frac{k^3}{3f}-kx\right)}\stackrel{\wedge}{\varphi}(k)\;dk\Big|\nn\\
&\leq&
\sup\limits_{\zeta\in\mathcal{A}}|\stackrel{\wedge}{\varphi}(\zeta)|\Big(\int\limits_{-\infty}^{-N}+\int\limits_{N}^{\infty}\Big)
\Big|e^{-i\left(\frac{(k-i\alpha)^3}{3f}-(k-i\alpha)x\right)}\Big|\;dk\nn\\
&\stackrel{\eqref{est}}{\leq}& \sup\limits_{\zeta\in\mathcal{A}}|\stackrel{\wedge}{\varphi}(\zeta)|\, e^{\frac{\alpha^3}{3f}}\,  e^{\alpha|x|} \int\limits_{-\infty}^\infty e^{-\frac{\alpha k^2}{f}} e^{|k|\,|x|}\;dk\nn\\
&\leq&\sup\limits_{\zeta\in\mathcal{A}}|\stackrel{\wedge}{\varphi}(\zeta)|\, e^{\frac{\alpha^3}{3f}}\,  e^{\alpha|x|} \int\limits_{-\infty}^\infty e^{-\frac{\alpha k^2}{f}} (e^{k\,|x|}+e^{-k\,|x|})\;dk\nn\\
&=& 2\sqrt{\frac{\pi f}{\alpha}}\; \sup\limits_{\zeta\in\mathcal{A}}|\stackrel{\wedge}{\varphi}(\zeta)|\, e^{\frac{\alpha^3}{3f}}\,  e^{\alpha|x|} e^{\frac{f|x|^2}{4\alpha}}\hspace{0.5cm}(x\in\mathbb{C})\,. \label{intinfty}
\ee

Finally combining \eqref{intinfty}, \eqref{intN-}, \eqref{intN+}, \eqref{intN} and \eqref{Phi} proves that the integral \eqref{intgamma} exists for all $x\in\mathbb{C}$ and obeys the estimate
\begin{align}
\Big| \int\limits_{\gamma_\alpha} e^{-i\left(\frac{k^3}{3f}-kx\right)}\stackrel{\wedge}{\varphi}(k)\;dk\Big| \leq&\  e^{N|x|}I_N
+ \alpha \sup\limits_{\zeta\in\mathcal{A}}|\stackrel{\wedge}{\varphi}(\zeta)|\, e^{\frac{\alpha^3}{3f}}\,  e^{\alpha|x|}e^{N|x|}\big(1+e^{\frac{\alpha^3}{f}}\big)\nn\\
& + 2\sqrt{\frac{\pi f}{\alpha}}\;\sup\limits_{\zeta\in\mathcal{A}}|\stackrel{\wedge}{\varphi}(\zeta)|\, e^{\frac{\alpha^3}{3f}}\,  e^{\alpha|x|}e^{\frac{f|x|^2}{4\alpha}}\hspace{0.5cm}(x\in\mathbb{C})\,.\label{finalest}\end{align}
In particular, by \eqref{intgamma} and \eqref{finalest}, \be |\widetilde{\psi}_{\alpha,f}(x)|\leq a e^{b|x|^2}\hspace{0.5cm}(x\in\mathbb{C})\,,\label{finalest2}\ee where $a,b<\infty$ are some real constants, depending on $\alpha,f,\varphi$ and $N$, but not on $x$.

We abbreviate $u(k,x):=e^{-i\left(\frac{k^3}{3f}-kx\right)}\stackrel{\wedge}{\varphi}(k)$. Then \be\psia(\cdot)=\int\limits_{\gamma_\alpha}u(k,\,\cdot\,)\;dk \nn\ee is holomorphic in all of $\mathbb{C}$, since $u(k,\,\cdot\,)$ is complex differentiable for all $k\in\gamma_\alpha$, and $\frac{1}{2}(\partial_{\textrm{Re\,}x}-i\partial_{\textrm{Im\,}x})u(k,x)$ is continuous on $\gamma_\alpha\times\mathbb{C}$. This proves the entireness of $\psia$. Then, by \eqref{finalest2}, the entire function $\psia$ is of order $\leq2$.\\\done

\noindent\bit Proof of independence of $\alpha$\eit:\quad Since for the path $\Gamma_{\pm R}:\ t\mapsto\,\pm R-it$\quad $(0<\alpha_1\leq t\leq \alpha_0)$ we have
\be \Big|\int\limits_{\Gamma_{\pm R}}e^{-i\left(\frac{k^3}{3f}-kx\right)}\stackrel{\wedge}{\varphi}(k)\;dk\Big|
&\stackrel{\eqref{est}}{\leq}&\sup\limits_{t\in[\alpha_a,\alpha_0]}|\stackrel{\wedge}{\varphi}(\pm R-it)|\int\limits_{\alpha_1}^{\alpha_0} e^{\frac{t^3}{3f}} e^{-\frac{tR^2}{f}} e^{|R|\,|x|} e^{|t|\,|x|}\;dt\nn\\
&=&o(1)\hspace{0.5cm}(R\to\infty,\ x\in\mathbb{C})\,,\nn\ee
it follows from the analyticity properties of $\stackrel{\wedge}{\varphi}$ that $\psia(x)$ $(x\in\mathbb{C})$ is independent of $\alpha\in(0,\alpha_0]$.\\\done

Thus it remains to show that $\psia$ and $\psi_f$ coincide. By an easy density argument, to identify $\psia$ with $\psi_f$ it suffices to show
\begin{align}
\lim\limits_{\alpha\downarrow 0}(g\,,\,\widetilde{\psi}_{\alpha,f})_{L^2}&=(g\,,\,\psi_f)_{L^2}\hspace{0.5cm}
(g\in C_0^\infty(\mathbb{R}))\,,\label{limit}\end{align} since $(g\,,\,\psia)_{L^2}$ is actually independent of $\alpha$. One finds
\be
\sqrt{2\pi}(g\,,\,\psia)_{L^2}&\stackrel{\eqref{intgamma}}{=}&\int\limits_\mathbb{R}\overline{g}(x)\int\limits_{\gamma_{\alpha}}
e^{-i\left(\frac{k^3}{3f}-kx\right)}\stackrel{\wedge}{\varphi}(k)
\;dk\;dx\nn\\
&=&\int\limits_\mathbb{R}\overline{g}(x)\Big(\int\limits_{-\infty}^{-N}+\int\limits_{N}^{\infty}\Big)
e^{-i\left(\frac{(k-i\alpha)^3}{3f}-(k-i\alpha)x\right)}\stackrel{\wedge}{\varphi}(k-i\alpha)\;dk\;dx\nn\\
&&+
\int\limits_{\mathbb{R}_x}\int\limits_{k\in[-N,N]} \overline{g}(x) e^{-i\left(\frac{k^3}{3f}-kx\right)}\stackrel{\wedge}{\varphi}(k)\;dk\;dx + o(1)\hspace{0.5cm}(\alpha\downarrow0)\,.\nn\\\label{217}
\ee
Setting \be F_\alpha(k,x):=\overline{g}(x) e^{-i\left(\frac{(k-i\alpha)^3}{3f}-(k-i\alpha)x\right)}\stackrel{\wedge}{\varphi}(k-i\alpha)\,, \label{F_alpha}\ee
by \eqref{est} and the uniform bound for $\stackrel{\wedge}{\varphi}$ in $\mathcal{A}$, one gets
\be |F_\alpha(k,x)|\leq \sup\limits_{\zeta\in\mathcal{A}}|\stackrel{\wedge}{\varphi}(\zeta)|\,e^{\frac{\alpha^3}{3f}}\,|g(x)| e^{-\frac{\alpha k^2}{f}} e^{|k|\,|x|} e^{\alpha|x|}\hspace{0.5cm}((k,x)\in\Omega)\,,\nn\ee where $\Omega=(\mathbb{R}\backslash[-N,N])\times\mathbb{R}$. Thus $F_\alpha\in L^1(\Omega, dk\,dx)$. Then Fubini's Theorem gives
\begin{align}
\int\limits_\mathbb{R}&\overline{g}(x)\Big(\int\limits_{-\infty}^{-N}+\int\limits_{N}^{\infty}\Big)
e^{-i\left(\frac{(k-i\alpha)^3}{3f}-(k-i\alpha)x\right)}\stackrel{\wedge}{\varphi}(k-i\alpha)\;dk\;dx\nn\\
&=\int\limits_{\mathbb{R}_x}\Big(\int\limits_{-\infty}^{-N}+\int\limits_{N}^{\infty}\Big)F_\alpha(k,x)\;dk\;dx\,.\label{Fubini}
\end{align}

By Cauchy estimates, the uniform boundedness of $\stackrel{\wedge}{\varphi}$ in an open neighborhood of $\mathcal{A}$ implies that its derivative $\stackrel{\wedge}{\varphi}'$ is uniformly bounded in a slightly smaller domain, and one has \be |\stackrel{\wedge}{\varphi}(k-i\alpha)-\stackrel{\wedge}{\varphi}(k)|\leq  \alpha\,\sup\limits_{\zeta\in\gamma}|\stackrel{\wedge}{\varphi}'(\zeta)|\,,\label{Cauchy_est}\ee where $\gamma$ denotes the path $k-i\alpha(1-t)$ $(t\in[0,1])$.
We claim that \be \mathcal{J}_\alpha:=\int\limits_{\mathbb{R}_x}\int\limits_{\mathbb{R}_k\backslash[-N,N]} \overline{g}(x) e^{-i\frac{(k-i\alpha)^3}{3f}} e^{ikx}(e^{\alpha x}-1)\stackrel{\wedge}{\varphi}(k-i\alpha)\;dk\;dx\to 0 \quad (\alpha\downarrow0)\,.\label{we_claim}\ee Indeed, by \eqref{phase} and the uniform bound on $\stackrel{\wedge}{\varphi}$ in $\mathcal{A}$,
\begin{align}
|\mathcal{J}_\alpha|
&\leq e^{\frac{\alpha^3}{3f}}\int\limits_{\mathbb{R}_x}\int\limits_{\mathbb{R}_k\backslash[-N,N]}|\overline{g}(x)|\,e^{-\frac{\alpha k^2}{f}} |e^{\alpha x}-1|\;|\stackrel{\wedge}{\varphi}(k-i\alpha)|\;dk\;dx\nn\\
&\leq e^{\frac{\alpha^3}{3f}}\sup\limits_{\zeta\in\mathcal{A}}|\stackrel{\wedge}{\varphi}(\zeta)|\,\Big(\int\limits_{\textrm{supp\,}g}|g(x)|\, |e^{\alpha x}-1|\;dx\Big)\ \Big(\int\limits_\mathbb{R}e^{-\frac{\alpha k^2}{f}}\;dk\Big)\,,
\label{223}\end{align} where \be \int\limits_\mathbb{R}e^{-\frac{\alpha k^2}{f}}\;dk=\frac{\sqrt{\pi f}}{\sqrt{\alpha}}\,,\hspace{0.5cm}
|e^{\alpha x}-1|\leq c\,\alpha|x|\hspace{0.5cm}(x\in\textrm{supp\,}g)\label{224}\ee for some constant $c$ depending on (the support of) $g$. Thus from \eqref{224} and \eqref{223} it follows that
\be |\mathcal{J}_\alpha|\leq C_{g,\stackrel{\wedge}{\varphi},f}\; e^{\frac{\alpha^3}{3f}}\sqrt{\alpha}\ \to\  0\hspace{0.5cm}(\alpha\downarrow0)\,,\ee where $C_{g,\stackrel{\wedge}{\varphi},f}$ is some constant (depending on $g$, $\stackrel{\wedge}{\varphi}$ and $f$).

Now combining \eqref{224}, \eqref{Cauchy_est} and \eqref{est} shows
\begin{align} \Big|\int\limits_{\mathbb{R}_x}&\int\limits_{\mathbb{R}_k\backslash[-N,N]}\overline{g}(x)e^{-i\frac{(k-i\alpha)^3}{3f}} e^{ikx}\big(\stackrel{\wedge}{\varphi}(k-i\alpha)-\stackrel{\wedge}{\varphi}(k)\big)\;dk\;dx\Big|\nn\\
&\leq e^{\frac{\alpha^3}{3f}}\sup\limits_{\zeta\in\mathcal{A}}|\stackrel{\wedge}{\varphi}'(\zeta)|
\int\limits_{\textrm{supp\,}g}|g(x)|\;dx\ \int\limits_{\mathbb{R}} \alpha e^{-\frac{\alpha k^2}{f}}\;dk\nn\\
&=O(\sqrt{\alpha})\hspace{0.5cm}(\alpha\downarrow0)\,.\label{226}\end{align}

By \eqref{F_alpha}, \eqref{we_claim} and \eqref{226} it follows that
\begin{align}
\lima \int\limits_{\mathbb{R}_x}\Big(\int\limits_{-\infty}^{-N}&+\int\limits_{N}^{\infty}\Big)F_\alpha(k,x)\;dk\;dx\nn\\
&= \lima \int\limits_{\mathbb{R}_x}\Big(\int\limits_{-\infty}^{-N}+\int\limits_{N}^{\infty}\Big)\overline{g}(x)
e^{ikx}e^{-i\frac{(k-i\alpha)^3}{3f}}\stackrel{\wedge}{\varphi}(k)\;dk\;dx\,.\label{227}
\end{align}
Note that $g(\cdot)$ and $e^{-i\frac{(\,\cdot\,-i\alpha)^3}{3f}}\stackrel{\wedge}{\varphi}(\cdot)$ (for $\alpha>0$) both are in $L^1(\mathbb{R})\cap L^2(\mathbb{R})$. Then \eqref{227}, \eqref{Fubini} and \eqref{217} imply
\begin{align} \lima(g\,,\,\psia)_{L^2}&=\lima \frac{1}{\sqrt{2\pi}}\int\limits_\mathbb{R}\int\limits_\mathbb{R}g(x)e^{ikx} e^{-i\frac{(k-i\alpha)^3}{3f}} \stackrel{\wedge}{\varphi}(k)\;dk\;dx\nn\\
&=\lima\int\limits_\mathbb{R}g(x) \big(e^{-i\frac{(\,\cdot\,-i\alpha)^3}{3f}}\stackrel{\wedge}{\varphi}\big)^\vee(x)\;dx\,,\label{228}
\end{align} where ${}^\vee:L^1(\mathbb{R})\cap L^2(\mathbb{R})\to L^\infty(\mathbb{R})\cap L^2(\mathbb{R})$.
By Plancherel's Theorem
\be\lima\,(g\,,\,\psia)_{L^2}\stackrel{\eqref{228}}{=}\lima\,(\stackrel{\wedge}{g}\,,\,e^{-i\frac{(\,\cdot\,-i\alpha)^3}{3f}}\stackrel{\wedge}{\varphi})_{L^2}
\label{229}\ee and \be(g\,,\,\psi_f)_{L^2}=(\stackrel{\wedge}{g}\,,\,\stackrel{\wedge}{\psi}_f)_{L^2}\stackrel{\eqref{psi_f_consequ}}{=}
(\stackrel{\wedge}{g}\,,\,e^{-i\frac{(\cdot)^3}{3f}}\stackrel{\wedge}{\varphi})_{L^2}\,.\label{230}\ee
Then \be \Big|\lima\,(g\,,\,\psia)_{L^2}-(g\,,\,\psi_f)_{L^2}\Big|&\stackrel{{\eqref{229}\atop\eqref{230}}}{=}& \lima \Big| \big(\stackrel{\wedge}{g}\,,\,(e^{-i\frac{(\,\cdot\,-i\alpha)^3}{3f}}-e^{-i\frac{(\cdot)^3}{3f}})\stackrel{\wedge}{\varphi}\big)_{L^2}\Big|\nn\\
&\stackrel{\eqref{phase}}{=}&\lima \Big|(\stackrel{\wedge}{g}\,,\,e^{-\frac{\alpha k^2}{f}-\frac{i\alpha^2k}{f}+\frac{\alpha^3}{3f}}\stackrel{\wedge}{\varphi})_{L^2}\Big|\nn\\
&\leq& \|\stackrel{\wedge}{g}\|_2\ \lima\,\|e^{-\frac{\alpha (\cdot)^2}{f}}e^{\frac{\alpha^3}{3f}}\stackrel{\wedge}{\varphi}\|_2\,,\nn\ee
and applying Lebesgue's Theorem on dominated convergence to $\lima\,\|e^{-\frac{\alpha (\cdot)^2}{f}}e^{\frac{\alpha^3}{3f}}\stackrel{\wedge}{\varphi}\|_2$ implies
\begin{align}
\lima\,(g\,,\,\psia)_{L^2}-(g\,,\,\psi_f)_{L^2}=0\,,
\nn\end{align} which is \eqref{limit}.\\\Done

\noindent\textbf{Proof of Proposition \ref{proposition:infinitely_many_zeros}:}\quad
Let $f>0$. We rewrite \eqref{F_f} as
\begin{align}
F_{f,\varphi}(z)=1-z-r_{f,\varphi}(z)\,,\hspace{0.5cm}r_{f,\varphi}(z):=(\varphi\,,\,(p^2+fx-z)^{-1}\varphi)_{L^2} \hspace{0.5cm} (z\in\mathbb{C}_+).\label{r_phi_f}
\end{align}
\bit Proof of entireness\eit:\quad Obviously, the functions $r_{f,\varphi}$ and $F_{f,\varphi}$ are analytic in $\mathbb{C}_+$. Lemma \ref{lemma:unitary_equivalence} gives
\begin{align}
r_{f,\varphi}(z)
&=(e^{-i\frac{p^3}{3f}}\varphi\,,\,(fx-z)^{-1}e^{-i\frac{p^3}{3f}}\varphi)_{L^2}
=(\psi_f\,,\,(fx-z)^{-1}\psi_f)_{L^2}\nn\\
&=\frac{1}{f}\int\limits_\mathbb{R}\frac{\overline{\psi}_f(\overline{x})\psi_f(x)}{x-\frac{z}{f}}\;dx\hspace{0.5cm}(z\in\mathbb{C}_+)\,,\label{resolventH(f)}
\end{align} where $\psi_f$ is given by \eqref{psi_f_consequ}. By Lemma \ref{lemma:psi_f_entire} and Lemma \ref{lemma:phi} the function $\psi_f$ has an entire extension. In \eqref{resolventH(f)} we can deform the path $\mathbb{R}$ into a contour $\gamma$ in $\overline{\mathbb{C}_-}$ such that the singularity $x=z/f$ for $\textrm{Im\,}z<0$ is enclosed by this contour $\gamma$ and the real line. Then using the Residue Theorem one obtains
\be r_{f,\varphi}(z)^{\mathbf{c},\Omega_f}=\frac{2\pi i}{f}\;\overline{\psi}_f\Big(\frac{\overline{z}}{f}\Big)\psi_f\Big(\frac{z}{f}\Big)+
(\varphi\,,\,(p^2+fx-z)^{-1}\varphi)_{L^2}\hspace{0.5cm}(z\in\mathbb{C}_-)\,.\label{cont_resolvent}\ee
For $z\in\mathbb{R}$ close to a fixed $x_0\in\mathbb{R}$ (more precisely, $|z-x_0|<\frac{\delta}{2}\cdot f$ for some fixed $\delta>0$), this extension is given by \be r_{f,\varphi}(z)^{\mathbf{c},\Omega_f}=\frac{2\pi i}{f}\;\overline{\psi}_f\Big(\frac{z}{f}\Big)\psi_f\Big(\frac{z}{f}\Big)+\frac{1}{f}\Big(\int\limits_{-\infty}^{\frac{x_0}{f}-\delta}
+\int\limits_{\mathcal{K}_\delta(\frac{x_0}{f})}+
\int\limits_{\frac{x_0}{f}+\delta}^{+\infty}\Big)\frac{\overline{\psi}_f(\overline{x})\psi_f(x)}{x-\frac{z}{f}}\;dx,\label{zreal}\ee where $\mathcal{K}_\delta(\frac{x_0}{f})$, denotes the clockwise oriented semicircle in the closed upper half complex plane with radius $\delta$ and center $x_0/f$.
Since $\psi_f$ is entire by Lemma \ref{lemma:psi_f_entire} and the integral in \eqref{zreal} exists for all $z\in\mathbb{R}$, it follows from \eqref{resolventH(f)}, \eqref{cont_resolvent} and \eqref{zreal} that the extension of $r_{f,\varphi}$ is entire. Thus, by \eqref{r_phi_f}, $F_{f,\varphi}^{\mathbf{c},\Omega_f}$ is entire. \\\done

\noindent\bit Proof of finite order\eit:\quad Fix $f>0$. For $|\textrm{Im\,}\frac{z}{f}|>1$ one gets, using the identity in \eqref{resolventH(f)},
\begin{align}
|(\varphi\,,\,(p^2+fx-z)^{-1}\varphi)_{L^2}|&=\frac{1}{f}|(\psi_f\,,\,(x-\frac{z}{f})^{-1}\psi_f)_{L^2}|\leq\frac{1}{f}
\int\limits_\mathbb{R}\frac{|\overline{\psi}_f(x)\psi_f(x)|}{|x-\frac{z}{f}|}\;dx\nn\\
&\leq \frac{1}{f}\int\limits_\mathbb{R}\frac{|\overline{\psi}_f(x)\psi_f(x)|}{|\textrm{Im\,}\frac{z}{f}|}\;dx
<\frac{1}{f}\int\limits_\mathbb{R}|\psi_f(x)|^2\;dx\nn\\
&=\frac{1}{f}\|\psi_f\|^2_2\,.\label{fo1}
\end{align}
For $|\textrm{Im\,}\frac{z}{f}|\leq1$ boundedness of the resolvent matrix element is shown by contour deformation and analytically continuing from $\{z\in\mathbb{C}\,|\,\frac{\textrm{Im\,}z}{f}>1\}$ into the strip $\sigma_f:=\{z\in\mathbb{C}\,|\,|\frac{\textrm{Im\,}z}{f}|\leq1\}$:
Denoting by $K_2(\textstyle\frac{\textrm{Re\,}z}{f})$ the counterclockwise oriented semicircle in the closed lower half complex plane with radius 2 and center $\textrm{Re\,}z/f$, we use the contour $(-\infty\,,\,\frac{\textrm{Re\,}z}{f}-2)\cup K_2(\frac{\textrm{Re\,}z}{f})\cup(\frac{\textrm{Re\,}z}{f}+2\,,\,+\infty)$ to get (using the analyticity of $r_{f,\varphi}$ established above)
\begin{align}
(\varphi\,,\,(p^2+fx-z)^{-1}\varphi)^{\mathbf{c},\sigma_f}_{L^2}&=\frac{1}{f}\Big(\int\limits_{-\infty}^{\frac{\textrm{Re\,}z}{f}-2}
+\int\limits_{K_2(\frac{\textrm{Re\,}z}{f})}
+\int\limits_{\frac{\textrm{Re\,}z}{f}+2}^{+\infty}\Big)\frac{\overline{\psi}_f(\overline{x})\psi_f(x)}{x-\frac{z}{f}}\;dx
\label{fo1a}\end{align} for all $z\in\mathbb{C}_+\cup\sigma_f$. For $z\in\sigma_f$, one finds
\begin{align}
\Big|\frac{1}{f}\int\limits_{\mathbb{R}\backslash(\frac{\textrm{Re\,}z}{f}-2,\frac{\textrm{Re\,}z}{f}+2)} \frac{\overline{\psi}_f(\overline{x})\psi_f(x)}{x-\frac{z}{f}}\;dx\Big|&\leq \frac{1}{f}\int\limits_{\mathbb{R}\backslash(\frac{\textrm{Re\,}z}{f}-2,\frac{\textrm{Re\,}z}{f}+2)}\frac{|\psi_f(x)|^2}{|x-\frac{z}{f}|}\;dx\nn\\
&\leq\frac{1}{2f}\int\limits_\mathbb{R}|\psi_f(x)|^2\;dx=\frac{1}{2f}\,\|\psi_f\|^2_2\,.\label{fo2}
\end{align} Since $|x-z/f|\geq1$ for $x\in K_2(\frac{\textrm{Re\,}z}{f})$ and $z\in\sigma_f$, one has
\be
\Big|\frac{1}{f}\int\limits_{K_2(\frac{\textrm{Re\,}z}{f})} \frac{\overline{\psi}_f(\overline{x})\psi_f(x)}{x-\frac{z}{f}}\;dx\Big|\leq
\frac{1}{f}\int\limits_{K_2(\frac{\textrm{Re\,}z}{f})}|\overline{\psi}_f(\overline{x})| |\psi_f(x)|\;|dx|\,.\label{star}\ee
By \eqref{psi_order2_est},
\begin{align}\frac{1}{f}\int\limits_{K_2(\frac{\textrm{Re\,}z}{f})}|\overline{\psi}_f(\overline{x})| |\psi_f(x)|\;|dx|
&\leq\frac{1}{f}\int\limits_{K_2(\frac{\textrm{Re\,}z}{f})}a e^{b|\overline{x}|^2} a e^{b|x|^2}\;|dx|\nn\\
&\leq\frac{2\pi}{f}  \max\limits_{x\in K_2(\frac{\textrm{Re\,}z}{f})}(a^2 e^{2b|x|^2})\nn\\
&\leq \frac{\widetilde{a}}{f}\,e^{\widetilde{b}\frac{|z|^2}{f^2}}\hspace{0.5cm}(\frac{|\textrm{Im\,}z|}{f}\leq1)
\label{fo3}
\end{align} for some finite real constants $\widetilde{a}$ and $\widetilde{b}$.
Finally, for $\textrm{Im\,}z/f< -1$, we use \eqref{cont_resolvent} and Lemma \ref{lemma:psi_f_entire} (more precisely, \eqref{psi_order2_est}) to get
\be
|r_{f,\varphi}^{\mathbf{c},\Omega_f}(z)|\leq A\;e^{B\frac{|z|^2}{f^2}}\label{fo4}
\ee for some finite constants $A$ and $B$ (possibly depending on $f$ and $\varphi$). Combining the estimates \eqref{fo1}, \eqref{fo3} together with \eqref{fo2}, and \eqref{fo4} proves that $r_{f,\varphi}^{\mathbf{c},\Omega_f}$ is of order $\leq2$. Thus, by \eqref{r_phi_f}, $F_{f,\varphi}^{\mathbf{c},\Omega_f}$ is of order $\leq2$.\\\done

\noindent \bit Proof of existence of infinitely many zeros\eit:\quad In order to prove that $F_{f,\varphi}^{\mathbf{c},\Omega_f}$ has infinitely many zeros it suffices to show (by Theorem \ref{theorem:infinitely_many}) that there exist no polynomials (in one variable) $P$ and $Q$ such that $F_{f,\varphi}^{\mathbf{c},\Omega_f}(z)=P(z)e^{Q(z)}$ for all $z\in\mathbb{C}$:

Since $p^2+fx$ is self-adjoint, one has $|r_{f,\varphi}(z)|\leq \textrm{const.\,} |\textrm{Im\,}z|^{-1}$,  $\textrm{Im\,}z> 0$. In particular,
\be \lim\limits_{\textrm{Im\,}z\to\infty}r_{f,\varphi}(z)=0\,.\label{r_to_0}\ee
Now assume that \be F_{f,\varphi}^{\mathbf{c},\Omega_f}(z)=P(z)e^{Q(z)}\hspace{0.5cm}(z\in\mathbb{C})\nn\ee for some polynomials $P$ and $Q$. Then
\be
0 = \lim\limits_{\textrm{Im\,}z\to\infty}F_{f,\varphi}^{\mathbf{c},\Omega_f}(z)-P(z)e^{Q(z)}  =  \lim\limits_{\textrm{Im\,}z\to\infty}1-z - P(z)e^{Q(z)}\,.\label{4}\ee
Equation \eqref{4} implies that $Q(z)$ is at most of degree one since otherwise the exponential would approach infinity rapidly along certain rays in the upper half plane.  In addition if $Q(z) = az +b$ with $a\ne 0$ the same reasoning implies $a=i\widetilde{a}$ with $\widetilde{a}>0$ which also contradicts \eqref{4} for $z\in\mathbb{C}_+$. Thus \be F_{f,\varphi}^{\mathbf{c},\Omega_f}(z)=1-z-r_{f,\varphi}^{\mathbf{c},\Omega_f}(z)=c\,P(z)\hspace{0.5cm}(z\in\mathbb{C})\label{6}\ee for some constant $c$. But, for $\textrm{Im\,}z>0$,
\be
\lim\limits_{\textrm{Im\,}z\to\infty}(-i\,\textrm{Im\,}z)r_{f,\varphi}^{\mathbf{c},\Omega_f}(i\,\textrm{Im\,}z)
\stackrel{\eqref{r_phi_f}}{=}\lim\limits_{\textrm{Im\,}z\to\infty}\int\limits_\mathbb{R}\frac{|\varphi(x)|^2 (-i\,\textrm{Im\,}z)}{p^2-fx-i\,\textrm{Im\,}z}\;dx=\|\varphi\|^2_2\,.\nn
\ee
Thus there exists no polynomial $P$ fulfilling \eqref{6} (except if $\varphi=0$). This finishes the proof of Proposition \ref{proposition:infinitely_many_zeros}.\\\Done

\subsection{Proof of Theorem \ref{theorem:resonances_instable}}\label{subsection:instability}
Our proof of Theorem \ref{theorem:resonances_instable} uses the expansions and estimates given in Proposition \ref{proposition:integrals}, Corollary \ref{corollary:psi}, Proposition \ref{proposition:resolventLHP}, Proposition \ref{proposition:resolvent_continued} and Corollary \ref{corollary:res_c} below.

\begin{proposition}\label{proposition:integrals}
\begin{lquote}Let $\varphi$ and $M$ be as in Theorem \ref{theorem:resonances_instable}. Let $\alpha\in(0,\min\{\tan\theta_0,k_0\})$.
Define the paths
\begin{align} \mathscr{C}_\pm &:\ \mathbb{R}_\pm\to\mathbb{C}\,,\ t\mapsto k(t):=\left\{\begin{array}{r@{\;,\quad}l} t-i\alpha&|t|>1
\\(\textrm{sgn\,}(t) - i\alpha)|t|&-1\leq t \leq 1\end{array}\right.\label{curve}\end{align} (which lie in the region of analyticity for $\stackrel{\wedge}{\varphi}$). Then, for $f>0$ sufficiently small,
\begin{align}\hspace{-1cm}\int\limits_{\mathscr{C}_+}e^{-\frac{i}{f}\left(\frac{k^3}{3}-kz\right)}
\stackrel{\wedge}{\varphi}(k)\;dk&=\Big(\sqrt{\frac{\pi f}{z^{1/2}}}e^{-i\pi/4}\stackrel{\wedge}{\varphi}(\sqrt{z})+O(f)\Big)e^{i\frac{2}{3f}z^{3/2}}
\,,\label{int_plus}\\
\hspace{-1cm}\int\limits_{\mathscr{C}_-}e^{-\frac{i}{f}\left(\frac{k^3}{3}-kz\right)}
\stackrel{\wedge}{\varphi}(k)\;dk&=\Big(\sqrt{\frac{\pi f}{z^{1/2}}}e^{i\pi/4}\stackrel{\wedge}{\varphi}(-\sqrt{z})+O(f)\Big)e^{-i\frac{2}{3f}z^{3/2}}+O(f)\label{int_minus}\\
&=\sqrt{\frac{\pi f}{z^{1/2}}}e^{i\pi/4}\stackrel{\wedge}{\varphi}(-\sqrt{z})e^{-i\frac{2}{3f}z^{3/2}}+O(f)\label{int_minus2}\end{align}
as $f\downarrow0$, \unif.
\end{lquote}
\end{proposition}
Note that by analyticity of $\stackrel{\wedge}{\varphi}$, Lemma \ref{lemma:psi_f_entire}, Lemma \ref{lemma:psi_f=psi_contor} and Lemma \ref{lemma:phi} and contour deformation, one has
\be \sqrt{2\pi}\,\psi_f(x)\stackrel{\eqref{intgamma}}{=}\int\limits_{\gamma_\alpha} e^{-i\left(\frac{k^3}{3f}-kx\right)}\stackrel{\wedge}{\varphi}(k)\;dk=\int\limits_{\mathscr{C}_-\cup\mathscr{C}_+} \hspace{-0.2cm}e^{-i\left(\frac{k^3}{3f}-kx\right)}\stackrel{\wedge}{\varphi}(k)\;dk\hspace{0.3cm}(x\in\mathbb{C})\,.\label{22.a}\ee Then as a corollary of Proposition \ref{proposition:integrals} and \eqref{22.a}, one gets

\begin{corollary}\label{corollary:psi}
\begin{lquote} Let $\varphi$ and $M$ be as in Theorem \ref{theorem:resonances_instable}. Let $\psi_f$ be given by \eqref{psi_f_consequ}. Then \begin{align} \sqrt{2\pi}\psi_f\Big(\frac{z}{f}\Big)=&\Big(\sqrt{\frac{\pi f}{z^{1/2}}}e^{-i\pi/4}\stackrel{\wedge}{\varphi}(\sqrt{z})+O(f)\Big)e^{i\frac{2}{3f}z^{3/2}}\nn\\
&+\Big(\sqrt{\frac{\pi f}{z^{1/2}}}e^{i\pi/4}\stackrel{\wedge}{\varphi}(-\sqrt{z})+O(f)\Big)e^{-i\frac{2}{3f}z^{3/2}}+O(f)\nn\end{align}
and
\begin{align}
\sqrt{2\pi}\;\overline{\psi}_f\Big(\frac{\overline{z}}{f}\Big)=&\Big(\sqrt{\frac{\pi f}{z^{1/2}}}e^{-i\pi/4}\overline{\stackrel{\wedge}{\varphi}}(-\overline{\sqrt{z}})+O(f)\Big)e^{i\frac{2}{3f}z^{3/2}}\nn\\
&+\Big(\sqrt{\frac{\pi f}{z^{1/2}}}e^{i\pi/4}\overline{\stackrel{\wedge}{\varphi}}(\overline{\sqrt{z}})+O(f)\Big)e^{-i\frac{2}{3f}z^{3/2}}+O(f)\nn
\end{align}
as $f\downarrow0$, \unif.
\end{lquote}
\end{corollary}

\begin{figure}[t]
                \centering
                \includegraphics[angle=0,width=0.8\textwidth]{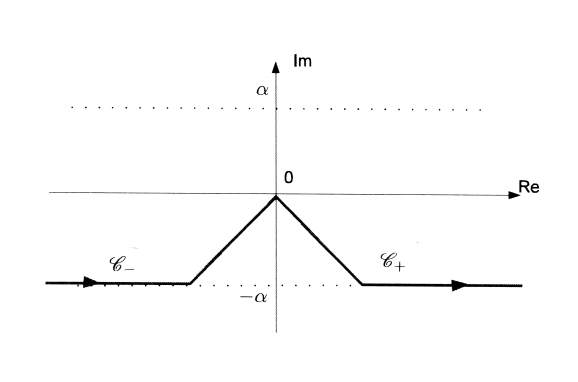}
                \caption{The contour $\mathscr{C}_-\cup\mathscr{C}_+$ defined in \eqref{curve}.}
                \label{fig:contour_C_pm}\end{figure}
\begin{figure}[h!]
                \centering
                \includegraphics[angle=0,width=0.8\textwidth]{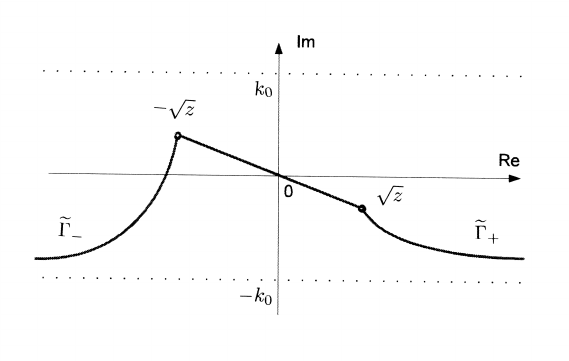}
                \caption{Deformed $\mathscr{C}_-\cup\mathscr{C}_+$ as used in the method of steepest descents.}
                \label{fig:deformed_C_pm}\end{figure}

\noindent\textbf{Proof of Proposition \ref{proposition:integrals}:}\quad
We break up each of the integrals in \eqref{int_plus} and \eqref{int_minus} into three parts after deforming the contour $\mathscr{C}_\pm$ as shown in Figure \ref{fig:deformed_C_pm}. There is one contribution handled by the method of steepest descents (see, e.g., \cite{E}) near $\pm\sqrt{z}$. There is one contribution integrating from $-\sqrt{z}$ to 0 (from 0 to $\sqrt{z}$, respectively). And finally there is an error term where we integrate from near $-\infty$ in $\mathbb{R}$ to near $-\sqrt{z}$ and similarly from near $+\sqrt{z}$ to near $+\infty$ in $\mathbb{R}$.\\
The bounds on $\sqrt{z}$ in the definition of $M$ in Theorem \ref{theorem:resonances_instable} make sure that for $z\in M$ the deformed integration contours used below lie in the region of analyticity for $\stackrel{\wedge}{\varphi}$.\smallskip\\
First we prove \eqref{int_plus}. For this we write
\be
\int\limits_{\mathscr{C}_+}e^{-\frac{i}{f}\left(\frac{k^3}{3}-kz\right)}
\stackrel{\wedge}{\varphi}(k)\;dk = I_1(z,f)+I_2(z,f)+I_3(z,f)\hspace{0.5cm}(z\in M\,,\ f>0)\,,\label{int_plus=sum}\ee where $I_j(z,f)$ $(j\in\{1,2,3\})$ are defined in \eqref{I_1}, \eqref{I_2} and \eqref{I_3} below.\medskip\\
\noindent\bit Integration from $0$ to $\sqrt{z}$\eit:\quad Defining
\be x:=\frac{z^{3/2}}{f}\,,\quad u:=s(1-\frac{s}{3})^{1/2}\label{x_and_u}\ee
gives, for $z\in M$ and $f>0$,
\begin{align}I_1(z,f)&:= \int\limits_{0}^{\sqrt{z}}e^{-\frac{i}{f}\left(\frac{k^3}{3}-kz\right)}
\stackrel{\wedge}{\varphi}(k)\;dk= \sqrt{z}\; e^{2ix/3}\int\limits_0^1e^{-ixs^2(1-s/3)}\stackrel{\wedge}{\varphi}((1-s)\sqrt{z})\;ds\label{I_1}\\
&= \sqrt{z}\; e^{2ix/3}\int\limits_0^{\sqrt{2/3}}e^{-ixu^2}\stackrel{\wedge}{\varphi}((1-s(u))\sqrt{z})\;\frac{ds(u)}{du}\;du\nn\\
&= \sqrt{z}\;\stackrel{\wedge}{\varphi}(\sqrt{z})e^{2ix/3}\int\limits_0^{\sqrt{2/3}}e^{-ixu^2}du  + E_1(z,f) e^{2ix/3}\,,\label{contribution1}
\end{align}
where
\be
E_1(z,f):=\frac{i\sqrt{z}}{x}\int\limits_0^{\sqrt{2/3}}\Big(\frac{d}{du}\,e^{-ixu^2}\Big)
\frac{\stackrel{\wedge}{\varphi}((1-s(u))\sqrt{z})\frac{ds(u)}{du}-\stackrel{\wedge}{\varphi}(\sqrt{z})}{2u}\;du\,.\nn
\ee
After integrating by parts one finds \be|E_1(z,f)|\leq cf\sup\big\{|\stackrel{\wedge}{\varphi}^{(n)}(t\sqrt{z})|\,\big|\,t\in(0,1]\,,\ n\leq2\big\}\label{E_1_estimate}\ee
for some $c>0$. We calculate
\begin{align}
\int\limits_0^{\sqrt{2/3}}e^{-ixu^2}\;du &= \int\limits_0^\infty e^{-ixu^2}\;du + O(|x|^{-1})= \frac{1}{2}\sqrt{\frac{\pi}{ix}}+O(|x|^{-1})\hspace{0.5cm}(|x|\to\infty)\nn\\
&\stackrel{\eqref{x_and_u}}{=} \frac{1}{2\sqrt{z}}\;\sqrt{\frac{\pi f}{z^{1/2}}}\,e^{-i\pi/4}+O(f)\hspace{0.5cm}(f\downarrow0)\,.\label{2.12}
\end{align}
Then using \eqref{2.12} and \eqref{E_1_estimate} in \eqref{contribution1} gives
\be
I_1(z,f) = \Big(\frac{1}{2} \sqrt{\frac{\pi f}{z^{1/2}}}\,e^{-i\pi/4}\stackrel{\wedge}{\varphi}(\sqrt{z})+O(f)\Big) e^{i\frac{2}{3f}\,z^{3/2}}\hspace{0.5cm}(f\downarrow 0)\,,\label{I_1estimate}
\ee \unif.\smallskip\\

\noindent \bit Integration near $+\sqrt{z}$\eit:\quad Define the integral
\be
I_2(z,f):=\int\limits_{\widetilde{\Gamma}_+}e^{-\frac{i}{f}\left(\frac{k^3}{3}-kz\right)}\stackrel{\wedge}{\varphi}(k)\;dk\hspace{0.5cm}(z\in M,\ f>0)
\label{I_2}
\ee
along a particular path $\widetilde{\Gamma}_+$ near $k=\sqrt{z}$. $\widetilde{\Gamma}_+$ is given as follows: We write $k=\zeta+\sqrt{z}$ which gives
\be
-\frac{2iz^{3/2}}{3}-i\Big(\frac{k^3}{3}-kz\Big)=-i\Big(\frac{\zeta^3}{3}+\sqrt{z}\zeta^2\Big).
\label{change_of_variables_k}\ee Using the method of steepest descents we want $\zeta$ to be on a curve, $\Gamma_+$, satisfying \be\textrm{Re\,}\Big(\frac{\zeta^3}{3}+\sqrt{z}\zeta^2\Big)=0.
\ee
We want $0\leq \textrm{Re\,}\zeta\leq\delta_1$ for some sufficiently small $\delta_1>0$ and $\textrm{Im\,}\zeta\leq0$.
With the abbreviations
\be \gamma:=\textrm{Re\,}\sqrt{z}\,,\quad  -\nu:=\textrm{Im\,}\sqrt{z}<0\,,\quad
x:=\textrm{Re\,}\zeta\,,\quad y:=\textrm{Im\,}\zeta\label{abbreviations}\ee
we obtain for $\zeta=x+iy$ on $\Gamma_+$
\be
y=-\gamma^{-1}(|z|^{1/2}-\nu)x+\sum\limits_{n=2}^\infty A_n x^n\hspace{0.5cm}(x\in[0,\delta_1])
\label{power1}
\ee
for some $A_n\in\mathbb{R}$. On this curve we have
\be
-i\Big(\frac{\zeta^3}{3}+\sqrt{z}\zeta^2\Big)=-b^2 x^2 +\sum\limits_{n=3}^\infty B_n x^n\,,\label{power2}
\ee
where
\be
b^2:=2\gamma^{-2}|z|(|z|^{1/2}-\nu)\nn
\ee
and $B_n$ is real. The power series in \eqref{power1} and \eqref{power2} converge for $|x|<\gamma$. (Note that for $z\in M$, $|z|^{1/2}-\nu>0$.) We have
\be
e^{-i\frac{2}{3f}z^{3/2}}\int\limits_{\widetilde{\Gamma}_+}e^{-\frac{i}{f}\left(\frac{k^3}{3}-kz\right)}
\stackrel{\wedge}{\varphi}(k)\;dk
\stackrel{\eqref{change_of_variables_k}}{=}\int\limits_{\Gamma_+}e^{-\frac{i}{f}
\left(\frac{\zeta^3}{3}+\sqrt{z}\zeta^2\right)}
\stackrel{\wedge}{\varphi}(\zeta+\sqrt{z})\;d\zeta\,, \label{aa}\ee where $\Gamma_+$ is given by \eqref{power1}.
Substituting \eqref{power1} and \eqref{power2} into \eqref{aa} and expanding $\stackrel{\wedge}{\varphi}(\zeta+\sqrt{z})$ in a power series we obtain
\be \int\limits_{\Gamma_+}e^{-\frac{i}{f}\left(\frac{\zeta^3}{3}+\sqrt{z}\zeta^2\right)}
\stackrel{\wedge}{\varphi}(\zeta+\sqrt{z})\;d\zeta
= \frac{1}{2} \sqrt{\frac{\pi f}{z^{1/2}}}\,e^{-i\pi/4}\stackrel{\wedge}{\varphi}(\sqrt{z})+O(f)\hspace{0.5cm}(f\downarrow0)\,,\label{use_of_power_exp1}\ee
and thus \be
I_2(z,f)=e^{i\frac{2}{3f}z^{3/2}}\Big(\frac{1}{2} \sqrt{\frac{\pi f}{z^{1/2}}}\,e^{-i\pi/4}\stackrel{\wedge}{\varphi}(\sqrt{z})+O(f)\Big)\hspace{0.5cm}(f\downarrow0)\,,\label{I_2estimate}
\ee \unif.\\

\noindent\bit Connecting $\widetilde{\Gamma}_+$ to infinity\eit:\quad We can give an exact expression for the path $\Gamma_+$ (defined by \eqref{power1}) as follows:
\be y=x\left(\frac{\nu}{\gamma+x}-\sqrt{\left(\frac{\nu}{\gamma+x}\right)^2+\frac{\gamma+\frac{x}{3}}{\gamma+x}}\right)\,,\nn
\ee
from which it follows that \be \left|\frac{y}{x}\right|\leq\sqrt{\frac{\gamma+\frac{x}{3}}{\gamma+x}}\label{y/x}\ee for all $\zeta=x+iy$ on $\Gamma_+$. We continue the path $\Gamma_+$ by starting at $\zeta(\delta_1)=:t_0-i\Theta_0$ (some $t_0,\Theta_0>0$). From \eqref{y/x}  with $x=t_0$ and $y=-\Theta_0$ we get $t_0>\Theta_0>0$. Thus, for $\zeta(t)=t-i\Theta_0$ $(t\geq t_0)$, we have
\begin{align}
\textrm{Re\,}\left(-i\left(\frac{\zeta^3(t)}{3}+\sqrt{z}\zeta^2(t)\right)\right)&=
\textrm{Re\,}\left(-i(t^2-\Theta_0^2-2it\Theta_0)\Big(\frac{t}{3}+\gamma-i(\nu+\frac{\Theta_0}{3})\Big)\right)\nn\\
&=-(t^2-\Theta_0^2)\Big(\nu+\frac{\Theta_0}{3}\Big)-2t\Theta_0\Big(\frac{t}{3}+\gamma\Big)\nn\\
&=:-G(t)<0\hspace{0.5cm}(t\geq t_0)\,.\nn
\end{align} The integral \begin{align}I_3(z,f)&:=e^{i\frac{2}{3f}\,z^{3/2}}\int\limits_{t_0-i\Theta_0}^{\infty-i\Theta_0}
e^{-\frac{i}{f}\left(\frac{\zeta^3}{3}+\sqrt{z}\zeta^2\right)}
\stackrel{\wedge}{\varphi}(\zeta+\sqrt{z})\;d\zeta\hspace{0.5cm}(z\in M\,,\ f>0)\label{I_3}\end{align} obeys the estimate
\begin{align}
\Big|\int\limits_{t_0-i\Theta_0}^{\infty-i\Theta_0}
e^{-\frac{i}{f}\left(\frac{\zeta^3}{3}+\sqrt{z}\zeta^2\right)}
\stackrel{\wedge}{\varphi}(\zeta+\sqrt{z})\;d\zeta\Big|&\leq \int\limits_{t_0}^\infty e^{-G(t)/f} |\stackrel{\wedge}{\varphi}(t+\sqrt{z}-i\Theta_0)|\;dt\nn\\
 &\leq e^{-c/f} \sup\limits_{t\geq t_0}|\stackrel{\wedge}{\varphi}(t+\sqrt{z}-i\Theta_0)|\label{I_3estimate}
\end{align}
for some $c>0$ independent of $f$, \unif. Note that $\Theta_0$ is of the order of $\delta_1$ so that $c$ can be taken independent of $\nu$. Note also that if $\delta_1$ is taken sufficiently small relatively to $\gamma$, this path is in a region of analyticity of $\stackrel{\wedge}{\varphi}$. \smallskip\\

Finally combining \eqref{I_3estimate}, \eqref{I_3}, \eqref{I_2estimate}, \eqref{I_1estimate} and \eqref{int_plus=sum} proves \eqref{int_plus}.\\\done

\noindent Next we prove \eqref{int_minus}. For this we write
\be
\int\limits_{\mathscr{C}_-}e^{-\frac{i}{f}\left(\frac{k^3}{3}-kz\right)}
\stackrel{\wedge}{\varphi}(k)\;dk = I_4(z,f)+I_5(z,f)-I_6(z,f)\hspace{0.5cm}(z\in M\,,\ f>0)\,,\nn\ee where $I_j(z,f)$ $(j\in\{4,5,6\})$ are defined in \eqref{I_4}, \eqref{I_5} and \eqref{I_6} below.\medskip\\

\noindent\bit Integration from $-\sqrt{z}$ to $0$\eit:\quad Using \eqref{x_and_u} we obtain, for $z\in M$ and $f>0$,
\begin{align} I_4(z,f)&:=
\int\limits_{-\sqrt{z}}^0 e^{-\frac{i}{f}\left(\frac{k^3}{3}-kz\right)}
\stackrel{\wedge}{\varphi}(k)\;dk\label{I_4}\\
&= \sqrt{z}\, e^{-2ix/3}\int\limits_0^1e^{ixs^2(1-s/3)}\stackrel{\wedge}{\varphi}(-(1-s)\sqrt{z})\;ds\nn\\
&= \sqrt{z}\, e^{-2ix/3}\int\limits_0^{\sqrt{2/3}}e^{ixu^2}\stackrel{\wedge}{\varphi}(-(1-s(u))\sqrt{z})\;\frac{ds(u)}{du}\,du\nn\\
&= \sqrt{z}\,\stackrel{\wedge}{\varphi}(-\sqrt{z})\, e^{-2ix/3}\int\limits_0^{\sqrt{2/3}}e^{ixu^2}\;du  + E_2(z,f) e^{-2ix/3}\,,
\label{contribution2}\end{align} where
\be
E_2(z,f):=\frac{-i\sqrt{z}}{x}\int\limits_0^{\sqrt{2/3}}\Big(\frac{d}{du}\,e^{ixu^2}\Big)
\frac{\stackrel{\wedge}{\varphi}(-(1-s(u))\sqrt{z})\frac{ds(u)}{du}-\stackrel{\wedge}{\varphi}(-\sqrt{z})}{2u}\;du\,.\nn
\ee
Note that $|e^{-2ix/3}|\leq1$, possibly much smaller than 1. On the other hand if $|\textrm{Im\,}z|\leq cf$ for some $c>0$, the term $|e^{-2ix/3}|$ may not be small. An integration by parts shows
\be|E_2(z,f)|\leq c |x|^{-1}\sup\big\{|\stackrel{\wedge}{\varphi}^{(n)}(-t\sqrt{z})|\,\big|\,t\in(0,1],\ n\leq2\big\}\,|e^{2ix/3}|,\label{E_2_estimate}\ee where by \eqref{x_and_u} $|x|^{-1}=O(f)$ as $f\downarrow 0$, uniformly for $z\in M$. The integral $\int_0^{\sqrt{2/3}}e^{ixu^2}du$ in \eqref{contribution2} is easily handled by Cauchy's Theorem: Using $v:=u\sqrt{x}$ we obtain
\begin{align}
\int\limits_0^{\sqrt{2/3}}e^{ixu^2}\;du &=\frac{1}{\sqrt{x}}\int\limits_0^{\sqrt{2x/3}} e^{iv^2}\;dv\nn\\
&= \frac{1}{\sqrt{x}}\int\limits_0^{\textrm{Re\,}\sqrt{2x/3}}e^{iv^2}\;dv
+\frac{1}{\sqrt{x}}\int\limits_{\textrm{Re\,}\sqrt{2x/3}}^{\sqrt{2x/3}}e^{iv^2}\;dv\,.\label{contribution2a}
\end{align}
We set \be a:=\textrm{Re\,}\sqrt{2x/3}>0\,,\quad -b:=\textrm{Im\,}\sqrt{2x/3}<0\,.\label{a,b}\ee
Since \be \int\limits_a^\infty e^{iv^2}\;dv=O(a^{-1})\hspace{0.5cm}(a\to\infty)\nn\ee
and \be \int\limits_0^\infty e^{iv^2}\;dv=\frac{1}{2}\sqrt{\pi} e^{i\pi/4}\nn\ee we estimate the integral $\int_0^{\textrm{Re\,}\sqrt{2x/3}}e^{iv^2}\;dv$ in \eqref{contribution2a} by
\be \int\limits_0^a e^{iv^2}\;dv=\frac{1}{2}\sqrt{\pi} e^{i\pi/4}+O(a^{-1})\hspace{0.5cm}(a\to\infty)\,.\label{int_0_a}\ee
With the substitution $v:=a-ib t$ we obtain
\be \int\limits_a^{a-ib} e^{iv^2}\;dv&=-ib\int\limits_0^1 e^{i(a^2-t^2 b^2-2itab)}\;dt\nn\ee and thus
\be\Big|\int\limits_a^{a-ib} e^{iv^2}\;dv\Big|&\leq b\int\limits_0^1 e^{2abt}dt=\frac{1}{2a}\,(e^{2ab}-1)\leq\frac{1}{2a}\,e^{2ab}\stackrel{\eqref{a,b}}{=}\frac{1}{2a}\,|e^{2ix/3}|\,.
\label{2.40}\ee
Note that, by \eqref{a,b} and \eqref{x_and_u},
\be 0<a^{-1}=\frac{\sqrt{3f}}{\sqrt{2}}\;\frac{1}{\textrm{Re\,}(z^{3/4})}=O(\sqrt{f})\hspace{0.5cm}(f\downarrow0)\,,\nn\ee uniformly for $z\in M$.

Finally combining \eqref{2.40}, \eqref{int_0_a}, \eqref{a,b}, \eqref{contribution2a}, \eqref{E_2_estimate} and \eqref{contribution2} gives  \be I_4(z,f)=
\Big(\frac{1}{2} \sqrt{\frac{\pi f}{z^{1/2}}}\,e^{i\pi/4}\stackrel{\wedge}{\varphi}(-\sqrt{z})+O(f)\Big) e^{-i\frac{2}{3f}\,z^{3/2}}+O(f)\label{I_4estimate}\ee as $f\downarrow0$, \unif.\medskip\\

\noindent\bit Integration near $-\sqrt{z}$\eit:\quad Define the integral
\be I_5(z,f):=\int\limits_{\widetilde{\Gamma}_-}e^{-\frac{i}{f}\left(\frac{k^3}{3}-kz\right)}\stackrel{\wedge}{\varphi}(k)\;dk\hspace{0.5cm}(z\in M\,,\ f>0)
\label{I_5}\ee
along a particular path $\widetilde{\Gamma}_-$ near $k=-\sqrt{z}$. We write $k=\zeta-\sqrt{z}$ which gives \be \frac{2iz^{3/2}}{3}-i\Big(\frac{k^3}{3}-kz\Big)=-i\Big(\frac{\zeta^3}{3}-\sqrt{z}\zeta^2\Big).
\label{change_of_variables_2}\ee Using the method of steepest descents we want $\zeta$ to be on a curve, $\Gamma_-$, satisfying \be\textrm{Re\,}\Big(\frac{\zeta^3}{3}-\sqrt{z}\zeta^2\Big)=0.\nn\ee With the abbreviations \eqref{abbreviations} we want $-\delta_2\leq x\leq 0$ for some sufficiently small $\delta_2$ and $y\leq0$, and we obtain for $\zeta=x+iy$ on $\Gamma_-$
\be y=\gamma^{-1}(|z|^{1/2}+\nu)x+\sum\limits_{n=2}^\infty \widetilde{A}_n x^n \hspace{0.5cm}(x\in[-\delta_2,0])
\label{power3}\ee for some $\widetilde{A}_n\in\mathbb{R}$. On this curve we have \be -i\Big(\frac{\zeta^3}{3}-\sqrt{z}\zeta^2\Big)=-\widetilde{b}^2 x^2 +\sum\limits_{n=3}^\infty \widetilde{B}_n x^n\,,\label{power4}\ee where
\be \widetilde{b}^2:=2\gamma^{-2}|z|(|z|^{1/2}+\nu)\nn\ee and $\widetilde{B}_n$ is real.
The power series in \eqref{power3} and \eqref{power4} converge for $|x|\leq\gamma$. One has
\be e^{i\frac{2}{3f}z^{3/2}}\int\limits_{\widetilde{\Gamma}_-}e^{-\frac{i}{f}\left(\frac{k^3}{3}-kz\right)}
\stackrel{\wedge}{\varphi}(k)\;dk \stackrel{\eqref{change_of_variables_2}}{=}\int\limits_{\Gamma_-}e^{-\frac{i}{f}\left(\frac{\zeta^3}{3}-\sqrt{z}\zeta^2\right)}
\stackrel{\wedge}{\varphi}(\zeta-\sqrt{z})\;d\zeta\,,\label{bb}\ee where $\Gamma_-$ is given by \eqref{power3}. Substituting \eqref{power3} and \eqref{power4} into \eqref{bb} and expanding $\stackrel{\wedge}{\varphi}(\zeta-\sqrt{z})$ in a power series we obtain
\be \int\limits_{\Gamma_-}e^{-\frac{i}{f}\left(\frac{\zeta^3}{3}-\sqrt{z}\zeta^2\right)}
\stackrel{\wedge}{\varphi}(\zeta-\sqrt{z})\;d\zeta
= \frac{1}{2} \sqrt{\frac{\pi f}{z^{1/2}}}\,e^{i\pi/4}\stackrel{\wedge}{\varphi}(-\sqrt{z})+O(f)\label{use_of_power_exp2}\ee as $f\downarrow0$, \unif, and thus
\be I_5(z,f)=\Big(\frac{1}{2} \sqrt{\frac{\pi f}{z^{1/2}}}\,e^{i\pi/4}\stackrel{\wedge}{\varphi}(-\sqrt{z})+O(f)\Big)e^{-i\frac{2}{3f}z^{3/2}}\hspace{0.5cm}(f\downarrow0)\,,
\label{I_5estimate}\ee \unif.\\

\noindent\bit Connecting $\widetilde{\Gamma}_-$ to infinity\eit:\quad We give an exact expression for the path $\Gamma_-$, defined in \eqref{power3}, as follows:
\be y=x\left(\frac{\nu}{\gamma-x}+\sqrt{\left(\frac{\nu}{\gamma-x}\right)^2+\frac{\gamma-\frac{x}{3}}{\gamma-x}}\right)\,.
\label{Gamma_-}\ee
We continue the path $\Gamma_-$ from its end point $\zeta(-\delta_2)=-t_0-i\Theta_0$ (some $t_0,\Theta_0>0$), so that $\zeta(t)=-t-i\Theta_0$, $t\geq t_0$. For this $\zeta(t)$ $(t\geq t_0)$ we calculate
\begin{align}
\textrm{Re\,}\left(-i\left(\frac{\zeta^3(t)}{3}-\sqrt{z}\zeta^2(t)\right)\right)&=
\textrm{Re\,}\left(-i(t^2-\Theta_0^2+2it\Theta_0)
\Big(-\Big(\frac{t}{3}+\gamma\Big)+i\Big(\nu-\frac{\Theta_0}{3}\Big)\Big)\right)\nn\\
&=(t^2-\Theta_0^2)\Big(\nu-\frac{\Theta_0}{3}\Big)-2t\Theta_0\Big(\frac{t}{3}+\gamma\Big)\nn\\
&= -t^2(\Theta_0-\nu)-2\gamma t\Theta_0-\Theta_0^2\Big(\nu-\frac{\Theta_0}{3}\Big)\nn\\
&=:-H(t)\label{H(t)}
\end{align} and \be H'(t)=2t(\Theta_0-\nu)+2\gamma\Theta_0\,,\quad H''(t)=2t(\Theta_0-\nu).
\label{H'}\ee
From the theory behind steepest descents it follows that $H(t_0)>0$ for $t_0>0$. If $\Theta_0>\nu$, then $H(t)$ is positive and increasing for $t\geq t_0$.

Equation \eqref{Gamma_-} with $y=-\Theta_0$ and $x=-t_0$ gives
\be
\Theta_0=t_0\left(\frac{\nu}{\gamma+t_0}+\sqrt{\left(\frac{\nu}{\gamma+t_0}\right)^2+\frac{\gamma+\frac{t_0}{3}}{\gamma+t_0}}\right)\,.\label{*}\ee
From \eqref{*} it follows (after some calculations) that $t_0>\gamma\nu|z|^{-1/2}$ implies $\Theta_0>\nu$. (Note that $\gamma\nu|z|^{-1/2}<\nu$ for $z\in M$.) Furthermore, since $(\gamma+\frac{t_0}{3})(\gamma+t_0)^{-1}>1/3$, it follows that \be \Theta_0>\frac{t_0}{\sqrt{3}}\,.\label{Theta_0>}\ee We now have to pick $t_0$ and $\Theta_0$ such that $\Theta_0-\nu>0$ and the path $-t-i\Theta_0-\sqrt{z}$ $(t\geq t_0)$ is in the region of analyticity of $\stackrel{\wedge}{\varphi}$.\medskip\\
\indent We fix $\varepsilon>0$ sufficiently small and first consider the case $\nu<\varepsilon$:\quad If $\nu<\varepsilon$, set \be t_0:=2\varepsilon\sqrt{3}\,.\label{t_0<}\ee Then
\be \Theta_0-\nu\stackrel{\eqref{Theta_0>}}{>}\frac{t_0}{\sqrt{3}}-\nu\stackrel{\eqref{t_0<}}{=}2\varepsilon-\nu>\varepsilon>0.\nn\ee Furthermore, since $\gamma\geq\varepsilon_2$ (by the definition of $M$ in Theorem \ref{theorem:resonances_instable}), equation \eqref{*} implies
\be \Theta_0<2\varepsilon(1+\sqrt{3}).\nn\ee Then, if $\varepsilon>0$ is small compared to $k_0$ and much smaller than $\varepsilon_2$, the path $-t-i\Theta_0-(\gamma-i\nu)$ $(t\geq t_0)$ is in the region of analyticity of $\stackrel{\wedge}{\varphi}$.\medskip\\
\indent We now consider the case $\nu\geq\varepsilon$:\quad If $\nu\geq\varepsilon$, we set \be\gamma=\alpha\nu\,,\label{gamma}\ee where the number $\alpha=\gamma/\nu$ is bounded above and below (by the definition of $M$ in Theorem \ref{theorem:resonances_instable} and because of $\nu\geq\varepsilon$).
We claim that there exist a $\lambda>0$ such that for \be t_0:=\lambda\gamma\label{t_0>}\ee one has $\Theta_0=\nu(1+\varepsilon)$. (Then, clearly, $\Theta_0-\nu>0$.) Indeed, dividing \eqref{*} by $\nu$ and then inserting \eqref{t_0>} and \eqref{gamma} gives \be \frac{\Theta_0}{\nu}=\frac{\lambda}{1+\lambda}   \sqrt{\left(\frac{\lambda}{1+\lambda}\right)^2+(\alpha\lambda)^2\frac{1+\frac{\lambda}{3}}{1+\lambda}}\ =:\ g(\lambda)\,.\nn\ee
Note that $g$ is continuous, $g(0)=0$ and $\lim_{\lambda\to\infty}g(\lambda)=\infty$. Thus, by the mean value theorem, there exists a $\lambda>0$ for which $g(\lambda)=1+\varepsilon$.\\ Choosing $\varepsilon>0$ such that $\Theta_0=\nu(1+\varepsilon)<k_0$ keeps $-t-i\Theta_0-(\gamma-i\nu)$ $(t\geq t_0)$ in the region of analyticity of $\stackrel{\wedge}{\varphi}$.

Now, by choosing $t_0$ and $\Theta_0$ in the way explained above, the path
$-t-i\Theta_0-\sqrt{z}$ ($t\geq t_0$, $z\in M$) is in the region of analyticity for $\stackrel{\wedge}{\varphi}$, $H(t)$ $(t\geq t_0)$ is positive and increasing, and we obtain, for $z\in M$ and $f>0$,
\begin{align}
&I_6(z,f):= e^{-i\frac{2}{3f}}z^{3/2} \int\limits_{-t_0-i\Theta_0}^{-\infty-i\Theta_0}e^{-\frac{i}{f}\left(\frac{\zeta^3}{3}-\sqrt{z}\zeta^2\right)}\stackrel{\wedge}{\varphi}(\zeta-\sqrt{z})\;d\zeta\,,
\label{I_6}\\
&|I_6(z,f)|\stackrel{\eqref{H(t)}}{\leq} \int\limits_{t_0}^\infty e^{-H(t)/f} |\stackrel{\wedge}{\varphi}(-t-i\Theta_0-\sqrt{z})|\;dt\,.\label{I_6_est1}\end{align}
Since $H''(t)>0$ $(t\geq t_0)$, we have \be H(t)>H(t_0)+(t-t_0) H'(t_0)\,.\label{2.54a}\ee Then, by use of \eqref{2.54a} and \eqref{H'} in \eqref{I_6_est1}, we get
\begin{align}|I_6(z,f)|\leq e^{-H(t_0)/f}\frac{1}{2\gamma\Theta_0} \sup\limits_{t\geq t_0}|\stackrel{\wedge}{\varphi}(-t-\sqrt{z}-i\Theta_0)|\hspace{0.5cm}(z\in M\,,\ f>0)\,.\label{I_6estimate-0}\end{align}
We see that from the method of steepest descents the exponent $\widetilde{b}^2x^2-\sum_{n=3}^\infty \widetilde{B}_nx^n$ from \eqref{power4} is increasing while $\widetilde{b}^2\geq 2\gamma$ so that $H(t_0)$ is bounded below independent of $\nu$. Then \be |I_6(z,f)| \leq e^{-c/f}\label{I_6estimate}\ee for some $c>0$ (independent of $f$) and all $f>0$, uniformly for $z\in M$.\medskip\\

We thus obtain \eqref{int_minus} by combining \eqref{I_6estimate}, \eqref{I_5estimate} and \eqref{I_4estimate}.
Estimate \eqref{int_minus2} easily follows from \eqref{int_minus} by noticing that $|e^{-i\frac{2}{3f}z^{3/2}}|\leq1$ $(z\in M,\ f>0)$.\\
\Done

We now consider the resolvent matrix element $(\varphi\,,\,(p^2+fx-z)^{-1}\varphi)_{L^2}$ for $z\in\mathbb{C}_-$:
\begin{proposition}\label{proposition:resolventLHP}
\begin{lquote} Let $\varphi$ and $M$ be as in Theorem \ref{theorem:resonances_instable}. Then
\begin{align}\hspace{-1cm}(\varphi\,&,\,(p^2+fx-z)^{-1}\varphi)_{L^2}=(\varphi\,,\,(p^2-z)^{-1}\varphi)_{L^2}\nn\\ & -\frac{i}{f}\int\limits_{-\infty}^0e^{-\frac{i}{f}\left(\frac{k^3}{3}-zk\right)}
\stackrel{\wedge}{\varphi}(k)\;dk\int\limits_{-\infty}^0e^{-\frac{i}{f}\left(\frac{k^3}{3}-zk\right)}
\overline{\stackrel{\wedge}{\varphi}}(-k)\;dk\nn\\
&+\frac{1}{z}\Big(  \int\limits_{-\infty}^0e^{-\frac{i}{f}\left(\frac{k^3}{3}-zk\right)}\overline{\stackrel{\wedge}{\varphi}}(-k)\;dk\stackrel{\wedge}{\varphi}(+0)+
\int\limits_{-\infty}^0e^{-\frac{i}{f}\left(\frac{k^3}{3}-zk\right)}\stackrel{\wedge}{\varphi}(k)\;dk\,\overline{\stackrel{\wedge}{\varphi}}(-0)
\Big)\nn\\
& + O(f)\hspace{0.5cm}(f\downarrow0)\,,\label{resexp}
\end{align} uniformly for $z\in M$. In particular,
\begin{align}
(\varphi\,,\,(p^2+fx-z)^{-1}\varphi)_{L^2}=&\ (\varphi\,,\,(p^2-z)^{-1}\varphi)_{L^2}\nn\\
&+\frac{\pi}{\sqrt{z}}\stackrel{\wedge}{\varphi}(-\sqrt{z})
\overline{\stackrel{\wedge}{\varphi}}(\overline{\sqrt{z}}) e^{-i\,\frac{4}{3f}\,z^{3/2}} + O(\sqrt{f})
\label{res-est}\end{align} as $f\downarrow0$, \unif.
\end{lquote}\end{proposition}

\begin{remark}\label{remark:phihat(0)}
\begin{lquote} For $\varphi\in\mathcal{D}_{\theta_0}$, $\stackrel{\wedge}{\varphi}$ is not necessarily continuous at zero; cf. Remark \ref{remark:1}.1. However, by \eqref{sup}, the left and right limits $\stackrel{\wedge}{\varphi}(\mp 0)$  and $\overline{\stackrel{\wedge}{\varphi}}(\mp 0)$ exist.
\end{lquote}
\end{remark}

\noindent\textbf{Proof of Proposition \ref{proposition:resolventLHP}:}\quad By contour deformation and Proposition \ref{proposition:integrals}, \begin{align} \int\limits_{-\infty}^0 e^{-\frac{i}{f}\left(\frac{k^3}{3}-zk\right)}\stackrel{\wedge}{\varphi}(k)\;dk
&=\Big(\sqrt{\frac{\pi f}{z^{1/2}}}e^{i\pi/4}\stackrel{\wedge}{\varphi}(-\sqrt{z})+O(f)\Big)e^{-i\frac{2}{3f}z^{3/2}}+O(f)\,,\label{nonbar}\\
\int\limits_{-\infty}^0 e^{-\frac{i}{f}\left(\frac{k^3}{3}-zk\right)}\overline{\stackrel{\wedge}{\varphi}}(-k)\;dk &= \overline{\int\limits_{0}^\infty e^{-\frac{i}{f}\left(\frac{k^3}{3}-\overline{z}k\right)}\stackrel{\wedge}{\varphi}(k)\;dk}\nn\\
&=\Big(\sqrt{\frac{\pi f}{z^{1/2}}}e^{i\pi/4}\overline{\stackrel{\wedge}{\varphi}}(\overline{\sqrt{z}})+O(f)\Big)e^{-i\frac{2}{3f}z^{3/2}}+O(f)
\label{bar}\end{align} as $f\downarrow0$, \unif. Then \eqref{res-est} is proven by inserting \eqref{nonbar} and \eqref{bar} into \eqref{resexp}.

It remains to prove the expansion \eqref{resexp}: For $z\in\mathbb{C}_-$ the resolvent matrixelement $(\varphi\,,\,(p^2+fx-z)^{-1}\varphi)_{L^2}$ is given by
\begin{align}
(\varphi\,,\,(p^2+fx-z)^{-1}\varphi)_{L^2}&=\frac{1}{f}(\psi_f\,,\,(x-z/f)^{-1}\psi_f)_{L^2}\nn\\ &=\frac{1}{f}\big(\stackrel{\wedge}{\psi}_f\,,\,((x-z/f)^{-1}\psi_f)^\wedge\big)_{L^2}
\nn\\
&= -\frac{i}{f}\,\iint\limits_{k_2>k_1} e^{\frac{i}{f}\left(\frac{k_2^3}{3}-zk_2\right)}e^{-\frac{i}{f}\left(\frac{k_1^3}{3}-zk_1\right)}
\overline{\stackrel{\wedge}{\varphi}}(k_2)\stackrel{\wedge}{\varphi}(k_1)\;dk_1\;dk_2\,.\label{15a}\end{align}
The last identity in \eqref{15a} can be seen as follows: One has
\begin{align} &\big(\stackrel{\wedge}{\psi}_f\,,\,((\,\cdot\,-z/f)^{-1}\psi_f)^\wedge\big)_{L^2}=\lim\limits_{R\to\infty}\int\limits_\mathbb{R} \overline{I_{\{|k|\leq R\}}(k)\stackrel{\wedge}{\psi}_f(k)}((\,\cdot\,-z/f)^{-1}\psi_f)^\wedge(k)\;dk\,,\nn\\
&\label{i1}\end{align} where $I_{\{|k|\leq R\}}$ denotes the indicator function on the set $\{k\in\mathbb{R}\,|\,|k|\leq R\}$. We claim that
\be ((\,\cdot\,-z/f)^{-1}\psi_f)^\wedge=\frac{1}{\sqrt{2\pi}}\left((\,\cdot\,-z/f)^{-1}\right)^{\wedge}*\stackrel{\wedge}{\psi}_f\ \in\  L^\infty(\mathbb{R})\quad(z\in\mathbb{C}_-,\ f>0)\,,\label{i2}\ee where $*$ denotes convolution. Indeed, for $z\in\mathbb{C}_-$, one finds
\begin{align}\mathcal{J}_z(\xi):=&\ \left((\,\cdot\,-z/f)^{-1}\right)^{\wedge}(\xi)=\frac{1}{\sqrt{2\pi}}\lim\limits_{R\to\infty} \int\limits_\mathbb{R} I_{\{|x|\leq R\}}(x)(x-z/f)^{-1} e^{-i\xi x}\;dx\nn\\
=&\  \frac{1}{\sqrt{2\pi}}(-i\pi) e^{-\frac{i}{f}z|\xi|}(1+\textrm{sign}(\xi))\,,\label{i3}\end{align} where
\be \textrm{sign}(\xi):=\left\{\begin{array}{r@{\,,\quad }l} 1&\xi>0
\\0&\xi=0\\-1 & \xi<0\end{array}\right.\,.\nn\ee Since $\mathcal{J}_z\in L^2(\mathbb{R})$ and $\stackrel{\wedge}{\psi}_f\in L^2(\mathbb{R})$, one has (see, e.g., \cite[Chapter IX.4]{RS2}) \begin{align}\begin{array}{l}
\mathcal{J}_z(k_2-\,\cdot\,)\stackrel{\wedge}{\psi}_f(\,\cdot\,)\in L^1(\mathbb{R}_{k_1},dk_1)\quad(k_2\in\mathbb{R})\,,\\
 (\mathcal{J}_z*\stackrel{\wedge}{\psi}_f)(\,\cdot\,):=\int\limits_\mathbb{R} \mathcal{J}_z(\,\cdot\,-k_1)\stackrel{\wedge}{\psi}_f(k_1)\;dk_1\ \in\ L^\infty(\mathbb{R}_{k_2},dk_2)\,,\end{array}\label{i5}\end{align} which proves \eqref{i2}. Now, combining \eqref{i1}, \eqref{i2} and \eqref{i3} gives
\be \big(\stackrel{\wedge}{\psi}_f\,,\,((\,\cdot\,-z/f)^{-1}\psi_f)^\wedge\big)_{L^2}=\frac{1}{\sqrt{2\pi}}\lim\limits_{R\to\infty}\int\limits_\mathbb{R} I_{\{|k_2|\leq R\}}(k_2)\overline{\stackrel{\wedge}{\psi}}_f(k_2)(\mathcal{J}_z*\stackrel{\wedge}{\psi}_f)(k_2)\;dk_2\,.\nn\\
\label{i6}\ee We use \eqref{i5} in \eqref{i6} and Fubini's Theorem to get \be &&\hspace{-1.5cm}\big(\stackrel{\wedge}{\psi}_f\,,\,((\,\cdot\,-z/f)^{-1}\psi_f)^\wedge\big)_{L^2}\nn\\
&=&\frac{1}{\sqrt{2\pi}}\lim\limits_{R\to\infty}\int\limits_\mathbb{R} \int\limits_\mathbb{R} I_{\{|k_2|\leq R\}}(k_2)\overline{\stackrel{\wedge}{\psi}}_f(k_2)\mathcal{J}_z(k_2-k_1)\stackrel{\wedge}{\psi}_f(k_1)\;dk_1\;dk_2\nn\\
&\stackrel{\eqref{i3}}{=}&\frac{1}{2\pi}\lim\limits_{R\to\infty}\iint\limits_{k_2>k_1}I_{\{|k_2|\leq R\}}(k_2)\overline{\stackrel{\wedge}{\psi}}_f(k_2)
(-2i\pi)e^{-\frac{i}{f}z(k_2-k_1)}\stackrel{\wedge}{\psi}_f(k_1)\;dk_1\;dk_2\nn\\
&\stackrel{\eqref{psi_f_consequ}}{=}&-i\iint\limits_{k_2>k_1}e^{i\frac{k_2^3}{3f}}\,\overline{\stackrel{\wedge}{\phi}}(k_2)e^{-\frac{i}{f}z(k_2-k_1)}
e^{-i\frac{k_1^3}{3f}}\stackrel{\wedge}{\phi}(k_1)\;dk_1\;dk_2\,,\nn
\ee which is \eqref{15a}.\\ Then, in \eqref{15a},
we split up the region of integration into the three components
\be \{(k_1\,,\,k_2)\in\mathbb{R}\times\mathbb{R}\,|\,k_2\geq k_1\}=\{k_2\geq k_1\geq 0\}\cup\{k_1\leq0 \wedge k_2\geq0\}\cup\{k_1\leq k_2\leq0\}\,.\nn\ee
Then a symmetry argument gives
\begin{align}
-\frac{i}{f}\,\iint\limits_{k_2>k_1}& e^{\frac{i}{f}\left(\frac{k_2^3}{3}-zk_2\right)}e^{-\frac{i}{f}\left(\frac{k_1^3}{3}-zk_1\right)}
\overline{\stackrel{\wedge}{\varphi}}(k_2)\stackrel{\wedge}{\varphi}(k_1)\;dk_1\;dk_2\nn\\
&=-\frac{i}{f}\,\int\limits_{-\infty}^0 e^{-\frac{i}{f}\left(\frac{k_1^3}{3}-zk_1\right)}
\stackrel{\wedge}{\varphi}(k_1)\;dk_1\,\int\limits_{-\infty}^0e^{-\frac{i}{f}\left(\frac{k_2^3}{3}-zk_2\right)}
\overline{\stackrel{\wedge}{\varphi}}(-k_2)\;dk_2
+ \mathcal{I}(z,f)\label{15}\end{align} for $z\in\mathbb{C}_-$, where
\begin{align}&\mathcal{I}(z,f):=-\frac{i}{f}\,\iint\limits_{k_2>k_1>0} e^{\frac{i}{f}\left(\frac{k_2^3}{3}-zk_2\right)}e^{-\frac{i}{f}\left(\frac{k_1^3}{3}-zk_1\right)}
\psi(k_1,k_2)\;dk_1\;dk_2\,,\label{I}\\
&\psi(k_1,k_2):=\overline{\stackrel{\wedge}{\varphi}}(k_2)\stackrel{\wedge}{\varphi}(k_1)+
\overline{\stackrel{\wedge}{\varphi}}(-k_1)\stackrel{\wedge}{\varphi}(-k_2)\label{psi(k,k)}\end{align} for $z\in\mathbb{C}_-$. (For later use note that $\psi(0,k_2)=\lim\limits_{k_1\downarrow0}\psi(k_1,k_2)$, $k_1,k_2>0$; cf. Remark \ref{remark:phihat(0)}.)

Next, our main aim is to show that, for $z\in M$, to lowest order in the expression \eqref{I} we can drop $(k_2^3-k_1^3)/3f$ in the exponent: Let $I_\varepsilon:=(\gamma-\varepsilon\,,\,\gamma+\varepsilon)$ for some $\varepsilon>0$ sufficiently small, where $\gamma\stackrel{\eqref{abbreviations}}{=}\textrm{Re\,}\sqrt{z}$ $(z\in M)$, and $\chi\in C_0^\infty(I_\varepsilon)$ with $\chi(k)=1$ for $k\in I_{\varepsilon/2}$.
Let \be \widetilde{\chi}(\cdot):=1-\chi(\cdot)\,,\hspace{0.5cm}\Theta(\cdot):=1_{\{k>0\}}(\cdot)
\,,\hspace{0.5cm}e_\pm(k):=e^{\pm \frac{i}{f}\left(\frac{k^3}{3}-zk\right)}\,.\label{chi_et_al}\ee
By \eqref{I} and \eqref{chi_et_al} one gets
\be\mathcal{I}(z,f)=\widetilde{J}(z,f)+J(z,f)\hspace{0.5cm}(z\in M\,,\ f>0)\,,\label{I.2}\ee
where
\begin{align}
&\widetilde{J}(z,f):=-\frac{i}{f}\,\iint\limits_{k_2>k_1>0} \Theta(k_2-k_1)\widetilde{\chi}(k_1)e_+(k_2)e_-(k_1)\psi(k_1,k_2)\;dk_1\;dk_2\,,\label{Jtilde}\\
&J(z,f):=-\frac{i}{f}\,\iint\limits_{k_2>k_1>0} \Theta(k_2-k_1)\chi(k_1)e_+(k_2)e_-(k_1)\psi(k_1,k_2)\;dk_1\;dk_2\label{J}\,.
\end{align}
Using
\be e_-(k_1)\stackrel{\eqref{chi_et_al}}{=}\frac{if}{k_1^2-z}\,\frac{\partial}{\partial k_1}\,e_-(k_1)\label{e1}\ee
in \eqref{Jtilde} one finds, for $z\in M$ and $f>0$,
\be
\widetilde{J}(z,f)=\iint\limits_{k_2>k_1>0} \Theta(k_2-k_1)\widetilde{\chi}(k_1)e_+(k_2)\frac{\psi(k_1,k_2)}{k_1^2-z}\Big(\frac{\partial}{\partial k_1}e_-(k_1)\Big)\;dk_1\;dk_2\,.
\ee
Then partial integration (against $k_1\in[0,k_2]$) gives
\begin{align}
\widetilde{J}(z,f)=&\int\limits_{k_2>0} \widetilde{\chi}(k_2)\frac{\psi(k_2,k_2)}{k_2^2-z}\;dk_2+\frac{\widetilde{\chi}(0)}{z}\int\limits_{k_2>0} e_+(k_2)\psi(0,k_2)\;dk_2\nn\\
&\ -\iint\limits_{k_2>k_1>0} \Theta(k_2-k_1)e_+(k_2)e_-(k_1)\Big( \frac{\partial}{\partial k_1}\,\frac{\widetilde{\chi}(k_1)\psi(k_1,k_2)}{k_1^2-z}\Big)\;dk_1\;dk_2\,,\label{17}
\end{align} where $\widetilde{\chi}(0)=1$. Integrating by parts once more in the last term of \eqref{17} gives
\begin{align} \widetilde{J}(z,f)&=-\frac{i}{f}\,\iint\limits_{k_2>k_1>0} \Theta(k_2-k_1)\widetilde{\chi}(k_1)e_+(k_2)e_-(k_1)\psi(k_1,k_2)\;dk_1\;dk_2\nn\\
&=\int\limits_{k_2>0}\widetilde{\chi}(k_2)\frac{\psi(k_2,k_2)}{k_2^2-z}\;dk_2 + \frac{1}{z}\int\limits_{k_2>0}e_+(k_2)\psi(0,k_2)\;dk_2 + O(f)\label{Jtilde_est}\end{align} as $f\downarrow0$. Thus by use of \eqref{Jtilde_est} in \eqref{I.2} we obtain
\be
\mathcal{I}(z,f)= J(z,f) + \int\limits_{k>0}\widetilde{\chi}(k)\frac{\psi(k,k)}{k^2-z}\;dk + \frac{1}{z}\int\limits_{k>0}e_+(k)\psi(0,k)\;dk + O(f)\label{I.3}\ee
as $f\downarrow0$. Combining \eqref{chi_et_al}, \eqref{J}, \eqref{e1} and \eqref{I.3} yields
\be
\mathcal{I}(z,f)&=&-\frac{i}{f}\,\iint\limits_{k_2>k_1>0} \Theta(k_2-k_1)\chi(k_1)\chi(k_2)e_+(k_2)e_-(k_1)\psi(k_1,k_2)\;dk_1\;dk_2 \nn\\
&&- \iint\limits_{k_2>k_1>0} \Theta(k_2-k_1)\frac{\chi(k_1)\widetilde{\chi}(k_2)}{k_2^2-z}\Big(\frac{\partial}{\partial k_2}  e_+(k_2)\Big) e_-(k_1)\psi(k_1,k_2)\;dk_1\;dk_2\nn\\
&&+ \int\limits_{k>0}\widetilde{\chi}(k)\frac{\psi(k,k)}{k^2-z}\;dk + \frac{1}{z}\int\limits_{k>0}e_+(k)\psi(0,k)\;dk + O(f)\label{19}\ee as $f\downarrow0$.
Integrating by parts twice in the second term on the r.h.s. of \eqref{19} gives
\be \mathcal{I}(z,f)= I_\chi(z,f)+ \int\limits_{k>0} (1-\chi^2(k)) \frac{\psi(k,k)}{k^2-z}\;dk +  \frac{1}{z}\int\limits_{k>0}e_+(k)\psi(0,k)\;dk + O(f) \label{20}\ee as $f\downarrow0$, where
\be I_\chi(z,f):=-\frac{i}{f}\,\iint\limits_{k_2>k_1>0} \Theta(k_2-k_1)\chi(k_1)\chi(k_2)e_+(k_2)e_-(k_1)\psi(k_1,k_2)\;dk_1\;dk_2\,.\label{21}\ee
(Recall that $\chi$ is supported around $\gamma$, away from zero.) In \eqref{21} we make the change of variables
\be k:=k_2-k_1\,,\quad y:=\frac{k_1+k_2}{2} \textrm{\quad such that\quad }k_1=y-\frac{k}{2}\textrm{\quad and\quad }k_2=y+\frac{k}{2}\,,\nn\ee
and a further change $k\mapsto fk$. We obtain, for $z\in M$ and $f>0$,
\begin{align} I_\chi(z,f)
&= -i\int\limits_{\mathbb{R}_y}\int\limits_{k=0}^\infty e^{i(y^2-z)k}e^{ik^3f^2/12}F(y,fk)\;dy\;dk\,,
\label{I_chi_1}\end{align} where
\be F(y,k):= \chi\Big(y-\frac{k}{2}\Big)\chi\Big(y+\frac{k}{2}\Big)\psi\Big(y-\frac{k}{2},y+\frac{k}{2}\Big)\hspace{0.5cm}(y,k\in\mathbb{R})\,.\label{F}\ee
Since $\chi\in C_0^\infty(\mathbb{R})$, one has $F\in C^\infty(\mathbb{R}^2,\mathbb{C})$ with
\be\textrm{supp\,}F\subset\{\xi\in\mathbb{R}^2\,|\,|\xi|\leq R\}\quad\textrm{for some $R>0$ sufficiently large.}\nn\ee  Thus
\be |F(y,fk)|\leq \|F(\cdot,\cdot)\|_\infty<\infty\hspace{0.5cm}(y\in\mathbb{R},\ k\in\mathbb{R},\ f>0)\label{F<infty}\ee and similarly
\be |\partial_{2}F(y,fk)|\leq\|\partial_{2}F(\cdot,\cdot)\|_\infty<\infty\hspace{0.5cm}(y\in\mathbb{R},\ k\in\mathbb{R},\ f>0)\,,\label{F_2<infty}\ee where $\partial_2$ denotes the partial derivative with respect to the second entry. We iteratively use the identity
\be \frac{e^{iky^2}}{y^n}=\frac{1}{2ik}\,\Big(\frac{\partial}{\partial y}+\frac{n+1}{y}\Big)\frac{e^{iky^2}}{y^{n+1}}\hspace{0.5cm}(n\in\mathbb{N})\label{identity}\nn\ee
to get
\be e^{iky^2}= \frac{1}{(2ik)^3}\Big(\frac{\partial}{\partial y}+\frac{1}{y}\Big)\Big(\frac{\partial}{\partial y}+\frac{2}{y}\Big)\Big(\frac{\partial}{\partial y}+\frac{3}{y}\Big)\frac{e^{iky^2}}{y^3}.\label{to_get}\ee
Then inserting \eqref{to_get} into \eqref{I_chi_1} gives, for $z\in M$ and $f>0$,
\begin{align} I_\chi&(z,f)
= -i\int\limits_{\mathbb{R}_y}\int\limits_{k=0}^1 e^{i(y^2-z)k}e^{ik^3f^2/12}F(y,fk)\;dk\;dy\nn\\
& -i\int\limits_{\mathbb{R}_y}\int\limits_{k=1}^\infty \frac{e^{ik^3 f^2/12}}{(2ik)^3}F(y,fk)\Big(\frac{\partial}{\partial y}+\frac{1}{y}\Big)\Big(\frac{\partial}{\partial y}+\frac{2}{y}\Big)\Big(\frac{\partial}{\partial y}+\frac{3}{y}\Big)\frac{e^{i(y^2-z)k}}{y^3}\;dk\;dy.
\label{split_int}\end{align}
By use of \eqref{F<infty} and \eqref{F_2<infty}, Taylor's Theorem with an expansion in $fk$ around zero gives
\be e^{ik^3f^2/12}F(y,fk)= F(y,0)+ O(f)\hspace{0.5cm}(f\downarrow0)\,,\label{Taylor_fk}\ee uniformly for $k\in[0,1]$ and $y\in\mathbb{R}$.
Analogously,
\be \frac{e^{ik^3 f^2/12}}{(2ik)^3}F(y,fk)=\frac{1}{(2ik)^3}\,F(y,0)+O(f)\hspace{0.5cm}(f\downarrow0)\,,\label{Taylor_fk_2}\ee uniformly for $y\in\mathbb{R}$ and $k\in(1,\infty)$. For $z\in M$, inserting \eqref{Taylor_fk_2} and \eqref{Taylor_fk} into \eqref{split_int} yields
\begin{align}
I_\chi&(z,f)=-i\int\limits_{\mathbb{R}}\int\limits_{k=0}^1 e^{i(y^2-z)k}F(y,0)\;dk\;dy + O(f)\nn\\
&- i\int\limits_{\mathbb{R}}\int\limits_{k=1}^\infty \frac{e^{i(y^2-z)k}}{y^3(2ik)^3}\Big(-\frac{\partial}{\partial y}+\frac{1}{y}\Big)\Big(-\frac{\partial}{\partial y}+\frac{2}{y}\Big)\Big(-\frac{\partial}{\partial y}+\frac{3}{y}\Big)F(y,0)\;dk\;dy +O(f)\nn\\
&= -i\int\limits_{\mathbb{R}}\int\limits_{k=0}^1 e^{i(y^2-z)k}F(y,0)\;dk\;dy - i\int\limits_{\mathbb{R}}\int\limits_{k=1}^\infty e^{i(y^2-z)k}F(y,0)\;dk\;dy +O(f)\label{int_2}\end{align} as $f\downarrow0$, uniformly for $z\in M$. Since
\begin{align}
&\int\limits_0^\infty e^{i(y^2-z)k}\;dk=\frac{i}{y^2-z}\hspace{0.5cm}(z\in\mathbb{C}_-\,,\ y\in\mathbb{R})\,,\nn\end{align}
it follows from \eqref{int_2} that
\begin{align}I_\chi(z,f)&=\int\limits_{\mathbb{R}}\frac{F(y,0)}{y^2-z}\;dy + O(f)\stackrel{\eqref{F}}{=}
\int\limits_{y>0}\frac{\chi^2(y)\psi(y,y)}{y^2-z}\;dy + O(f)
\label{I_est}\end{align}
as $f\downarrow0$, uniformly for $z\in M$. Then combining \eqref{I_est} and \eqref{20} gives
\begin{align}
\mathcal{I}(z,f)=&\int\limits_{k>0} \frac{\psi(k,k)}{k^2-z}\;dk + \frac{1}{z}\int\limits_{k>0} e_+(k)\psi(0,k)\;dk + O(f)\hspace{0.5cm}(f\downarrow0)\,,
\label{calI}\end{align} uniformly for $z\in M$, where $\psi(0,k)=\lim_{k_1\downarrow0}\psi(k_1,k)$ and
\be\int\limits_{k>0} \frac{\psi(k,k)}{k^2-z}\;dk&\stackrel{\eqref{psi(k,k)}}{=}&\int\limits_{0}^\infty\frac{\overline{\stackrel{\wedge}{\varphi}}(k)\stackrel{\wedge}{\varphi}(k)+
\overline{\stackrel{\wedge}{\varphi}}(-k)\stackrel{\wedge}{\varphi}(-k)}{k^2-z}\;dk
=\int\limits_{-\infty}^\infty\frac{|\stackrel{\wedge}{\varphi}(k)|^2}{k^2-z}\;dk\nn\\
&=& (\varphi\,,\,(p^2-z)^{-1}\varphi)_{L^2}\hspace{0.5cm}(z\in\mathbb{C}_-)\,.\label{int=res}
\ee By \eqref{chi_et_al}, \eqref{psi(k,k)} and a change of variables $k\mapsto -k$, one finds
\begin{align}
\int\limits_{k>0} e_+(k)\psi(0,k)\;dk=&\int\limits_{-\infty}^0 e^{-\frac{i}{f}\left(\frac{k^3}{3}-zk\right)}\overline{\stackrel{\wedge}{\varphi}}(-k)\;dk\stackrel{\wedge}{\varphi}(+0)\nn\\
&+ \int\limits_{-\infty}^0 e^{-\frac{i}{f}\left(\frac{k^3}{3}-zk\right)}\stackrel{\wedge}{\varphi}(k)\;dk\,\overline{\stackrel{\wedge}{\varphi}}(-0)\,.\label{2139}
\end{align} Then combining \eqref{2139}, \eqref{int=res}, \eqref{calI}, \eqref{15} and \eqref{15a} proves \eqref{resexp}.\\\Done

For the continued resolvent matrix element one has:
\begin{proposition}\label{proposition:resolvent_continued}
\begin{lquote} Let $\varphi$ and $M$ be as in Theorem \ref{theorem:resonances_instable}. Then
\begin{align}\hspace{-1cm} (\varphi\,,&\,(p^2+fx-z)^{-1}\varphi)_{L^2}^{\mathbf{c},\Omega_f}=\frac{i}{f}
\int\limits_{\Gamma'}e^{-\frac{i}{f}\left(\frac{k^3}{3}-kz\right)}
\overline{\stackrel{\wedge}{\varphi}}(-\overline{k})\;dk
\int\limits_{\Gamma'}e^{-\frac{i}{f}\left(\frac{k^3}{3}-kz\right)}
\stackrel{\wedge}{\varphi}(k)\;dk\nn\\
&-\frac{i}{f}
\int\limits_{k<0}e^{-\frac{i}{f}\left(\frac{k^3}{3}-kz\right)}
\overline{\stackrel{\wedge}{\varphi}}(-k)\;dk
\int\limits_{k<0}e^{-\frac{i}{f}\left(\frac{k^3}{3}-kz\right)}
\stackrel{\wedge}{\varphi}(k)\;dk\nn\\
& +\frac{1}{z}\Big(  \int\limits_{k<0}e^{-\frac{i}{f}\left(\frac{k^3}{3}-zk\right)}\overline{\stackrel{\wedge}{\varphi}}(-k)\;dk\stackrel{\wedge}{\varphi}(+0)+
\int\limits_{k<0}e^{-\frac{i}{f}\left(\frac{k^3}{3}-zk\right)}\stackrel{\wedge}{\varphi}(k)\;dk\,\overline{\stackrel{\wedge}{\varphi}}(-0)
\Big)\nn\\
&+(\varphi\,,\,(p^2-z)^{-1}\varphi)_{L^2} + O(f)\hspace{0.5cm}(f\downarrow0),\label{24}
\end{align} uniformly for $z\in M$, where the contour $\Gamma'$ is the distorted $\mathscr{C}_-\cup\mathscr{C}_+$ as used in the proof of Proposition \ref{proposition:integrals}. ($\mathscr{C}_-\cup\mathscr{C}_+$ is defined in \eqref{curve}.)
\end{lquote}
\end{proposition}

\noindent\textbf{Proof of Proposition \ref{proposition:resolvent_continued}:}\quad Combining \eqref{r_phi_f}, \eqref{cont_resolvent}, \eqref{22.a} and \eqref{resexp} proves \eqref{24}.\\\Done

\noindent\textbf{Proof of Theorem \ref{theorem:resonances_instable}:}\quad By Proposition \ref{proposition:resonance=solution}, the resonances of $H_\varphi(f)$ are the zeros of $F_{f,\varphi}^{\mathbf{c},\Omega_f}$ in the open lower half complex plane. By \eqref{r_phi_f}, these are the solutions in $\mathbb{C}_-$ of
\be F_{f,\varphi}^{\mathbf{c},\Omega_f}(z)=1-z-(\varphi\,,\,(p^2+fx-z)^{-1}\varphi)_{L^2}^{\mathbf{c},\Omega_f}=0.\label{zeros}\ee From the expansion \eqref{24}, \eqref{22.a}, the estimates given in Corollary \ref{corollary:psi}, \eqref{nonbar} and \eqref{bar}, it follows that for $z=r$
\begin{align}
(\varphi\,,\,(p^2+fx-z)^{-1}\varphi)_{L^2}^{\mathbf{c},\Omega_f}=&\Big(\frac{\pi}{\sqrt{z}} \stackrel{\wedge}{\varphi}(\sqrt{z}) \overline{\stackrel{\wedge}{\varphi}}(-\overline{\sqrt{z}})+O(\sqrt{f})\Big) e^{i\frac{4}{3f}z^{3/2}} +O(1)e^{i\frac{2}{3f}z^{3/2}}\,,
\label{25}\end{align}
as $f\downarrow0$, uniformly for $z\in M$, where we have used that $(\varphi\,,\,(p^2-z)^{-1}\varphi)_{L^2}=O(1)$ $(f\downarrow0)$. Thus (because of \eqref{geq_delta}), if $-\textrm{Im\,}z> c_0f$ for $c_0>0$ sufficiently large, the equation \eqref{zeros}
cannot have a solution $z=r$ in the  region $M$ considered in the theorem. This finishes the proof of Theorem \ref{theorem:resonances_instable}.\\\Done

The following corollary is a consequence of \eqref{22.a}, Corollary \ref{corollary:psi}, \eqref{nonbar}, \eqref{bar} and Proposition \ref{proposition:resolvent_continued}:\\

\begin{corollary}\label{corollary:res_c}
\begin{lquote} Let $\varphi$ and $M$ be as in Theorem \ref{theorem:resonances_instable}.
Then
\begin{align}\hspace{-1cm} (\varphi\,,\,(p^2+fx-z)^{-1}\varphi)_{L^2}^{\mathbf{c},\Omega_f}=&\frac{\pi}{\sqrt{z}}
\stackrel{\wedge}{\varphi}(\sqrt{z}) \overline{\stackrel{\wedge}{\varphi}}(-\overline{\sqrt{z}})e^{i\frac{4}{3f}z^{3/2}}\nn\\
&+\frac{i\pi}{\sqrt{z}}\Big(\stackrel{\wedge}{\varphi}(\sqrt{z})\overline{\stackrel{\wedge}{\varphi}}(\overline{\sqrt{z}})+
\stackrel{\wedge}{\varphi}(-\sqrt{z})\overline{\stackrel{\wedge}{\varphi}}(-\overline{\sqrt{z}})\Big)
\nn\\
&+(\varphi\,,\,(p^2-z)^{-1}\varphi)_{L^2}+O(\sqrt{f})\hspace{0.5cm}(f\downarrow0)\,,\label{26}
\end{align} uniformly for $z\in M$.
\end{lquote}
\end{corollary}

\begin{remark}
\begin{lquote}
We have written \eqref{24} with a view toward computation (cf. Section \ref{section:numerical_analysis}). It is better than \eqref{26} in the sense that the error is of order $O(f)$ rather than $O(\sqrt{f})$.
\end{lquote}
\end{remark}

\subsection{Proof of Theorem \ref{theorem:AC}}\label{subsection:AC_proof}

\noindent\textbf{Proof of Theorem \ref{theorem:AC}:}\quad It suffices to show \be
\|K(f,\theta)-K(0,\theta)\|\to0\hspace{0.5cm}(f\downarrow0)
\label{K-K}\ee for all $\theta\in S_{\theta_0}$: By the second resolvent equation it follows from \eqref{K-K} that $K(f,\theta)\to K(0,\theta)$ as $f\downarrow0$ in norm resolvent sense for all $\theta\in S_{\theta_0}$ with $\textrm{Im\,}\theta>0$. In particular, the corresponding spectral projection (Riesz projection) of $K(f,\theta)$ converges in norm to the one of $K(0,\theta)$. In this sense the resonances of $H_\varphi(\,\cdot\,,f)$ are stable.\medskip\\

We shall now prove \eqref{K-K}. If $f=0$, then
\be K(0,\theta)=-i\partial_t+H(\theta)\,,\hspace{0.5cm}
H(\theta):=\begin{pmatrix} e^{-2\theta}p^2 & U(\theta)\varphi \\[1.5ex] (U(\overline{\theta})\varphi\,,\,\cdot\,)_{L^2} & 1 \end{pmatrix}\,,
\label{K0th}\ee which follows from \eqref{K}, \eqref{VHV} and \eqref{UhU}. Analogously to \cite[Lemma 2.7]{Y}, the spectrum of $K(0,\theta)$ is
\begin{align}&\sigma(K(0,\theta))=\{\omega n+z\,|\,z\in\sigma\left(H(\theta)\right)\,,\ n\in\mathbb{Z}\}\,,\nn\end{align}
and $K(0,\theta)$ has nonreal eigenvalues (which are resonances of $H_\varphi(\,\cdot\,,0)$) at $\{\omega n+r_0\,|\,n\in\mathbb{Z}\}$.
It follows from Proposition \ref{proposition:existence_resonance} and dilation analyticity (see \cite{AC})
that
\begin{align}
\sigma\left(H(\theta)\right)&=e^{-i2\textrm{Im\,}\theta}[0,\infty)\cup\{\textrm{nonreal eigenvalues of }H(\theta)\}\nn\\
&\hspace{2.5cm}\cup\{\textrm{real eigenvalues of }H(\theta)\}\nn\\
&=e^{-i2\textrm{Im\,}\theta}[0,\infty)\cup\{\textrm{resonances of }H_\varphi(\,\cdot\,,0)\}\nn\\
&\hspace{2.5cm}\cup\{\textrm{(real) eigenvalues of }H_\varphi(\,\cdot\,,0)\}\,.\nn
\end{align}

By use of \eqref{K0th}, \eqref{VHV}, \eqref{UhU} and \eqref{K}, we calculate \be K(f,\theta)-K(0,\theta)=\begin{pmatrix}\frac{f^2}{2\omega^2}(1+\cos(2\omega t)) & U(\theta)(T(t,f)-1)\varphi \\[1.5ex] (U(\overline{\theta})(T(t,f)-1)\varphi\,,\,\cdot\,)_{L^2} & 0 \end{pmatrix}\nn\ee
and get
\begin{align}
\|K(f,\theta)-K(0,\theta)\|\leq \frac{f^2}{\omega^2}+\sup\limits_{t\in\mathbb{R}}(\| U(\theta)(T(t,f)-1)\varphi\|_2 +\| U(\overline{\theta})(T(t,f)-1)\varphi\|_2)\nn\\
\label{Kdifference}\end{align} for all $\theta\in S_{\theta_0}$. Furthermore,
\begin{align}
\big(U(\theta)T(t,f)\varphi\big)(x)\stackrel{\eqref{unitary_group}}{=}e^{\theta/2}\big(T(t,f)\varphi\big)(e^\theta x)\hspace{0.5cm}(\theta\in S_{\theta_0}\,\ f>0,\ t\in\mathbb{R})
\nn\end{align}
and
\be
(T(t,f)\varphi)(e^\theta x) &\stackrel{\eqref{T}}{=}& \Big(e^{i2f\omega^{-2}\sin(\omega t)p}e^{-if\omega^{-1}\cos(\omega t)\,\cdot}\varphi(\cdot)\Big)(e^\theta x)\nn\\
&=& \Big(e^{-if\omega^{-1}\cos(\omega t)(\,\cdot\,+2f\omega^{-2}\sin(\omega t))}\varphi(\,\cdot\,+2f\omega^{-2}\;\sin(\omega t))\Big)(e^\theta x)\nn\\
&=& e^{-if\omega^{-1}\cos(\omega t)\big(e^{\theta}x+2f\omega^{-2}\sin(\omega t)\big)}\varphi(e^{\theta}x+2f\omega^{-2}\;\sin(\omega t))\nn\\
&=& e^{-if^2\omega^{-3}\sin(2\omega t)}\,e^{-if\omega^{-1}\cos(\omega t)e^\theta x}\varphi(e^{\theta}x+2f\omega^{-2}\;\sin(\omega t))\,,\nn\\
\label{T_phi}\ee
where we have used that $\{e^{i\alpha p}\}_{\alpha\in\mathbb{R}}$ is just the group of translations; in particular \be (e^{i2f\omega^{-2}\sin(\omega t)p}g)(y)=g(y+2f\omega^{-2}\sin(\omega t))\hspace{0.5cm}(g\in L^2(\mathbb{R})\,,\ y\in\mathbb{R})\,.\nn\ee
Since $U(\theta)=U(\textrm{Re\,}\theta)U(i\textrm{Im\,}\theta)$ and $\|U(\textrm{Re\,}\theta)\|=1$ for all  $\theta\in S_{\theta_0}$,  equation \eqref{T_phi} implies
\begin{align} \|U(\theta)&(T(t,f)-1)\varphi\|_2=\|U(i\textrm{Im\,}\theta)(T(t,f)-1)\varphi\|_2\nn\\[1.1ex]
=&\;\big\| e^{-i\frac{\textrm{Im\,}\theta}{2}}\big(e^{-if^2\omega^{-3}\sin(\omega t)}e^{-if\omega^{-1}\cos(\omega t)e^{i\textrm{Im\,}\theta}(\cdot)}\varphi(e^{i\textrm{Im\,}\theta}\cdot\,+2f\omega^{-2}\sin(\omega t))\nn\\
&\hspace{7cm}-\varphi(e^{i\textrm{Im\,}\theta}\cdot\,)\big)\big\|_2\nn\\
=&\;\|e^{-if^2\omega^{-3}\sin(\omega t)}e^{-if\omega^{-1}\cos(\omega t)e^{i\textrm{Im\,}\theta}(\cdot)}\varphi(e^{i\textrm{Im\,}\theta}\cdot\,+2f\omega^{-2}\sin(\omega t))-\varphi(e^{i\textrm{Im\,}\theta}\cdot\,)\|_2\,.
\label{U_norm}\end{align} for all $\theta\in S_{\theta_0}$, $f>0$, $t\in\mathbb{R}$.
Setting
\begin{align} a_f(x)&:=f^2\omega^{-3}\sin(\omega t)+f\omega^{-1}\cos(\omega t)\cos(\textrm{Im\,}\theta)x\,,\label{af}\\
b_f(x)&:=f\omega^{-1}\cos(\omega t)\sin(\textrm{Im\,}\theta)x\label{bf}
\end{align}
gives
\begin{align}
e^{-if^2\omega^{-3}\sin(\omega t)}e^{-if\omega^{-1}\cos(\omega t)e^{i\textrm{Im\,}\theta}x}&=e^{-i f\omega^{-1}\cos(\omega t)(\cos(\textrm{Im\,}\theta)+i\sin(\textrm{Im\,}\theta))x}\nn\\
&=e^{-ia_f(x)}e^{b_f(x)}\,.\label{short_a_b}
\end{align}
Thus, by \eqref{short_a_b}, equation \eqref{U_norm} becomes
\begin{align}\|U(\theta)(T(t,f)-1)\varphi\|_2\ =&\ \big\|\ea\eb \big[\varphi(e^{i\textrm{Im\,}\theta}\cdot\,+2f\omega^{-2}\sin(\omega t))-\varphi(e^{i\textrm{Im\,}\theta}\cdot\,)\nn\\
&\ +\varphi(e^{i\textrm{Im\,}\theta}\cdot\,)\big]-\varphi(e^{i\textrm{Im\,}\theta}
\cdot\,)\big\|_2\nn\\
\leq&\ \big\|\ea\eb \big[\varphi(e^{i\textrm{Im\,}\theta}\cdot\,+2f\omega^{-2}\sin(\omega t))-\varphi(e^{i\textrm{Im\,}\theta}\cdot\,)\big]\,\big\|_2\nn\\
&+\|(\ea\eb-1)\varphi(e^{i\textrm{Im\,}\theta}\cdot\,)\|_2\nn\\
=&\ \big\|\eb \big[\varphi(e^{i\textrm{Im\,}\theta}\cdot\,+2f\omega^{-2}\sin(\omega t))-\varphi(e^{i\textrm{Im\,}\theta}\cdot\,)\big]\big\|_2\nn\\
&+\|(\ea\eb-1)\varphi(e^{i\textrm{Im\,}\theta}\cdot\,)\|_2\,.\label{U-norm2}
\end{align}
Next we will prove that the r.h.s. of \eqref{U-norm2} converges to zero as $f\downarrow0$. We estimate the first summand by
\begin{align}
\big\|\eb &\big[\varphi(e^{i\textrm{Im\,}\theta}\cdot\,+2f\omega^{-2}\sin(\omega t))-\varphi(e^{i\textrm{Im\,}\theta}\cdot\,)\big]\big\|_2\nn\\
\leq& \sup\limits_{|\widetilde{\beta}|\leq 2f/\omega^2}\big\|e^{f\omega^{-1}|\cos(\omega t)|\,|\sin(\textrm{Im\,}\theta)|\,|\,\cdot\,|}\big[\varphi(e^{i\textrm{Im\,}\theta}\cdot\,+\widetilde{\beta})-
\varphi(e^{i\textrm{Im\,}\theta}\cdot\,)\big]\,\big\|_2\nn\\
\leq& \sup\limits_{|\widetilde{\beta}|\leq 2f/\omega^2}\big\|e^{f\omega^{-1}|\sin\theta_0|\,|\,\cdot\,|}\big[\varphi(e^{i\textrm{Im\,}\theta}\cdot\,+\widetilde{\beta})-
\varphi(e^{i\textrm{Im\,}\theta}\cdot\,)\big]\,\big\|_2\label{other}
\end{align} for $0<\theta_0<\pi/2$.
Obviously,
\be\lim\limits_{f\downarrow0}\sup\limits_{|\widetilde{\beta}|\leq 2f/\omega^2}\big\|e^{f\omega^{-1}|\sin\theta_0|\,|\,\cdot\,|}\big[\varphi(e^{i\textrm{Im\,}\theta}\cdot\,+\widetilde{\beta})-
\varphi(e^{i\textrm{Im\,}\theta}\cdot\,)\big]\,\big\|_2=0\,.\label{limsup}\ee
The second summand also converges to zero as $f\downarrow0$:
It follows from \eqref{af} and \eqref{bf} that
\be |(e^{-ia_f(x)}e^{b_f(x)}-1)\varphi(e^{i\textrm{Im\,}\theta}x)|\to 0\hspace{0.5cm} (f\downarrow0)\nn\ee
for all $x\in\mathbb{R}$. Furthermore,
\be
|(e^{-ia_f(x)}e^{b_f(x)}-1)\varphi(e^{i\textrm{Im\,}\theta}x)|^2 &\stackrel{{\eqref{af}\atop\eqref{bf}}}{\leq}&
|(e^{f\omega^{-1}|\cos(\omega t)|\,|\sin(\textrm{Im\,}\theta)|\,|x|}+1)\varphi(e^{i\textrm{Im\,}\theta}x)|^2\nn\\
&\leq& |(e^{f_0\omega^{-1}|\sin\theta_0|\,|x|}+1)\varphi(e^{i\textrm{Im\,}\theta}x)|^2\nn\\
&=:&g(x)\nn
\ee
and $g\in L^1(\mathbb{R})$, since by assumption \eqref{phi_uniformly} the functions $\varphi(e^{\theta}\,\cdot\,)$ and $e^{f_0\omega^{-1}|\sin\theta_0|\,|\,\cdot\,|}\varphi(e^{\theta}\cdot+2\omega^{-2}\beta)$ both are in $L^2(\mathbb{R})$ for all $\beta\in[-f_0,f_0]$ and $\theta\in S_{\theta_0}$. Thus by dominated convergence
\be\|(e^{-ia_f}e^{b_f}-1)\varphi(e^{i\textrm{Im\,}\theta}\cdot)\|_2\to 0\hspace{0.5cm}(f\downarrow0)\,.\label{domi}\ee
Combining \eqref{domi}, \eqref{limsup}, \eqref{other} and \eqref{U-norm2} shows
\be \|U(\theta)(T(t,f)-1)\varphi\|_2\to 0\hspace{0.5cm}(f\downarrow0)\nn\ee for all $t\in\mathbb{R}$, $\theta\in S_{\theta_0}$. This implies
\be \sup\limits_{t\in\mathbb{R}}(\| U(\theta)(T(t,f)-1)\varphi\|_2 +\| U(\overline{\theta})(T(t,f)-1)\varphi\|_2)\to0\hspace{0.5cm}(f\downarrow0)\label{supnorm}\ee for all $\theta\in S_{\theta_0}$. Using \eqref{supnorm} in
\eqref{Kdifference} proves \eqref{K-K}.\\\Done

\section{Numerical analysis of the Friedrichs model}\label{section:numerical_analysis}
The numerical data in this section have been obtained by use of Mathematica 8.\smallskip\\

Let \be\varphi(x):=\mu\,e^{-x^2/2}\hspace{0.5cm}(x\in\mathbb{R},\ \textrm{$\mu>0$ sufficiently small}).\label{phi}\ee Obviously, $\varphi\in S(\mathbb{R})$. ($S(\mathbb{R})$ denotes the Schwartz space.) Furthermore, $\varphi$ is a fixed point of the Fourier transform, i.e., $\varphi=\stackrel{\wedge}{\varphi}$. Thus $\varphi\in\mathcal{T}_{k_0}$ for all $k_0>0$. For this particular $\varphi$ the regions defined in \eqref{Omega} are \be \Omega_f=\mathbb{C}_-\quad (f\geq0)\,.\label{Omega_numerics}\ee

\subsubsection*{Resonances of $H_\varphi$ for small $\mu>0$:}
\begin{figure}[h!]
                \centering
                \includegraphics[angle=0,width=0.8\textwidth]{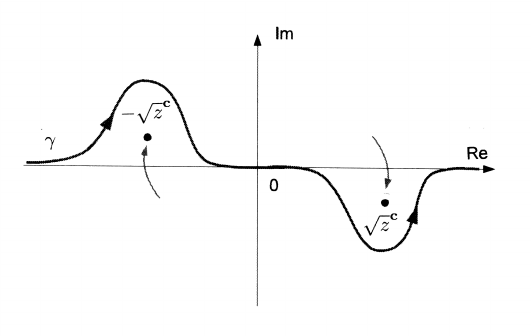}
                \caption{Contour deformation used in order to get \eqref{int_cont1}}
                \label{fig:deform_gamma}\end{figure}
According to Proposition \ref{proposition:resonance=solution}, the resonances of $H_\varphi$, defined in \eqref{H}, are precisely the solutions of \be F^{\mathbf{c},\Omega_0}_{0,\varphi}(z)=1-z-(\varphi\,,\,(p^2-z)^{-1}\varphi)^{\mathbf{c},\Omega_0}_{L^2}=0\label{zeros=resonances2}\ee
in $\mathbb{C}_-$. For the resolvent matrix element one finds
\begin{align}(\varphi\,,\,(p^2-z)^{-1}\varphi)_{L^2}&=(\stackrel{\wedge}{\varphi}\,,\,(k^2-z)^{-1}\stackrel{\wedge}{\varphi})_{L^2}
=\mu^2\int\limits_{\mathbb{R}}\frac{e^{-k^2}}{k^2-z}\;dk\nn\\
&=\mu^2\int\limits_{\mathbb{R}}\frac{e^{-k^2}}{(k-\sqrt{z})(k+\sqrt{z})}\;dk\nn\\
&=\frac{\mu^2}{2\sqrt{z}}\,\int\limits_{\mathbb{R}}e^{-k^2}\Big(\frac{1}{k-\sqrt{z}}-\frac{1}{k+\sqrt{z}}\Big)\;dk\hspace{0.5cm}
(z\in\mathbb{C}_+)\,.
\label{resolventH}
\end{align} The integrand in (\ref{resolventH}) has singularities at $\pm\sqrt{z}$. Now in \eqref{resolventH} we deform the integration path $\mathbb{R}$ into a path $\gamma$ as shown in Figure \ref{fig:deform_gamma} without changing the integral. Then continuation gives
\begin{align} \Big(\int\limits_{\gamma}\frac{e^{-k^2}}{k\mp \sqrt{z}}\;dk\Big)^{\mathbf{c},\Omega_0}&= \pm 2\pi i\, \textrm{Res}\Big(\frac{e^{-k^2}}{k\mp\sqrt{z}},\pm\sqrt{z}\Big)+\int\limits_{\mathbb{R}}\frac{e^{-k^2}}{k\mp\sqrt{z}}\;dk
\hspace{0.5cm}(z\in\mathbb{C}_-)\,,
\label{int_cont1}\end{align} where the residues are
\begin{align} \textrm{Res}\Big(\frac{e^{-k^2}}{k\mp\sqrt{z}},\pm\sqrt{z}\Big)&=e^{-z}\hspace{0.5cm}(z\in\mathbb{C}_-)\,.
\label{Res}\end{align}
Inserting \eqref{Res} and \eqref{int_cont1} into \eqref{resolventH} yields
\begin{align}(\varphi\,,\,(p^2-z)^{-1}\varphi)^{\mathbf{c},\Omega_0}_{L^2}&= \mu^2\,2\pi i\, \frac{e^{-z}}{\sqrt{z}}+\mu^2\int\limits_{\mathbb{R}}\frac{e^{-k^2}}{k^2-z}\;dk\hspace{0.5cm}(z\in\mathbb{C}_-)\,,
\label{resolvent_cont1}\end{align} and
\begin{align}\int\limits_{\mathbb{R}}\frac{e^{-k^2}}{k^2-z}\;dk&= \frac{-\pi e^{-z}}{\sqrt{z}}\Big(i+\frac{\textrm{erf}(i\sqrt{z})}{i}\Big)\hspace{0.5cm}(z\in\mathbb{C}_-)\,,\label{int_cont2}
\end{align} where
\be\textrm{erf}(x):=\frac{2}{\sqrt{\pi}}\int\limits_0^x e^{-t^2}\;dt\hspace{0.5cm}(x\in\mathbb{R})\,.\label{erf}\ee The error function $\textrm{erf}(\cdot)$ has an entire extension (see, e.g., \cite[Chapter 2.4]{O}).

\begin{figure}[h!]
                \centering
                \includegraphics[angle=0,width=0.8\textwidth]{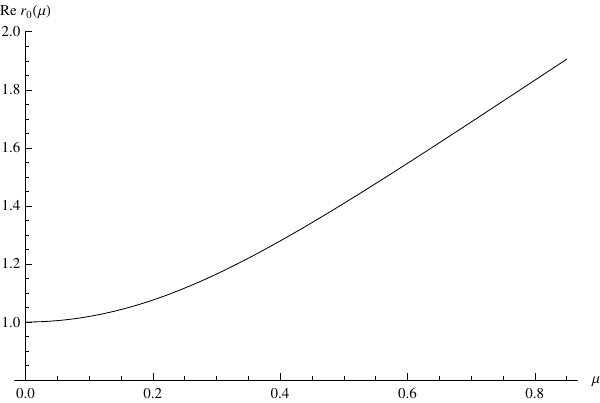}
                \caption{Real part of the resonance $r_0(\cdot)$ of $H_{\varphi}$}
                \label{fig:Re_mu}\end{figure}

\begin{figure}[h!]
                \centering
                \includegraphics[angle=0,width=0.8\textwidth]{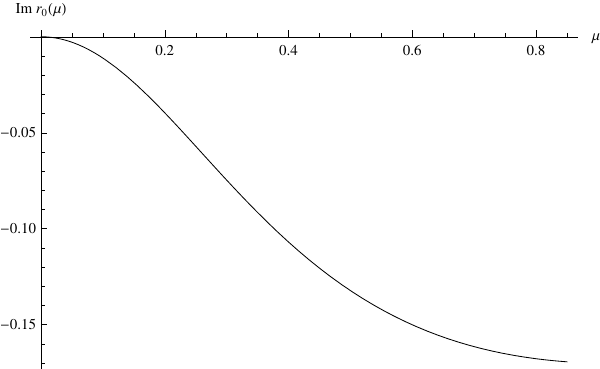}
                \caption{Imaginary part of the resonance $r_0(\cdot)$ of $H_{\varphi}$}
                \label{fig:Im_mu}\end{figure}

\begin{figure}[h!]
                \centering
                \includegraphics[angle=0,width=0.8\textwidth]{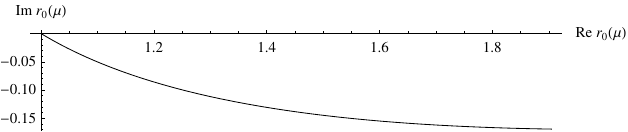}
                \caption{The resonance $r_0(\mu)$ $(0\leq\mu\leq 0.85)$ of $H_{\varphi}$ in the complex plane}
                \label{fig:E_mu_in_C}\end{figure}

Finally, by combining \eqref{phi}, \eqref{zeros=resonances2}, \eqref{resolvent_cont1} and \eqref{int_cont2}, the resonances of $H_\varphi$ are the solutions of
\begin{align} 0&=1-z-\mu^2\,\left(2\pi i\, \frac{e^{-z}}{\sqrt{z}}-\frac{\pi\,e^{-z}}{\sqrt{z}}\,\Big(i+\frac{\textrm{erf}(i\sqrt{z})}{i}\Big)\right)\hspace{0.5cm}
(z\in\mathbb{C}_-).\label{zeros_f=0}
\end{align}
By Proposition \ref{proposition:existence_resonance}, for $\mu>0$ sufficiently small there is exactly one resonance of $H_\varphi$ in a sufficiently small complex neighborhood of 1 (the embedded eigenvalue of $H_0$ defined in \eqref{H_0}).
Let $r_0(\mu)$, $\mu>0$ sufficiently small, denote this unique resonance near the embedded eigenvalue 1 of $H_0$. Solving\footnote{by use of the  \begin{tt}FindRoot\end{tt}-algorithm of Mathematica 8 (which is basically Newton's method)} \eqref{zeros_f=0} for $\mu>0$ sufficiently small reveals the behavior of the resonance $r_0(\mu)$ of $H_\varphi$ as $\mu\downarrow0$ shown in Figures \ref{fig:Re_mu} through \ref{fig:E_mu_in_C}. One finds that the resonance of $H_\varphi$ converges to the original embedded eigenvalue 1 of $H_0$.\enlargethispage{0.3cm}

\noindent For later use we compute
\be r_0\big(\mu=\frac{1}{10}\big)= 1.01905- 0.0111115\,i =: r_0\,.\label{r_0}\ee

\subsubsection*{Resonances of $H_\varphi(f)$:}

Fix $\mu>0$ sufficiently small. Again by Proposition \ref{proposition:resonance=solution} the resonances of $H_\varphi(f)$ $(f>0)$ are the solutions of
\be
F_{\varphi,f}^{\mathbf{c},\Omega_f}(z)=1-z-r_{\varphi,f}^{\mathbf{c},\Omega_f}(z)=0\hspace{.5cm}(z\in\mathbb{C}_-)\,,\label{zeros=resonances3}\ee
where
\be
r_{\varphi,f}^{\mathbf{c},\Omega_f}(z)\stackrel{\eqref{r_phi_f}}{=}(\varphi\,,\,(p^2+fx-z)^{-1}\varphi)^{\mathbf{c},\Omega_f}_{L^2}\hspace{0.3cm}
(z\in\mathbb{C}_-)\,,\hspace{0.5cm}
\varphi(x)\stackrel{\eqref{phi}}{=}\mu e^{-x^2/2}\,.\nn\ee
For this particular $\varphi$ one has
\begin{align}
&r_{\varphi,f}^{\mathbf{c},\Omega_f}(z)=\frac{2\pi i}{f}\,\psi_f^2\Big(\frac{z}{f}\Big)
+ (\varphi\,,\,(p^2+f x-z)^{-1}\varphi)_{L^2}\hspace{0.5cm}(z\in\mathbb{C}_-)\,,\label{cont_resolvent_f}\end{align} where $\psi_f$ is the entire extension of
\be \psi_f(x)\stackrel{{\eqref{psi_f_consequ}\atop\eqref{phi}}}{=}\frac{\mu}{\sqrt{2\pi}}\int\limits_\mathbb{R}e^{-i\left(\frac{k^3}{3f}-kx\right)}
e^{-\frac{k^2}{2}}\;dk \hspace{0.5cm}(x\in\mathbb{R},\ f>0).\label{psi_f_phi}\ee

Note that $\psi_f(z/f)$ is a (highly) oscillatory integral. In order to numerically solve\footnote{again by use of the \begin{tt}FindRoot\end{tt}-algorithm of Mathematica 8} \eqref{zeros=resonances3}, we have substantially used the expansion for $r_{\varphi,f}^{\mathbf{c},\Omega_f}(z)=(\varphi\,,(p^2+f x-z)^{-1}\varphi)^{\mathbf{c},\Omega_f}_{L^2}$ given in Section \ref{subsection:instability}, Proposition \ref{proposition:resolvent_continued}. It follows from \eqref{24} that the error produced by using this expansion in \eqref{zeros=resonances3} is of order $O(\mu^2 f)=O(10^{-2}f)$ as $f\downarrow0$. In addition there is a numerical error produced by Mathematica which is less than $10^{-a}+|r|\,10^{-p}$, where $a=p=53\,\log_{10}(2)\approx16$ (machine precision) and $r$ is the computed zero. (The parameters $a$ and $p$ are called \enquote{accuracy} and \enquote{precision}.) Note that $|r|$ is very close to one.

\begin{figure}[h!]
                \centering
                \includegraphics[angle=0,width=0.8\textwidth]{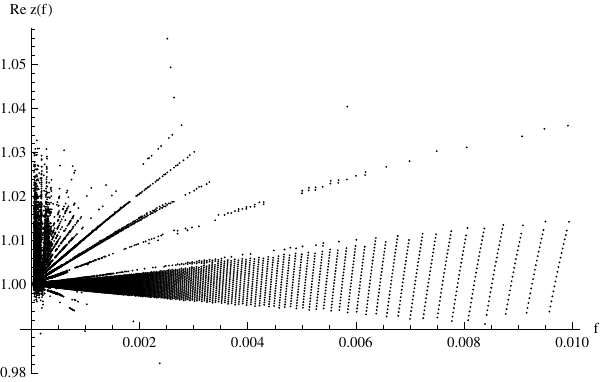}
                \caption{Real part of resonances $r(f)$ of $H_{\varphi}(f)$ vs. $f$; for $\mu=1/10$}
                \label{fig:mod1Re}\end{figure}

\begin{figure}[h!]
                \centering
                \includegraphics[angle=0,width=0.8\textwidth]{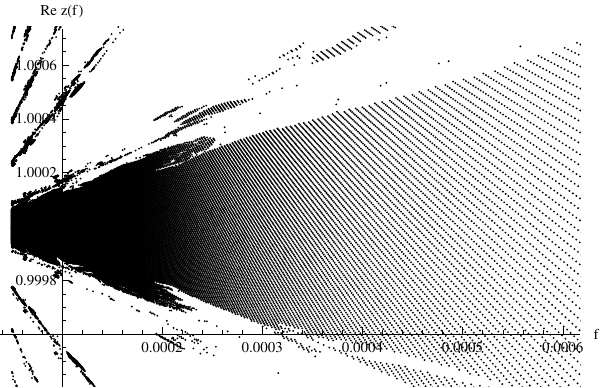}
                \caption{Real part of resonances $r(f)$ of $H_{\varphi}(f)$ vs. $f$ on a finer scale; for $\mu=1/10$}
                \label{fig:mod1Re_larger_scale}\end{figure}

\begin{figure}[h!]
                \centering
                \includegraphics[angle=0,width=0.8\textwidth]{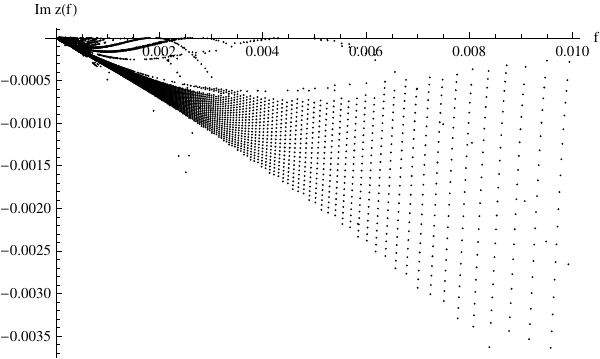}
                \caption{Imaginary part of resonances $r(f)$ of $H_{\varphi}(f)$ vs. $f$; for $\mu=1/10$ }
                \label{fig:mod1Im}\end{figure}

\begin{figure}[h!]
                \centering
                \includegraphics[angle=0,width=0.8\textwidth]{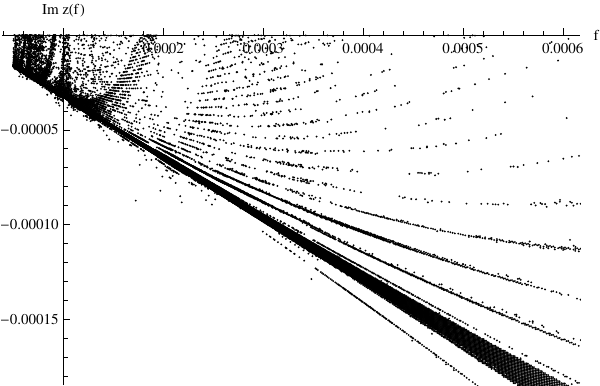}
                \caption{Imaginary part of resonances $r(f)$ of $H_{\varphi}(f)$ vs. $f$ on a finer scale; for $\mu=1/10$ }
                \label{fig:mod1Im_larger_scale}\end{figure}

\subsubsection*{Interpretation of the Figures \ref{fig:mod1Re} -- \ref{fig:mod1Im_larger_scale}:}

In Figures \ref{fig:mod1Re} through \ref{fig:mod1Im_larger_scale} one sees that there exist many resonances of $H_\varphi(f)$ in a neighborhood of the pre-existing resonance $r_0\stackrel{\eqref{r_0}}{=}1.01905- 0.0111115\,i $ of $H_\varphi(0)$. This is in good accordance with the result of Corollary \ref{corollary:infinitely_many_resonances}.

 Figure \ref{fig:mod1Im} and Figure \ref{fig:mod1Im_larger_scale} show that the imaginary part of the resonances $r(f)$ of $H_\varphi(f)$ in a neighborhood of the pre-existing resonance $r_0$ converges to zero as $f\downarrow0$. In particular, $r(f)\not\to r_0$ as $f\downarrow0$. Thus the pre-existing resonance $r_0$ is unstable. This behavior confirms the instability result of Theorem \ref{theorem:resonances_instable}. According to Figure \ref{fig:mod1Re} and Figure \ref{fig:mod1Re_larger_scale}, the real part of the resonances $r(f)$ seems to converge to the original embedded eigenvalue 1 of $H_0$ as $f\downarrow0$. At present it is not clear to the authors how to analytically prove (or disprove) this convergence of the real part.

A priori, the graphs in Figures \ref{fig:mod1Im} and \ref{fig:mod1Im_larger_scale} can be interpreted in two different ways which are hard to distinguish numerically:
\begin{enumerate}
\item In Figure \ref{fig:mod1Im} and Figure \ref{fig:mod1Im_larger_scale} each \enquote{ray} tending to zero is a \enquote{bundle} of imaginary parts, all close to each other. Analogously, in Figure \ref{fig:mod1Re} and Figure \ref{fig:mod1Re_larger_scale} each \enquote{ray} (possibly tending to 1) is a \enquote{bundle} of real parts, all close to each other.
\item In Figure \ref{fig:mod1Im} and Figure \ref{fig:mod1Im_larger_scale} each \enquote{ray} tending to zero is one (highly) oscillatory function of $f$, representing the imaginary part of a single resonance. Analogously, in Figure \ref{fig:mod1Re} and \ref{fig:mod1Re_larger_scale} each \enquote{ray} (possibly tending to 1) is one oscillatory function of $f$, representing the real part of a single resonance.
\end{enumerate}
 It is not clear what kind of objects (in terms of $f$) the real and imaginary parts of the resonances are: Branches of multi-valued analytic functions? Strata?

\newpage

\newpage

\begin{appendix}\label{appendix}

\section{Proof of Corollary \ref{corollary:infinitely_many_resonances}}\label{appendix:infinitely_many_resonances}

By Proposition \ref{proposition:resonance=solution}, the real zeros of $F_{f,\varphi}^{\mathbf{c},\Omega_f}$ $(f>0)$ are precisely the eigenvalues of the self-adjoint operator $H_\varphi(f)$. Thus by Proposition \ref{proposition:infinitely_many_zeros} (and Definition \ref{defi:resonance}) it suffices to show that $H_\varphi(f)$ has at most finitely many eigenvalues:

Let $f>0$, $\varphi\neq0$. Let $\lambda$ be an eigenvalue of $H_\varphi(f)$. Then, for some $\mathscr{H}\ni\begin{pmatrix}u\\ c\end{pmatrix}\neq0$
\be \begin{pmatrix} p^2+fx-\lambda & M_\varphi \\ (\varphi,\,\cdot\,)_{L^2} & 1-\lambda\end{pmatrix} \begin{pmatrix}u\\ c\end{pmatrix} = \begin{pmatrix}0\\0\end{pmatrix}\,.\nn\ee In particular, \begin{align} &(p^2+fx-\lambda)u+c\varphi=0\,,\label{ev1}\\
&(\varphi,u)_{L^2}+c(1-\lambda)=0\,.\label{ev2}\end{align} Since $p^2+fx$ does not have any eigenvalues, equation \eqref{ev1} implies $c\neq0$. W.l.o.g. one can set $c=1$. Then \eqref{ev1} implies that
\be u=-(p^2+fx-\lambda)^{-1}\varphi\label{A3}\ee and $\varphi$ is in the domain of $(p^2+fx-\lambda)^{-1}$. Inserting \eqref{A3} into \eqref{ev2} (with $c=1$) gives \be 1-\lambda-(\varphi\,,\,(p^2+fx-\lambda)^{-1}\varphi)_{L^2}=0\,.\label{A5a}\ee By Lemma \ref{lemma:unitary_equivalence} and \eqref{psi_f_consequ} one has \be (\varphi\,,\,(p^2+fx-\lambda)^{-1}\varphi)_{L^2}=\frac{1}{f}\big(\psi_f\,,\,(x-\frac{\lambda}{f})^{-1}\psi_f\big)_{L^2}\,,\label{matrixelement}\ee where the function $\psi_f$ is entire by Lemma \ref{lemma:psi_f_entire}.
Then
\be \|(p^2+fx-\lambda)^{-1}\varphi\|_2^2=\frac{1}{f}\|(x-\frac{\lambda}{f})^{-1}\psi_f\|_2^2
=\frac{1}{f}\int\limits_{\mathbb{R}}\frac{|\psi_f(x)|^2}{|x-\frac{\lambda}{f}|^2}\;dx<\infty\nn\ee implies that \be \psi_f\Big(\frac{\lambda}{f}\Big)=0\,.\label{psi(lambda)=0}\ee
Thus the set of all eigenvalues $\lambda$ of $H_\varphi(f)$ must be discrete, since otherwise \eqref{psi(lambda)=0} would imply that $\psi_f(x)=0$ for all $x\in\mathbb{R}$.

Next we will prove that for eigenvalues $\lambda$ of $H_\varphi(f)$ the resolvent matrix element \eqref{matrixelement} is bounded in the sense of \eqref{last}, as a function of $\lambda$, varying over the set of eigenvalues. Then \eqref{A5a} (combined with the set of eigenvalues being discrete) implies that there are at most finitely many eigenvalues of $H_\varphi(f)$.

Therefor we write
\begin{align} |(\psi_f\,,\,(x-\frac{\lambda}{f})^{-1}\psi_f)_{L^2}|&=\Big|\Big(\int\limits_{\mathbb{R}\backslash[\frac{\lambda}{f}-1,\frac{\lambda}{f}+1]}+
\int\limits_{\frac{\lambda}{f}-1}^{\frac{\lambda}{f}+1}\Big) \frac{|\psi_f(x)|^2}{x-\frac{\lambda}{f}}\;dx\Big|.\label{A8}\end{align} Clearly,
\be\Big|\int\limits_{\mathbb{R}\backslash[\frac{\lambda}{f}-1,\frac{\lambda}{f}+1]}\frac{|\psi_f(x)|^2}{x-\frac{\lambda}{f}}\;dx\Big|
\leq\int\limits_\mathbb{R}|\psi_f(x)|^2\;dx =\|\psi_f\|^2_2\nn\ee and
\be \int\limits_{\frac{\lambda}{f}-1}^{\frac{\lambda}{f}+1}\frac{|\psi_f(x)|^2}{|x-\frac{\lambda}{f}|}\;dx
\stackrel{\eqref{psi(lambda)=0}}{=}\int\limits_{-1}^1\frac{|\psi_f(x+\frac{\lambda}{f})-\psi_f(\frac{\lambda}{f})|^2}{|x|}\;dx\,.\nn\ee
Using $|e^{ikx}-1|\leq c|x|^{\varepsilon/4}\langle k\rangle^{\varepsilon/4}$, one finds
\begin{align} \Big|\psi_f\Big(x+\frac{\lambda}{f}\Big)-\psi_f\Big(\frac{\lambda}{f}\Big)\Big|&=\frac{1}{\sqrt{2\pi}}\Big|\int\limits_\mathbb{R} e^{ik\frac{\lambda}{f}}(e^{ixk}-1)\stackrel{\wedge}{\psi}_f(k)\;dk\Big|\nn\\
&\leq c |x|^{\varepsilon/4}\int\limits_\mathbb{R} \langle k\rangle^{-\frac{1}{2}-\frac{\varepsilon}{4}} |\langle k\rangle^{\frac{1}{2}+\frac{\varepsilon}{2}}\stackrel{\wedge}{\psi}_f(k)|\;dk\label{A10}
\end{align}  for $x\in\mathbb{R}$, small $f,\varepsilon>0$ and some $c<\infty$. By Schwarz' inequality one has
\begin{align}
\Big|\int\limits_\mathbb{R} \langle k\rangle^{-\frac{1}{2}-\frac{\varepsilon}{4}} |\langle k\rangle^{\frac{1}{2}+\frac{\varepsilon}{2}}\stackrel{\wedge}{\psi}_f(k)|\;dk\Big|&=\big|\big(\langle \,\cdot\,\rangle^{-\frac{1}{2}-\frac{\varepsilon}{4}}\,,\,|\langle \,\cdot\,\rangle^{\frac{1}{2}+\frac{\varepsilon}{2}}\stackrel{\wedge}{\psi}_f|\big)_{L^2}\big|\nn\\
&\leq \|\langle \,\cdot\,\rangle^{-\frac{1}{2}-\frac{\varepsilon}{4}}\|_2\;\|\langle \,\cdot\,\rangle^{\frac{1}{2}+\frac{\varepsilon}{2}}\stackrel{\wedge}{\psi}_f\|_2\,,\label{A11}
\end{align} where
\be \|\langle \,\cdot\,\rangle^{\frac{1}{2}+\frac{\varepsilon}{2}}\stackrel{\wedge}{\psi}_f\|_2=\|\langle p\rangle^{\frac{1}{2}+\frac{\varepsilon}{2}}\psi_f\|_2\,.\label{A13}\ee
Then, by \eqref{A11} and \eqref{A10},
\begin{align}\int\limits_{-1}^1\frac{|\psi_f(x+\frac{\lambda}{f})-\psi_f(\frac{\lambda}{f})|^2}{|x|}\;dx&\leq
c'\|\langle p\rangle^{\frac{1}{2}+\frac{\varepsilon}{2}}\psi_f\|_2^2 \int\limits_{-1}^1 |x|^{-1+\frac{\varepsilon}{2}}\;dx\nn\\
&=\frac{4c'}{\varepsilon}\|\langle p\rangle^{\frac{1}{2}+\frac{\varepsilon}{2}}\psi_f\|_2^2\,.\label{A14}
\end{align}
Combining \eqref{A8} through \eqref{A14} proves
\be
|(\psi_f\,,\,(x-\frac{\lambda}{f})^{-1}\psi_f)_{L^2}|
&\leq& \|\psi_f\|_2^2+ \frac{4c'}{\varepsilon}\|\langle p\rangle^{\frac{1}{2}+\frac{\varepsilon}{2}}\psi_f\|_2^2\label{last2}
\ee for some $c'<\infty$. By \eqref{psi_f_consequ} one has \be\|\psi_f\|_2=\|\varphi\|_2\,,\quad\|\langle p\rangle^{\frac{1}{2}+\frac{\varepsilon}{2}}\psi_f\|_2=\|\langle p\rangle^{\frac{1}{2}+\frac{\varepsilon}{2}}\varphi\|_2\,.\label{A13}\ee
Thus, by combining \eqref{matrixelement} and \eqref{A13}, the estimate \eqref{last2} is equivalent to
\be |(\varphi,(p^2+fx-\lambda)^{-1}\varphi)_{L^2}|\leq\frac{1}{f}\Big(\|\varphi\|_2^2+ \frac{4c'}{\varepsilon}\|\langle p\rangle^{\frac{1}{2}+\frac{\varepsilon}{2}}\varphi\|_2^2\Big)\,.\label{last}\ee \Done We remark that the estimate \eqref{last} also follows using Mourre's method; see \cite{PSS}.

\section{Boundedness of dilation analytic vectors in a sector}\label{dilation analytic vectors}
In this appendix, $A$ denotes the closure of $(px +xp)/2\upharpoonright S(\mathbb{R})$ in $L^2(\mathbb{R})$. The self-adjoint operator $A$ is the infinitesimal generator of the unitary group of dilations on $L^2(\mathbb{R})$ (introduced in \eqref{unitary_group}), $U(\theta)=e^{i\theta A}$ $(\theta\in\mathbb{R})$.
\begin{lemma}\label{boundedness}
\begin{lquote}
Let $f \in \mathcal{D}(A^2)$. Then $f$ is differentiable on $\mathbb{R}\backslash\{0\}$ and
\be |f(x)| \le (2\sqrt{|x|})^{-1}\|(A^2 +1)f\|_2. \label{A_est}\ee
\end{lquote}
\end{lemma}

\noindent\textbf{Proof:}\quad
We define $L^2(\mathbb{R}_+):=\{I_{[0,\infty]}\phi\,|\,\phi\in L^2(\mathbb{R})\}$, where $I_{[0,\infty]}$ denotes the indicator function on the interval $[0,\infty]$. Note that $U(\theta) = e^{i\theta  A}$ $(\theta\in\mathbb{R})$ maps $L^2(\mathbb{R}_+)$ into itself.
Define
\be W:\ L^2(\mathbb{R}_+)\to L^2(\mathbb{R})\,,\quad W\phi(\cdot):=e^{(\cdot)/2}\phi(e^{(\cdot)}).\label{W}\ee
Then $W$ is unitary and
\be(WU(\theta)W^{-1}g)(u) = g(u +\theta)\hspace{1cm}(g\in L^2(\mathbb{R}),\ u,\theta\in\mathbb{R}).\nn\ee
Thus if $f \in \mathcal{D}(A^2)$, the function $WI_{[0,\infty]}f$ is in $H_2(\mathbb{R})$ (the Sobolev space of second order) and is thus a differentiable function.  It follows that $f(x)$ is differentiable for $x>0$. An analogous proof works for $x<0$. This shows the differentiability. Next we will prove \eqref{A_est}. We compute for $f \in L^2(\mathbb{R})$
\begin{align}
((1 + iA)^{-1}f)(x) + ((1 - iA)^{-1}f)(x) & = 2((1 + A^2)^{-1}f)(x)  \notag \\
&  =  \int_{-\infty}^{\infty} e^{-|\theta|} e^{\theta/2} f(e^{\theta} x) d\theta\,.\nn
\end{align}
Thus
\begin{align}
\left |2((1 + A^2)^{-1}f)(x)\right | & \leq\left(\int_{-\infty}^{\infty}e^{-2|\theta|}d\theta\right)^{1/2}  \left(\int_{-\infty}^{\infty} e^{\theta}|f(e^{\theta}x)|^2d\theta\right)^{1/2}\nn\\
&=   \left(\int_{-\infty}^{\infty} e^{\theta}|f(e^{\theta+ u}x/|x|)|^2d\theta\right)^{1/2} \label{A5}\\
&=   |x|^{-1/2}\left(\int_{-\infty}^{\infty}\big|f(yx/|x|)|^2dy\right)^{1/2}  \nn\\
& \le |x|^{-1/2} \|f\|_2\,,\label{B3tilde}
\end{align} where in \eqref{A5} the substitution $|x| = e^u$ was used.
The lemma thus follows.  \\
\Done

\begin{proposition}\label{proposition:B2}
\begin{lquote}
If $f$ is dilation analytic in angle $\theta_0$ in the sense that $e^{\theta_0|A|}f\in L^2(\mathbb{R})$, then $f$ is analytic in the union of sectors $\Gamma_{\theta_0}^\pm := \{z \in\mathbb{C}\,|\, z = \pm |z|e^{i\phi},\  |z| > 0,\ \phi\in\mathbb{R}, |\phi| < \theta_0\}$
and
\be |f(e^{i\phi}x)| \le  (2\sqrt{|x|})^{-1}\|(A^2 +1)e^{|\phi| |A|}f\|_2\hspace{0.5cm}(|\phi| < \theta_0,\ x\in\mathbb{R}\backslash\{0\})\,.\label{estPropB} \ee
\end{lquote}
\end{proposition}

\noindent\textbf{Proof:}\quad
Using the unitary operator $W$ (defined in \eqref{W}) it is easily seen that if $f$ is dilation analytic in angle $\theta_0$, $f$ has a version which is analytic in the set $\Gamma_{\theta_0}:=\Gamma_{\theta_0}^+\cup\Gamma_{\theta_0}^-$. In addition the analytic extension of $U(\theta)f$ for $|\textrm{Im\,}\theta| < \theta_0$ is given by $e^{\theta/2}f(e^{\theta}x)$. Inserting $f=(A^2+1)^{-1}(A^2+1)f$ into \eqref{B3tilde} proves \eqref{estPropB}.\\\Done

\begin{remark}\label{remark:also_phi_hat}
\begin{lquote}
Since $A$ is the closure of $(px+xp)/2\upharpoonright S(\mathbb{R})$ in $L^2(\mathbb{R})$ and $\mathscr{F}A^n\mathscr{F}^{-1}=(-A)^n$ $(n\in\mathbb{N})$, where $\mathscr{F}$ denotes the Fourier transform, the estimates in Lemma \ref{boundedness} and Proposition \ref{proposition:B2} also hold for $\mathscr{F}f=\stackrel{\wedge}{f}$ instead of $f$.
\end{lquote}
\end{remark}

\begin{remark}\label{remark:B2}
\begin{lquote}Let $\mathcal{D}_{\theta_0}$ be the linear space of dilation analytic vectors in the sense of Definition \ref{defi:dilation_analytic_vector}. We remark that $f\in\mathcal{D}_{\theta_0}$ is equivalent to $f$ being an analytic vector for the self-adjoint operator $A$, i.e.,
\be f\in\mathcal{D}(e^{\theta A})\quad\textrm{and}\quad e^{\theta A}f=\sum\limits_{n=0}^\infty\frac{\theta^n}{n!} A^n f\hspace{0.5cm}(\theta\in\mathbb{C}\,,\ |\theta|<\theta_0)\,.\nn\ee For a thorough discussion of analytic vectors we refer the reader to \cite{Nelson} or, e.g., \cite{Weid}.
\end{lquote}
\end{remark}

\end{appendix}

\noindent\textbf{Acknowledgements}:\quad J.R. is grateful to the DFG for making her stay in Charlottesville possible and to Markus Klein for some helpful  discussions!

\end{document}